\documentclass[sn-mathphys,Numbered]{sn-jnl}

\usepackage{graphicx}%
\usepackage{multirow}%
\usepackage{amsmath,amssymb,amsfonts}%
\usepackage{amsthm}%
\usepackage{mathrsfs}%
\usepackage[title]{appendix}%
\usepackage{xcolor}%
\usepackage{textcomp}%
\usepackage{manyfoot}%
\usepackage{booktabs}%
\usepackage{algorithm}%
\usepackage{algorithmicx}%
\usepackage{algpseudocode}%
\usepackage{listings}%

\usepackage{soul}
\usepackage{url}
\usepackage[small]{caption}
\usepackage{graphicx}
\usepackage{mathtools}
\usepackage{booktabs}
\usepackage{comment}
\usepackage{tablefootnote}

\usepackage{stfloats}

\usepackage{siunitx}
\usepackage{adjustbox}
\usepackage{subcaption}
\usepackage{blindtext}
\usepackage{multirow}
\usepackage{diagbox}

\usepackage{amsmath}
\usepackage{amssymb}
\usepackage{amsfonts}
\usepackage[colorinlistoftodos]{todonotes}
\usepackage{xcolor}
\usepackage{xspace}
\usepackage{color, colortbl}

\usepackage[normalem]{ulem} 
\usepackage{booktabs} 
\usepackage{array}
\usepackage{makecell}

\usepackage{natbib}
\usepackage{enumitem}
\setlist{nolistsep} 

\usepackage{tikz,xcolor}
\definecolor{lime}{HTML}{A6CE39}
\DeclareRobustCommand{\orcidicon}
{
    \begin{tikzpicture}
    \draw[lime, fill=lime] (0,0) circle [radius=0.16] 
    node[white] {{\fontfamily{qag}\selectfont \tiny ID}};    \draw[white, fill=white] (-0.0625,0.095) circle [radius=0.007];    
    \end{tikzpicture}
    \hspace{0mm}}
\foreach \x in {A, ..., Z}{%
    \expandafter\xdef\csname orcid\x\endcsname{\noexpand\href{https://orcid.org/\csname orcidauthor\x\endcsname}{\noexpand\orcidicon}}
}

\newcommand{\ece}{early childhood education\xspace}
\newcommand{\ai}{artificial intelligence\xspace}

\newcommand{\irobiq}{iRobiQ\xspace}

\begin{document}

\title[Article Title]{The Key Artificial Intelligence Technologies in Early Childhood Education: A Review}


\author[1,2]{\fnm{Yi} \sur{Honghu}}\email{25010020@pxu.edu.cn}

\author*[3]{\fnm{Liu} \sur{Ting}\hspace{-1.5mm}\orcidC{}\hspace{-1mm}}\email{t.liu@vu.nl}

\author*[4]{\fnm{Lan} \sur{Gongjin}\hspace{-1.5mm}\orcidA{}\hspace{-1mm}}\email{langj@sustech.edu.cn}

\affil[1]{\orgdiv{School of Education (Teachers Colleage)}, \orgname{Guangzhou University}, \orgaddress{\city{Guangzhou}, \postcode{510006}, \country{China}}}

\affil[2]{\orgdiv{School of Preschool Education}, \orgname{Pingxiang University}, \orgaddress{\city{Pingxiang}, \postcode{337055}, \country{China}}}

\affil*[3]{\orgdiv{Department of Computer Science}, \orgname{ VU University
Amsterdam}, \orgaddress{\city{Amsterdam}, \postcode{1081HV}, \country{the Netherlands}}}

\affil*[4]{\orgdiv{Department of Computer Science and Engineering}, \orgname{Southern University of Science and Technology}, \orgaddress{
\city{Shenzhen}, \postcode{518055}, 
\country{China}}}


\abstract{
Artificial Intelligence (AI) technologies have been applied in various domains, including early childhood education (ECE). 
Integration of AI educational technology is a recent significant trend in ECE. 
Currently, there are more and more studies of AI in ECE.
To date, there is a lack of survey articles that discuss the studies of AI in ECE.
In this paper, we provide an up-to-date and in-depth overview of the key AI technologies in ECE that provides a historical perspective, summarizes the representative works, outlines open questions, discuss the trends and challenges through a detailed bibliometric analysis, and provides insightful recommendations for future research.
We mainly discuss the studies that apply AI-based robots and AI technologies to ECE, including improving the social interaction of children with an autism spectrum disorder.
This paper significantly contributes to provide an up-to-date and in-depth survey that is suitable as introductory material for beginners to AI in ECE, as well as supplementary material for advanced users.
}




\maketitle

\section{Introduction}
\label{sec:introduction}

In recent years, Artificial Intelligence (AI) is revolutionizing the way of human life \cite{williams2018popbots}, in particular the education domain. 
In the AI era, using AI technology in education is a significant trend.
AI educational technologies play an important role in modern education by providing unique learning experiences to students and improving their learning \cite{prentzas2013artificial}. 
AI approaches have been incorporated into educational technologies for improving interaction with students and employed in an educational context to satisfy various educational needs \cite{roblyer2007integrating}. 
AI educational technologies provide students and teachers benefits compared to traditional education, including inspiring students' motives for creative activities \cite{prentzas2013artificial}, avoiding shame and embarrassment in front of their peers, assisting teachers to provide students with personalized self-adaptive learning \cite{xu2020dilemma} and remove repetitive tasks and value-add to meaningful teaching tasks.

Early childhood is a wonderful time to spark children's interest in AI \cite{su2022artificial}.
Among all stages of education, early childhood education (ECE) is particularly important because they first learn how to interact with others including peers and teachers, and begin to develop interests that will inspire them throughout their lives.
ECE generally sets the stage for future educational success in secondary education and higher education \cite{gammage2006early}. 
For the formative years of ECE, creating a modern environment where children learn fundamental skills is crucial for their future success. 
Integration of AI concepts in ECE is significantly crucial to how children realize AI in future \cite{su2022artificial}.
Although there are a few existing survey articles that reviewed the studies of AI in related education (e.g., music education, medical education), there is a lack of survey articles that discuss the studies of AI in ECE. 
In this paper, we summarize the representative studies of popular intelligent robots, and main AI technologies in ECE, analyses challenges and trends by bibliometrics, and provides insightful recommendations for future research. 
This paper significantly contributes to provide an up-to-date and in-depth survey that is suitable as introductory material for beginners to AI in ECE, as well as supplementary material for advanced users.

\subsection{Related Review}

In the early literature, Holland Simon \cite{holland2000artificial} reviews the principal approaches of AI in music education.
Drigas et al. \cite{drigas2012artificial,drigas2011review} discuss the studies of AI in special education over the last decade (2001–2010).
In 2013, a survey article \cite{prentzas2013artificial} claims to review the studies of AI methods in early childhood educational technology, which mainly concerns computer-based learning systems rather than focusing on AI in ECE.

\begin{table*}[!ht] \centering \footnotesize
\renewcommand{\arraystretch}{1.0}
\setlength\tabcolsep{3pt} 
\begin{tabular} {l l l c c c} \toprule
Article & methods & topics & Trends & Challenges  & Bibliometrics \\ \midrule
\cite{prentzas2013artificial} & Computer-based & Early childhood education & $-$ & $-$ & $-$ \\
\cite{holland2000artificial} & AI & Music education & $-$ & $-$ & $-$ \\
\cite{drigas2012artificial} \cite{drigas2011review} & AI & Special education & $-$ & $-$ & $-$ \\
\cite{huijnen2017robots} & Robots & ASD & $-$ & $-$ & $-$ \\
\cite{chassignol2018artificial} & AI & AIEd trends & \checkmark  & $-$ & $-$  \\
\cite{zawacki2019systematic} & AI Applications & Higher education & $-$ & \checkmark & $-$ \\
\cite{zaidi2019review} & AI \& Online learning & AIEd market & \checkmark  & \checkmark & $-$ \\
\cite{chan2019applications} & AI & Medical education & $-$ & \checkmark & $-$ \\
\cite{chen2020artificial} & AI & AIEd & $-$ & $-$ & $-$  \\ 
\cite{sapci2020artificial} & AI training \& tools & Medical education & $-$ &  $-$ & $-$ \\
\cite{chen2020application} & AI & AIEd & \checkmark & $-$ & \checkmark  \\
\cite{ahmad2020artificial} & AI & \makecell[l]{Market \& outlook} & \checkmark  & \checkmark & \checkmark \\
\cite{buchanan2021predicted} & AI health technologies & Nursing education & $-$ & $-$ & $-$ \\
\cite{maghsudi2021personalized} & AI/ML & Personalized education & $-$ & \checkmark & $-$ \\
\textbf{Ours} & AI-based & \textbf{Early childhood education} & \pmb{\checkmark} & \pmb{\checkmark} & \pmb{\checkmark} \\ \midrule
\end{tabular}
\caption{Overview of previous review articles of AI in education (AIEd) and the position of ours (this review). $\checkmark$ ($-$) represents that the review paper has (not) discussed the content. }
\label{tab:related-work}
\end{table*}

Recently, Huijnen et al. \cite{huijnen2017robots} review the studies about AI-based robots that are used in therapy and education for children with Autism Spectrum Disorder (ASD).
Chassignol et al. \cite{chassignol2018artificial} identify the prospective impact of AI technologies on the studying process and predict the trends in education.
Chan and Zary \cite{chan2019applications} review the current AI applications and the challenges of implementing AI in medical education.
Zawacki et al. \cite{zawacki2019systematic} review the studies of AI applications in higher education.
Zaidi et al. \cite{zaidi2019review} provides an overview of the effectiveness of online learning and AI in education.
Chen et al. \cite{chen2020artificial} provide an overview to assess the impact of AI in education.
In \cite{sapci2020artificial}, the authors present a review to evaluate the current state of AI training and the use of AI tools to enhance the learning experience for medical and health informatics students.
Chen et al. \cite{chen2020application} review the studies of AI in education in terms of applications and theory gaps.
In \cite{buchanan2021predicted}, the authors summarize the current studies and predict the influences of AI health technologies on nursing education over the next 10 years and beyond.
Maghsudi et al. \cite{maghsudi2021personalized} provide a brief review of the state-of-the-art studies to observe the challenges of AI/ML-based personalized education and discuss potential solutions.

Especially, Ahmad et al. \cite{ahmad2020artificial} review the studies of AI in secondary education and higher education by highlighting the future scope and market opportunities for AI in education, the existing tools, research trends, current limitations, and pitfalls of AI in education. 
Although these existing survey articles review the studies of AI in education including the domains of special education, music education, medical education, and nursing education, there is a lack of survey articles that review the studies of AI in \ece. 
The comparison of the previous surveys of AI in education and the position of our work is shown in \autoref{tab:related-work} which presents the covered topics, whether the research trends and challenges and bibliometrics are included.
In particular, only two articles \cite{chen2020application,ahmad2020artificial} of them provide detailed bibliometrics analysis and many existing review articles \cite{holland2000artificial,drigas2012artificial,drigas2011review,prentzas2013artificial,huijnen2017robots,chen2020artificial,sapci2020artificial,buchanan2021predicted} of them do not even discuss trends and challenges.
Although the article \cite{ahmad2020artificial} discusses trends and challenges with bibliometrics, it focuses on the future scope and market opportunities of AI in secondary education and higher education.
This survey fills a gap in the existing review articles of AI in education.

\subsection{Contributions}

We emphasize that a comprehensive and solid overview of AI in \ece topics should review the representative articles, discuss challenges, provide detailed bibliometrics analysis, and observe trends for further research.
As shown in \autoref{tab:related-work}, although trends and challenges are crucial in an overview, a few current articles analyse both trends and challenges. 
In particular, bibliometrics is rarely applied to analyse the existing studies of AI in ECE.
The main contributions of the work are summarized as follows:
 
\begin{enumerate} [noitemsep]
    \item Delineate the picture of AI in ECE.
    \item Provide a complete/detailed overview of the existing literature on AI in ECE.
    \item Identify the challenges and analyze the research trends for future developments by bibliometrics.
\end{enumerate}

\begin{figure}[!ht] \centering \small 
    \includegraphics[width=0.55\textwidth,trim={5 50 5 16},clip]{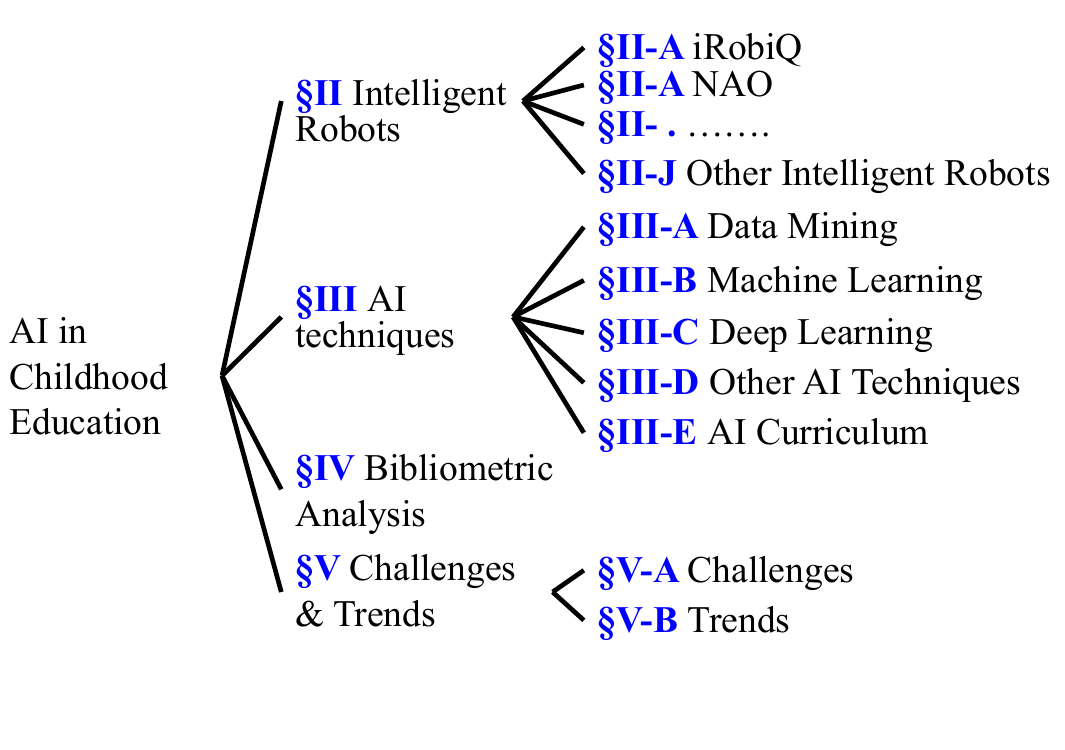}
    \caption{Taxonomy of this survey on AI for \ece.}
    \label{fig:taxonomy}
\end{figure}

This paper provides an up-to-date and in-depth overview that is suitable as introductory material for beginners to AI in ECE, as well as supplementary material for advanced users.
\autoref{fig:taxonomy} shows the taxonomy of this survey, where the picture of AI in ECE is clearly presented for readers to quickly figure out the structure of this survey.
Specifically, the rest of this paper is organized as follows. 
In \autoref{sec:robots}, we present the popular intelligent robots that are used for ECE.
The main AI technologies such as data mining, machine learning, and deep learning for AI in ECE are summarized in \autoref{sec:aitechniques}.
We address the bibliometric analysis in \autoref{sec:bibliometric}.
Challenges and trends are discussed in \autoref{sec:challenges_trends}. 
Finally, we conclude this paper in \autoref{sec:conclusion}.
\vspace{-0.2cm}
 \vspace{-0.3cm}

\section{Bibliometric Analysis}
\label{sec:bibliometric}


Bibliometrics is the use of statistical methods to analyse books, articles and other publications. 
It is an effective way to measure the information of publication in the scientific community \cite{iftikhar2019bibliometric}.
However, current survey papers rarely use bibliometrics to analyse the studies, particularly AI in ECE. 
In general, bibliometric data can be obtained from various databases such as PubMed, Web of Science, and Google Scholar.
In this paper, we choose the Scopus database for literature retrieval.
Scopus is the largest abstract/citation database with peer-reviewed literature and covers a wider journal range, of help both in keyword searching and citation analysis \cite{falagas2008comparison}, which is released by Elsevier. 
Importantly, Elsevier provides a Python library \footnote{\url{https://github.com/pybliometrics-dev/pybliometrics}} to retrieve the data for the expected topics from the Scopus database \cite{rose2019pybliometrics}.
In this work, we review the articles about AI in ECE by retrieving the format of "title, abstract, and keywords" in the Scopus database.

\begin{figure*}[!ht] \centering
    \includegraphics[width=0.75\textwidth,trim={5 10 20 30},clip]{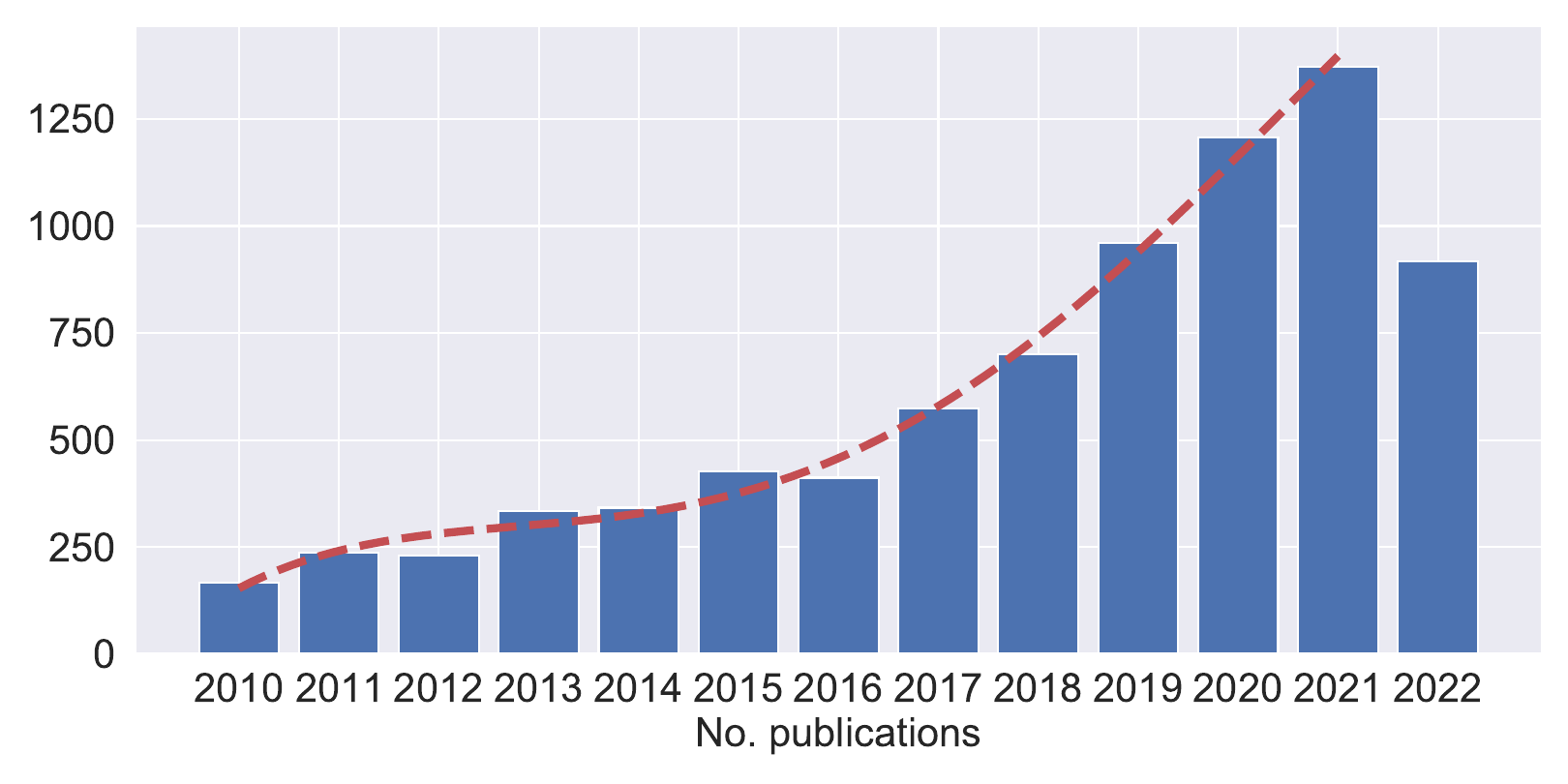}
        \caption{The number of publications per year (from 2010 to 2022) on the topic of AI in ECE. The dashed red lines are the fit of the number of publications. Scopus database returned 9014 results until 10/10/2022.}\label{fig:years}
\end{figure*}

We retrieve the literature on this topic by combining keywords by the search code of title-abs-key(("artificial intelligence" OR "AI" OR "machine learning" OR "deep learning" OR "data mining" OR "virtual reality" OR "natural language processing" OR "robot" OR "robots") AND ("early childhood education" OR "autism" OR "autism spectrum disorder" OR "ASD" OR "preschool" OR "kindergarten")). 
To observe the trends clearly, we retrieve the data for each year over ten years from 2010 to date.
The number of publications over the years (from 2010 to 2022) is shown in \autoref{fig:years}. 
The dashed red lines are the cubic polynomial fit of the number of publications.
The number of publications shows a clear growth trend from 2010 to 2021, while an insignificant drop in 2012 and 2016.
In particular, it shows a significant increase in recent years from 2017 to 2021 which is the outbreak period of AI. 
We therefore conclude that more and more studies apply AI technologies to early childhood education. 
Note that although the number of publications shows a significantly increasing, the total number of studies of AI in ECE is still small.

\vspace{-10pt}
\begin{figure*}[!ht] \centering
    \includegraphics[width=0.6\textwidth,trim={423 15 80 20},clip]{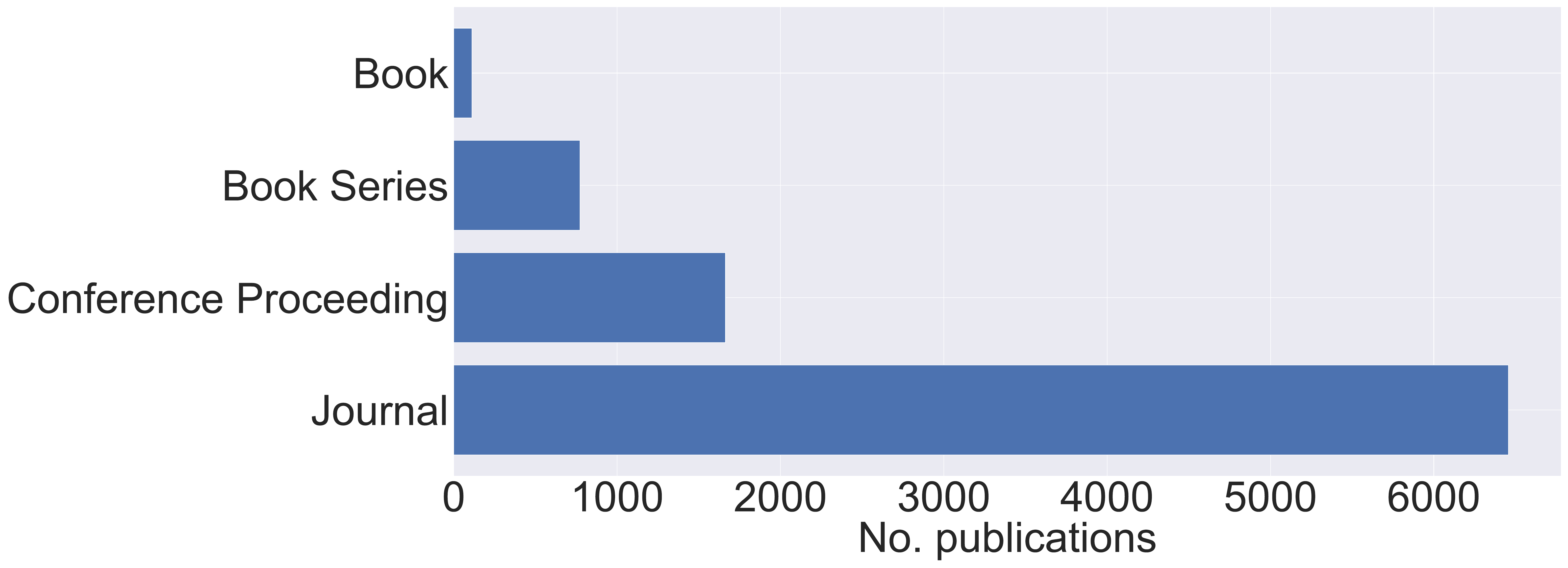} 
        \caption{Aggregation type of publications on the topic of AI in ECE (from 2010 to 10/10/2022).}\label{fig:aggregation}
\end{figure*} 

\begin{figure*}[!ht] \centering
\includegraphics[width=0.98\textwidth,trim={10 15 20 20},clip]{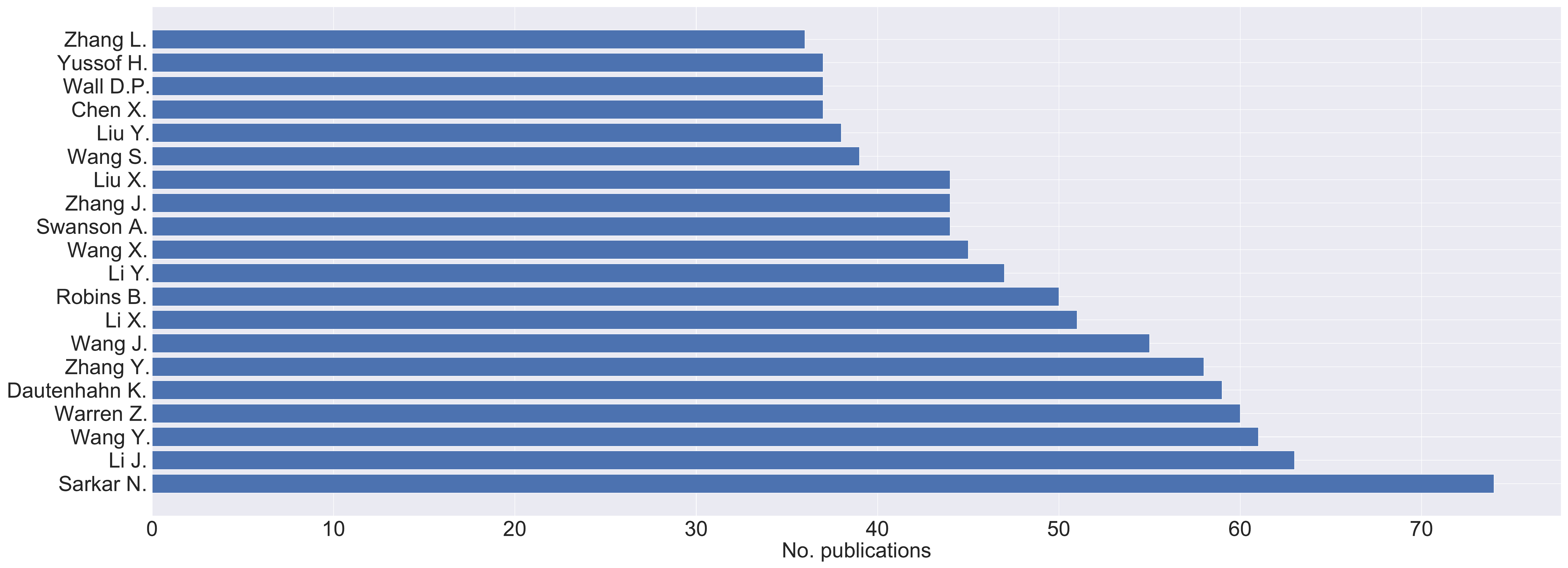}
\caption{Top 20 authors and their number of publications on the topic of AI in ECE (from 2010 to 10/10/2022).}
\label{fig:author}
\end{figure*}

Data aggregation is a crucial step in the literature review, which can be used to provide statistical analysis for collated research data and to create summarized data.
We retrieve the aggregation type of publications from 2010 to 2022, as shown in \autoref{fig:aggregation}. 
The results show that the related works are mainly reported in journals and conference proceedings.
Specifically, there are more than 6000 publications and more than 1000 publications reported in journals and conference proceedings respectively. 
We notice that most publications of AI in ECE are reported in journals for scientific theory presentation rather than technologies and academic exchange at conferences.
We assume that current studies investigate AI in ECE at the level of scientific research rather than practical applicaitons, which is consistent with our conclusion in \autoref{subsec:discussionAI}.
To provide readers with a guideline for following the studies from a researcher who focuses on ECE, we retrieve the publication’s authors.
The top 20 authors and their number of publications on the topic of AI in ECE are shown in \autoref{fig:author}.
All these 20 authors published more than 30 studies on the related topic of AI in ECE. 
We notice that most of them focus on the applications of AI-based intelligent robots rather than AI in ECE. 
Currently, there are a few researchers who focus on AI in ECE.

\begin{figure*}[!ht] \centering
\includegraphics[width=0.98\textwidth,trim={10 15 20 30},clip]{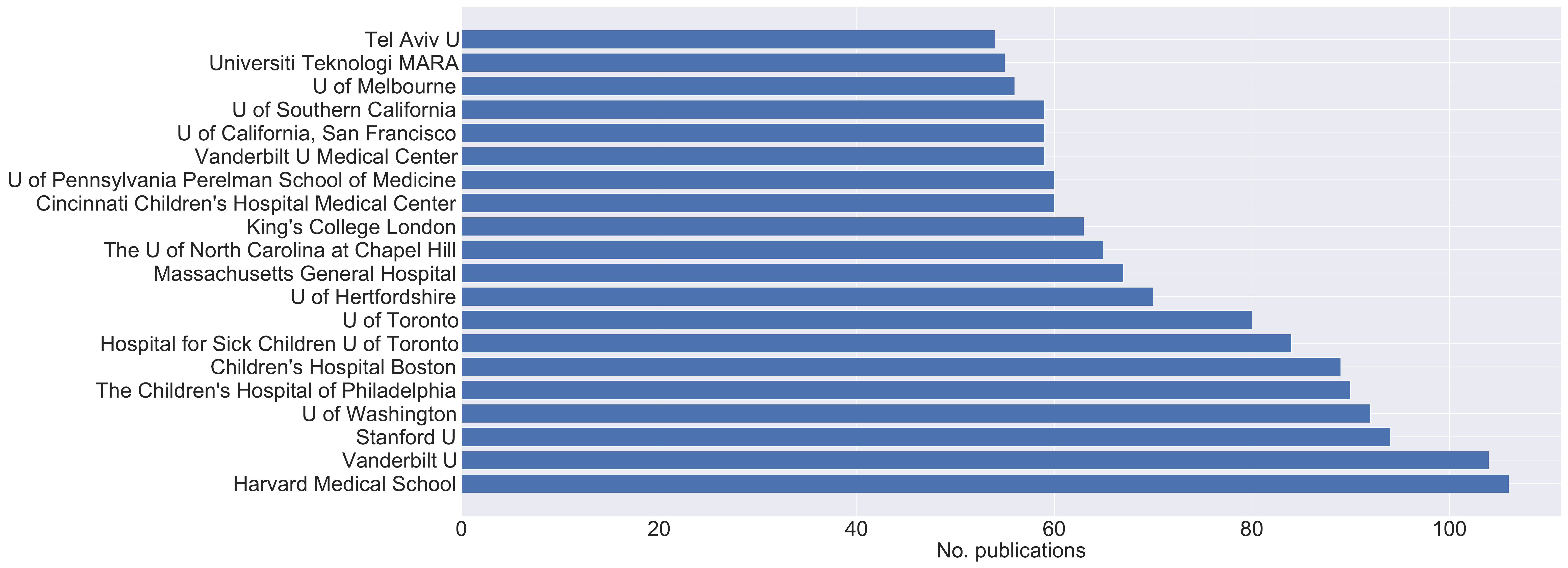}
\caption{Top 20 affiliations and their number of publications on the topic of AI in ECE (from 2010 to 10/10/2022).}
\label{fig:affiliations}
\end{figure*}

\begin{figure*}[!ht] \centering
\includegraphics[width=0.98\textwidth,trim={10 15 30 40},clip]{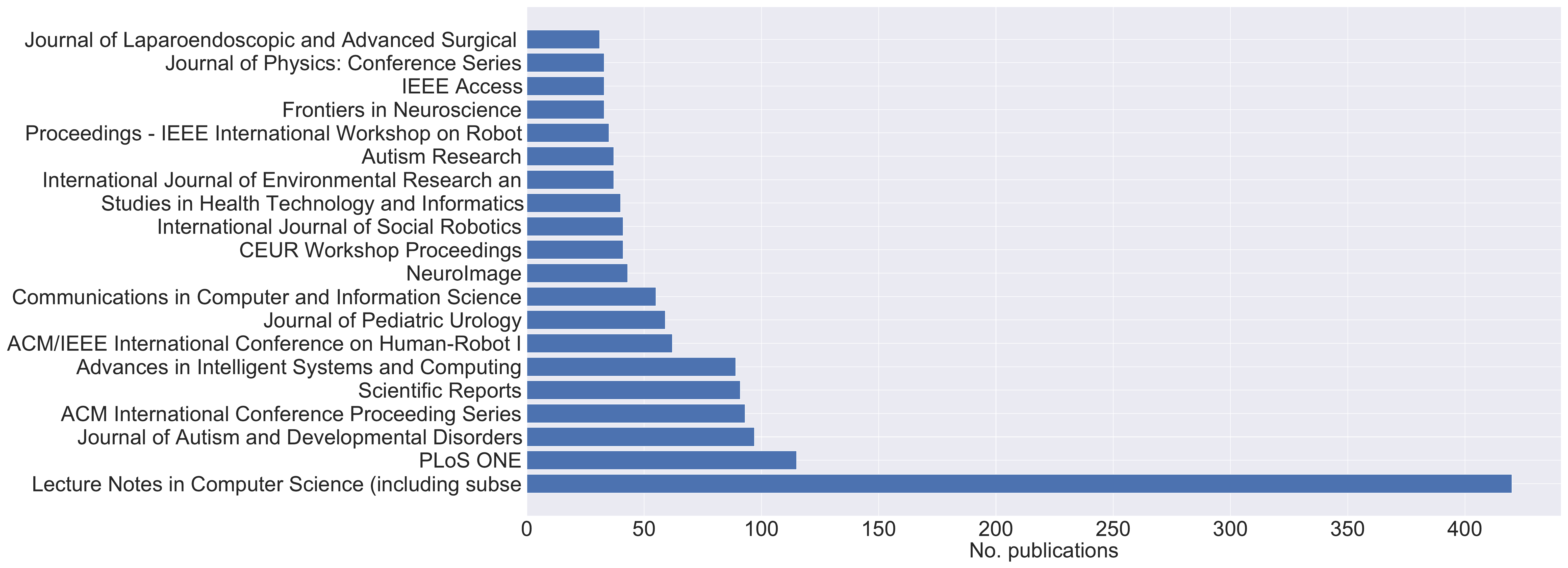}
\caption{The top 20 venues and their number of publications on the topic of AI in ECE (from 2010 to 10/10/2022).}
\label{fig:venues}
\end{figure*}

Moreover, we retrieve the affiliations to reveal the top affiliations.
The top 20 affiliations and their number of publications on the topic of AI in ECE are shown in \autoref{fig:affiliations}.
In particular, Harvard Medical School published the most articles which are more than 100.
Vanderbilt University also published more than 100 articles. 
Importantly, most of these top 20 affiliations are medical schools or hospitals, including Harvard Medical School, The Children's Hospital of Philadelphia, Children's Hospital Boston, Hospital for Sick Children University of Toronto, Massachusetts General Hospital, Cincinnati Children's Hospital Medical Center, University of Pennsylvania Perelman School of Medicine, and Vanderbilt University Medical Center.
The document source discovers the ways that share research results in a specific domain. 
There are several venues, including conferences and journals, where researchers share their studies with others and build their reputations. 
The top 20 venues and their number of publications on the topic of AI in ECE are shown in \autoref{fig:venues}.
The related studies of AI in ECE are basically published in the journals and conferences of the computer science domain such as the conferences and journals in ACM and IEEE society. 
In addition, there are many studies are reported in multidisciplinary and mega journals such as PLoS ONE, Scientific Reports, and IEEE Access.

\vspace{-5pt}
\vspace{-5pt}

\section{Intelligent Robots for ECE}
\label{sec:robots}

\begin{figure}[!ht] \centering \small
    \includegraphics[width=0.68\textwidth,trim={5 0 5 5},clip]{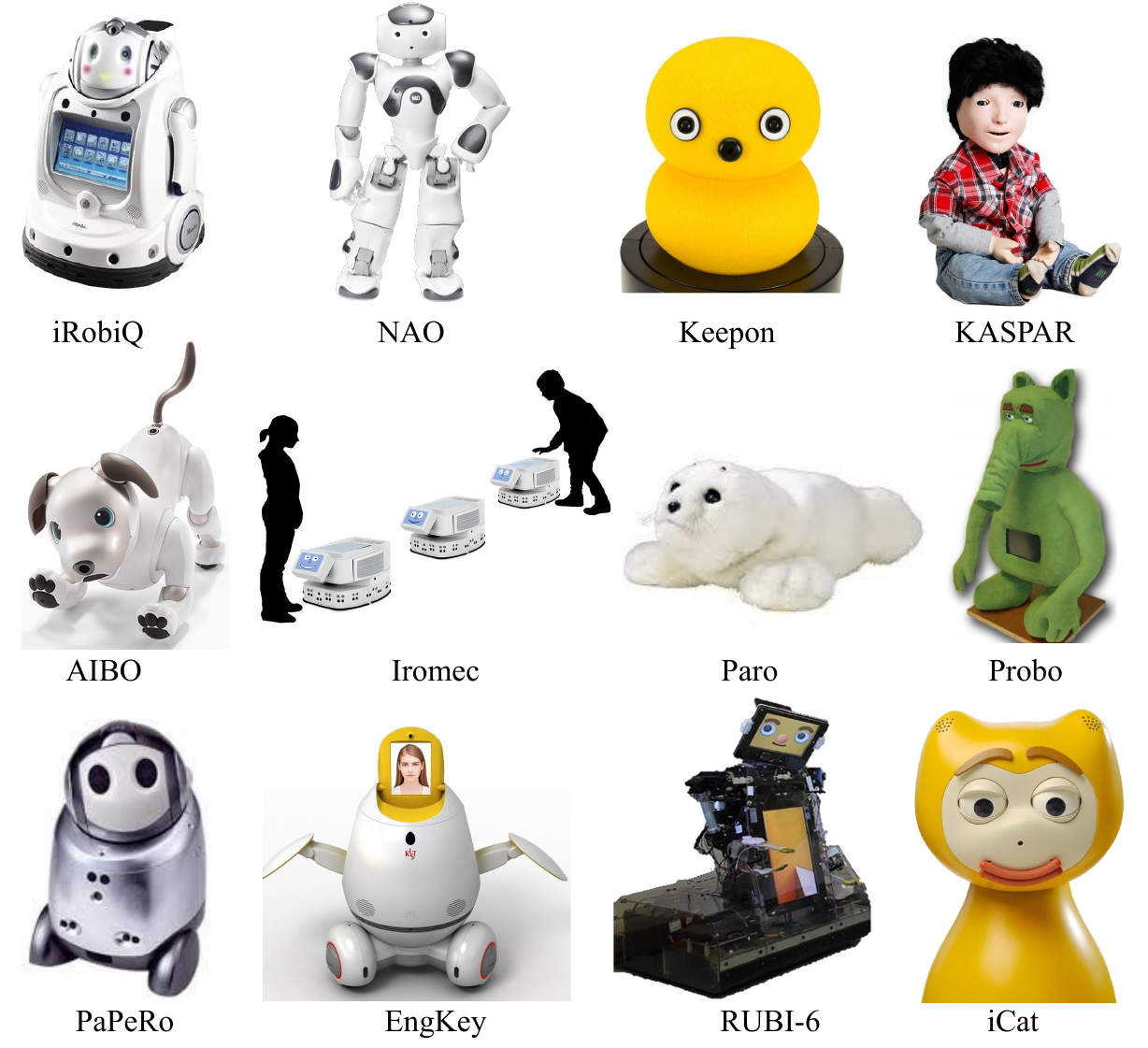}
    \caption{The popular intelligent robots in ECE.}
    \label{fig:robots}
\end{figure}

Intelligent robots have demonstrated good performance in children-related areas, such as ECE.
Huijnen et al. \cite{huijnen2017robots} present an overview of the studies about AI-based robots that are used in therapy and education for children with ASD.
The results indicate that intelligent robots are possible valuable tools in the education or therapy of autistic children.
Ismail et al. \cite{ismail2018application} analyzed the applications of various robots for improving social and communication skills among autistic children. 
The PhD thesis \cite{haibin2012development} discussed the popular design approaches and issues of social robots for children, introduced several representative social robotics and reviewed several representative facial expression-based emotion recognition algorithms of social robotics.
Furthermore, a robotic nanny was designed in a pilot study to evaluate whether the children liked the appearances and functions of the robotic nanny and collect the parent's opinions on the remote user interface designs. 
Results show that most children and parents express great interest in the robotic nanny and provide comparatively positive evaluations.
In this section, we provide a comprehensive review of the existing studies of representative intelligent robots in ECE.
The popular intelligent robots are shown in \autoref{fig:robots}.

\subsection{iRobiQ} \label{subsec:irobiq}
\irobiq, an advanced intelligent robot, is equipped with many functions such as broadcasting sound, facial emotion, fall protection, and collision protection.
The previous studies on educational robots mainly focus on the preference of robot appearance \cite{koay2007living}, functions, the reaction of a robot (utterance and touching) \cite{austermann2008good}, the contents, and the effectiveness of educational robots.
Especially, many studies have been actively conducted on \irobiq.
The findings show that children perceive intelligent robots as learning helpers and playmates \cite{hyun2010usability}.

There are many studies that applied \irobiq to investigate children's perceptions of AI-based things.
Hyun et al. \cite{hyun2012young} explored children's perception of \irobiq, the assistive intelligent robot.
The findings show that although children perceive \irobiq as closer to a human than other artificial things, they perceive \irobiq as a hybrid compound entity, and perceive a robot as an existence close to humans only when it performed functions similar to the functions of human cognition.
In \cite{han2015examining}, \irobiq is used to examine children’s perceptions toward the computer- and robot-mediated augmented reality systems for interactive and participatory dramatic activities.
Eunja and Hyunmin \cite{hyun2009characteristics} applied \irobiq to investigate children's behavior when they were given free access to intelligent devices at kindergarten.
The results demonstrate that children are capable of intelligent robot utilization under instructions. 
Hyun et al. \cite{hyun2010relationships} investigated biological, mental, social, moral, and educational perceptions of children with \irobiq.
The results suggest that intelligent robots should be placed in the classroom for ECE.

Intelligent robots are educational tools with the potential to enhance language and literacy skills in children \cite{neumann2020social}.
There are many studies that applied \irobiq to improve childhood language learning.
Lee and Hyun \cite{lee2015intelligent} developed language-intervention content on \irobiq as a special education agent to promote language interaction for children with speech disorders.
The results show that intelligent robots perform a significant effect in assisting speech disorder treatment. 
In \cite{hsiao2015irobiq}, \irobiq was used as a language teaching/learning tool to improve children’s reading ability and learning behavior. 
Fifty-seven children from prekindergarteners were divided into an experimental group of 30 children using \irobiq and a control group of 27 children using a tablet-PC. 
The results show that the experimental group performs better reading ability than that of the control group. 

\subsection{NAO}

NAO is an open and easy-to-handle intelligent humanoid robot platform with a comprehensive and functional design \cite{gouaillier2009mechatronic}.
Ahmad et al. \cite{ahmad2016children} present a long-term study conducted at a school with twelve children by playing a snakes and ladders game with an NAO.
The study investigated children's views on various adaptation behaviours such as emotion, memory, and personality of intelligent robots in education for maintaining and creating long-term engagement and acceptance. 
The results show that children reacted positively toward the use of intelligent robots in education. 
Rosi et al. \cite{rosi2016use} investigate if an intelligent robot could improve the efficacy of a game-based, nutritional education intervention.
The results show a significant increase in the nutritional knowledge of children involved in a game-based, single-lesson, educational intervention. 
Vrochidou et al. \cite{vrochidou2018social} investigated the effectiveness of intelligent robots in ECE through an innovative game-based activity for teaching/learning numeracy.
The results indicated that children were motivated by the presence of NAO and performed a better understanding of the mathematical concepts.

Ioannou et al. \cite{ioannou2015pre} studied children's behaviour with NAO to explore how robots attract children’s attention and interest.
The results demonstrate that children show various interest in the particular behaviours of intelligent robots such as NAO dancing and interacting with NAO easily.
Depe{\v{s}}ov{\'a} et al. \cite{depevsova2018search} addressed the examples of implementing interesting solutions with NAO for teaching, particularly ECE. 
Teachers can apply NAO to improve ECE with desired solutions.
In \cite{alkhalifah2015using}, NAO is used to design an intelligent robot system that facilitates the kindergarten education process by providing quizzes to test the children’s understanding.
The system enables teachers to continuously review the student’s progress through an application interface.


The existing studies indicate that intelligent robots are effective education agents for the therapy of children with ASD.
In \cite{suzuki2017nao}, NAO is used to investigate whether the use of intelligent robot dancing can be applied to facilitate the dance therapy of children with ASD. 
The results show that NAO can be employed effectively in therapy for children with ASD.
Amanatiadis et al. \cite{amanatiadis2017interactive} applied NAO to robot-assisted special education through specially designed social interaction games.
The results demonstrated that robot-assisted treatment improves children's behaviour, and therefore an engagement of intelligent robots in special education is encouraged.
The thesis \cite{gao2016improvements} presents the improvements of NAO Robots in the education system.
The experimental results show that the autism behaviour of the children with ASD in NAO-based education is greatly reduced than the normal class session.
Most children with Down syndrome have not received special education. 
Jim{\'e}nez et al. \cite{jimenez2019recognition} propose the use of the humanoid robot NAO for the education of children with Down syndrome.
The operation of the Humanoid robot NAO with sensors such as tactile, cameras, and microphones attracts the children's attention to solve the issue that children with Down syndrome lose their attention very quickly, which builds interaction between robots and children. 
The results indicate that the use of an NAO would be a great improvement for the education of children with Down syndrome which differs from traditional education.


\subsection{KASPAR}

KASPAR is a child-sized humanoid robot that equips with minimal sensing capability and was capable of being controlled remotely for child-robot interaction, which is designed as a social companion to improve the behaviour of autistic children for simpler and more comfortable social interaction \cite{wood2017iterative,wood2021developing}.
Wainer et al. \cite{wainer2010collaborating} investigate whether children have difficulties communicating and engaging in social activities, and would perform more collaborative behaviours after playing and interacting with intelligent robots.
The results suggest that children’s intermediaries with KASPAR bring better entertainment and collaboration with a human.
Similarly, Wainer et al. \cite{wainer2014using} present a novel implementation of a collaborative game involving KASPAR playing games with autistic children.
The results demonstrate the proof-of-concept of using an intelligent robot to encourage collaboration among autistic children.
In \cite{wainer2014pilot}, a pilot study in which a novel experimental setup involving a KASPAR participating in a collaborative was implemented and tested with autistic children.
The results show that children perform more engagement in playing and better collaborative behaviours with their partners.

Huijnen et al. \cite{huijnen2017implement} present an overview to systematically describe intelligent robots in ECE and therapy interventions for autistic children in general and KASPAR in particular.
Zaraki et al. \cite{zaraki2018novel} presented an intelligent robot-based development of social skills and capabilities in children with Autism Spectrum Disorders as well as typically developing children. 
Specifically, a reinforcement learning algorithm with verbal expression of KASPAR’s choice uncertainty is developed to facilitate understanding by children.
The experimental results show that Kaspar only made a few unexpected associations, mostly due to exploratory choices, and eventually reached minimal uncertainty and all children expressed enthusiasm.
Costa et al. \cite{costa2013your} investigated a novel scenario for robot-assisted play to increase body awareness, which teaches autistic children about identifying the human body.
The results show that intelligent robots can be considered a tool for prolonging children's attention spans.
In the further study, Costa et al. \cite{costa2015using} studied a human-robot interaction focusing on tactile interaction, in which autistic children interacted with KASPAR.
The results show that autistic children spent increasing time on the intelligent robot through the appropriate physical interaction using KASPAR.

\subsection{Keepon} 

Keepon is a small creature-like robot designed to conduct nonverbal interactions with children \cite{kozima2005using,kozima2009keepon}. 
Costescu et al. \cite{costescu2015reversal} investigated the role of Keepon in a cognitive flexibility task performed by children with ASD.
The results show that children with ASD tend to enjoy interacting with Keepon.
In \cite{kozima2007robot}, Keepon is placed in a playroom and tele-controlled, where children show a wide range of spontaneous actions such as speaking to Keepon. 
Kozima et al. \cite{kozima2007social} applied Keepon to conduct a longitudinal observation of unconstrained child-robot interaction at a daycare center for autistic children. 
The results indicate that autistic children generally have difficulty in interpersonal communication, and are able to approach Keepon with a sense of curiosity and security. 

Kozima and Nakagawa \cite{kozima2006social} applied Keepon to explore the possible use of interactive robots in communication care for children, especially children with special needs. 
The children naturally and spontaneously showed various communicative actions to Keepon.
Peca et al. \cite{peca2016infants} investigated if children regard an unknown intelligent robot as a communicative partner through learning from a brief conversation between a human adult and Keepon. 
The results indicate that children perform a significantly higher level of initiations with the interactive robot than in the non-active robot condition.

\subsection{AIBO} 

AIBO is a series of robotic dogs designed and manufactured by Sony, which increase children's motivation and offer a more joyful perception to children during the learning process. 
Batliner et al. \cite{batliner2004you} applied AIBO to process children’s speech, emotional speech, human-robot communication, cross-linguistics, and reading.
Intelligent robots such as AIBO are used to analyze children's attitudes for interaction with morphological different devices in terms of appearance and behaviour \cite{bartlett2004dogs}. 
The results show that children are attached to the word ‘dog’ reflecting a conceptualization that robots look like dogs (in particular AIBO) and are closer to living dogs than other devices.
Fior et al. \cite{fior2010children} examine how children regard an intelligent robot after a brief interaction.
The results indicate that most children enjoy friendships with intelligent robots such as AIBO by showing positive affiliation and social activities.
A suitable learning environment with a proper learning program is essential for children \cite{wei2011joyful}.
The results demonstrate that joyful learning systems with flexible, mobile and joyful intelligent robots are useful for ECE. 

\subsection{IROMEC}

IROMEC is designed with a modular robot companion tailored towards engaging in social interaction for children with disabilities, which provides children with new opportunities to prevent isolation and develop new skills.
The study \cite{van2017can} explored the application of IROMEC for children with severe physical disabilities in rehabilitation and special education.
The results show that the application of intelligent robots like IROMEC for children with severe physical disabilities is positive and worthwhile, but usability and feasibility are crucial.
Similarly, Heuvel et al. \cite{van2017robots} used the IROMEC robot to improve rehabilitation and special education for children with severe physical disabilities. 
The existing use cases of IROMEC provide the potential to support childhood playing with severe physical disabilities, especially movement functions, learning and applying knowledge, communication/interpersonal interactions and relationships. 

Ferrari et al. \cite{ferrari2009therapeutic} investigated the role of IROMEC in therapy and education for children with special needs, and addressed the therapeutic and educational objectives related to autistic children.
The results show that intelligent robots are helpful for children to learn new developmental skills.
Robins et al. \cite{robins2012scenarios} developed a novel set of ten scenarios for robot-assisted games for children with special needs. 
These scenarios are a meaningful extension of IROMEC for investigating how intelligent robots become social mediators from solitary to collaborative activities with peers and carers/teachers.

\subsection{iCat}

iCat was developed by the Dutch research firm Philips and designed to be a generic companion robotic platform for studying social human-robot interaction with mechanically rendered facial expressions \cite{van2005icat}.
iCat is generally used to interact with children by wearing a wireless sensor for measuring children's electrodermal activity.
The results yield significant correlations regarding how children perceive the interaction and their emotions. 
Leite et al. \cite{leite2012modelling} investigated the children's interaction with iCat which recognises and responds to children's emotions. 
The findings suggest that iCat empathic behaviour affected positively children's cognition of intelligent robots.
Shahid et al. \cite{shahid2010child} investigated how children interact with intelligent robots, where iCat and a child collaborated to play a simple card guessing game. 
The results show that children enjoyed playing a game with iCat and felt happier afterwards.


Palestra et al. \cite{palestra2017artificial} proposed an AI system for robot-assisted interaction based on a behavioural treatment protocol for autistic children, in which iCat is used to elicit specific behaviours in autistic children. 
The results show that a social robot as the mediator is successful in the robot-assisted treatment of autistic children. 
Kasimoglua et al. \cite{kasimoglu2020robotic} introduced iCat for techno-psychological distraction technologies to reduce children's anxiety and improve their behaviour during dental treatment.
The results indicate that intelligent robots successfully mitigate dental anxiety and stress.

\subsection{Paro}

PARO is an interactive robot with a seal appearance \cite{sharkey2014paro}, which has five types of sensors of tactile, light, audition, temperature, and posture sensors.
Cifuentes et al. \cite{cifuentes2020social} applied PARO to a psychiatric ward of children and adolescents for relaxation and an improvement in communication.
Crossman et al. \cite{crossman2018influence} provide the scientific demonstration that intelligent robots such as PARO are useful for improving children’s mental health. 
The results show that PARO is useful to improve children’s mood through a range of stressful circumstances, such as medical procedures and school assessments.
Pipitpukdee and Phantachat \cite{pipitpukdee2011study} use PARO to mitigate the behaviour problems of Thai Autistic children such as low motivation and communication problems.
The results demonstrate that PARO could be applied to increase communication and motivation. 
Shibata and Coughlin \cite{shibata2014trends} conducted PARO in childhood activities and therapy for both cognitive and physical rehabilitation.
The results show that PARO performs clear therapeutic effects on children, particularly those with cognitive impairments.

\subsection{Probo} 

Probo is designed to be a natural interaction while employing human-like social cues and communication modalities \cite{goris2010probo,goris2008huggable}. 
Chevalier et al. \cite{chevalier2017dialogue} proposed specific strategies for human-robot interaction dialogues and interactions for school-aged autistic children. 
Vanderborght et al. \cite{vanderborght2012using} applied Probo to provide robot-assisted therapy for autistic children. 
The results show that the social performance of autistic children improves when using the intelligent robot Probo as a medium for social storytelling than when a human reader tells the stories.
Probo imitates children’s faces to encourage children to notice facial expressions in a play-based game.

\subsection{Other Intelligent Robots} 

There are also other intelligent robots that apply AI to ECE, such as PaPeRo \cite{osada2006scenario}, EngKey \cite{yun2011engkey}, and RUBI \cite{movellan2007rubi}.

\textit{PaPeRo} is a popular intelligent robot because of its cute appearance and facial recognition system.
Han et al. \cite{han2005educational} compared the effects of traditional media-assisted learning with the effects of robot-assisted learning.
The results suggest that robot-assisted learning is more effective for children’s learning concentration, learning interest and academic achievement than traditional media-assisted learning.
Kawata et al. \cite{kawata2008field} proposed a method to promote the intermixing of parents and children, and developed the network-based interactive child-watch system.
Parents in remote locations could get detailed information about their children's activities and expressions when desired via asynchronous communication.

\textit{Engkey} is a friendly and accessible avatar tele-education robot with a dumpy egg-shaped appearance that has been deployed in South Korean classrooms to teach students English via telepresence \cite{sharkey2016should}.
Kim et al. \cite{kim2012gesture} proposed a robot motion programming method by applying a three-level robot motion hierarchical structure and a gesture variation method.
Children show great interest in learning English with Engkey.

\textit{RUBI} was designed for accelerating the development of robots that interact with people \cite{movellan2007rubi}, which is generally used to be an intelligent sociable robot as a tool in ECE \cite{johnson2012design}.
Movellan et al. \cite{movellan2009sociable} used RUBI to improve children's vocabulary skills. 
The results show that RUBI conducts a significant improvement in the average vocabulary score of the children.
Malmir et al. \cite{malmir2013home} explored whether RUBI could autonomously data-mine the children's behavior and provide insights into the preferred activities by children.
The results indicate that RUBI could apply the facial expression data to accurately predict the children's preference for different activities.


\subsection{Discussion}


The summary of these popular intelligent robots for ECE, including their specifications and the studies on these intelligent robots, are addressed in \autoref{tab:robots}.
Many intelligent robots have been applied in early childhood education. 
In general, the existing studies applied artificial intelligence technologies to these intelligent robots for better interaction with children, particularly children with interaction disorders such as ASD or Down syndrome.
For example, deep learning-based computer vision is used to recognise the children's facial emotions and present facial emotions to children for building friendly interaction with children \cite{del2017computer,kamble2021face,haibin2012development,vanderborght2012using,malmir2013home}.
In addition, many intelligent robots \cite{van2005icat,hsiao2015irobiq,osada2006scenario,goris2010probo} provide the function of facial emotion recognition and facial expressions.
Furthermore, intelligent robots and artificial intelligence are used to improve childhood language learning and literacy skills in children \cite{neumann2020social}.
Intelligent robots and artificial intelligence technologies perform a significant effect in assisting speech disorders treatment and children’s reading ability and learning behavior \cite{lee2015intelligent,hsiao2015irobiq,neumann2020social}. 
To support the development of these AI technologies in ECE, these intelligent robots are generally equipped with the hardware of cameras, speakers and microphones. 
In summary, AI-based intelligent robots significantly contribute to early childhood education, particularly the education of children with interaction disorders.

\begin{table*}[!ht] \centering \footnotesize
\renewcommand{\arraystretch}{0.9} \setlength\tabcolsep{1pt}
\begin{tabular}{l l l l l m{3.6cm}} \toprule
\makecell{AI Robots} & Studies & Developers &  Hardware & Features & Remarks \\ \midrule
iRobiQ & \makecell[l]{\cite{hyun2009characteristics,hyun2012young,han2015examining} \\ \cite{neumann2020social,lee2015intelligent,hsiao2015irobiq} \\ \cite{kasimoglu2020robotic}}& Yujin Robot & \makecell[l]{$\bullet$ cameras \\ $\bullet$ speaker \\ $\bullet$ touch screen} & \makecell[l]{$\bullet$ emotions \\ $\bullet$ speech \\ $\bullet$ actions} & 5 facial expressions, intelligent avoiding obstacles for house surveillance, natural language process. \\ \midrule
Probo & \makecell[l]{\cite{goris2008huggable,goris2010probo,chevalier2017dialogue,vanderborght2012using}}& \makecell[l]{Vrije \\ Universiteit \\ Brussel} & \makecell[l]{$\bullet$ digital cameras \\ $\bullet$ microphones \\ $\bullet$ touch sensors} & \makecell[l]{$\bullet$ speech \\ $\bullet$ emotions \\ $\bullet$ gestures} & A research platform to study cognitive human-robot interaction. \\ \midrule
NAO & \makecell[l]{\cite{ahmad2016children,ioannou2015pre,vrochidou2018social,alkhalifah2015using,depevsova2018search,rosi2016use,gao2016improvements,suzuki2017nao,amanatiadis2017interactive,jimenez2019recognition}} & \makecell[l]{Aldebaran \\ Robotics} & \makecell[l]{$\bullet$ cameras \\ $\bullet$ speaker \\ $\bullet$ microphone } & \makecell[l]{$\bullet$ walk \& dance \\ $\bullet$ talk \\ $\bullet$ play games} & An autonomous, programmable humanoid robot used in numerous academic institutions for education. \\ \midrule
KASPAR & \makecell[l]{\cite{wood2017iterative,wood2021developing,wainer2010collaborating,wainer2014using} \\ \cite{huijnen2017implement,zaraki2018novel,costa2013your,costa2015using}} & \makecell[l]{University of \\ Hertfordshire} & \makecell[l]{$\bullet$ cameras \\ $\bullet$ tactile sensors} & \makecell[l]{$\bullet$ speech \\ $\bullet$ emotions \\ $\bullet$ actions} & A child-sized humanoid robot specifically developed for childhood interaction. \\ \midrule
Keepon & \makecell[l]{\cite{kozima2005using,kozima2009keepon,costescu2015reversal} \\ \cite{kozima2007social,kozima2006social,peca2016infants}} & NICT & \makecell[l]{$\bullet$ cameras \\ $\bullet$ microphone } & \makecell[l]{$\bullet$ dance mode \\ $\bullet$ touch mode} & Keepon is a small robot designed to study childhood interaction. \\ \midrule
AIBO & \makecell[l]{\cite{batliner2004you,bartlett2004dogs,fior2010children,wei2011joyful}}& Sony & \makecell[l]{$\bullet$ cameras \\ $\bullet$ microphone \\ $\bullet$ speaker} & \makecell[l]{$\bullet$ object detection \\ $\bullet$ SLAM \\ $\bullet$ Self-charge} & An AI-based robot, which is used extensively in education and academia \\ \midrule
IROMEC & \makecell[l]{\cite{van2017robots,van2017can,ferrari2009therapeutic,robins2012scenarios}} & IROMEC & \makecell[l]{$\bullet$ microphone \\ $\bullet$ screen \\ $\bullet$ other sensors}& \makecell[l]{$\bullet$ emotions \\ $\bullet$ sound recognition \\ $\bullet$ actions} & An interactive social mediator robot to be a playmate for children with disabilities \\ \midrule
iCat & \makecell[l]{\cite{kasimoglu2020robotic} \\ \cite{van2005icat,shahid2010child,leite2012modelling,palestra2017artificial}} & Philips & \makecell[l]{$\bullet$ cameras \\ $\bullet$ speakers \\ $\bullet$ stereo microphones} &  \makecell[l]{$\bullet$ face recognition \\ $\bullet$ facial expressions \\ $\bullet$ control devices} & A companion robotic platform for studying social human-robot interaction. \\ \midrule
Paro & \makecell[l]{\cite{pipitpukdee2011study,sharkey2014paro,shibata2014trends,crossman2018influence,cifuentes2020social}} & AIST & \makecell[l]{$\bullet$ microphone \\ $\bullet$ multimodal sensors} & \makecell[l]{$\bullet$ speech \\ $\bullet$ voice sampling \\ $\bullet$ heating system} & An advanced interactive robot with cute seal shape for children therapeutic. \\ 
\bottomrule
\end{tabular}
\caption{The summary of the popular intelligent robots for ECE, including their specifications and the related studies.}
\label{tab:robots}  
\end{table*}


Although intelligent robots and AI technologies have been applied to early childhood education, the applications of AI in ECE are still weak and need to be further investigated.
The current studies on intelligent robots generally apply weak AI technologies such as the question-answer chatbot in ECE \cite{lee2015intelligent,hsiao2015irobiq}.
Although intelligent robots are perceived by children as closer to a human than other artificial things, they perceive intelligent robots as a hybrid compound entity and perceive a robot as an existence close to humans only when it performed functions similar to the functions of human cognition \cite{hyun2012young}.
The AI-based intelligent robots are still far to provide children with a better interaction even the human-like cognition for ECE. 
Therefore, intelligent robots need to integrate state-of-the-art AI technologies in ECE, such as the AI-powered language model ChatGPT, to generate human-like interaction with children.
In summary, intelligent robots should apply state-of-the-art AI technologies to improve the AI applications in early childhood education and extend novel applications to assist early childhood education, particularly the education of children with interaction disorders.

\section{Key AI technologies in ECE}
\label{sec:aitechniques}

Many AI technologies have been widely applied in ECE. 
In this section, we review the studies of the main AI technologies including data mining, machine learning, deep learning, virtual reality, computer vision, and natural language processing in ECE.


Data mining is a popular AI technique that has been widely applied in ECE.
Data mining technologies are used to extract useful information from a vast amount of data, which discover new, accurate, and useful patterns in the data.
Chang \cite{chang2007study} applied a data mining approach to find hidden knowledge and obtain useful information as a reference for decision-making and evaluation in children's healthcare.
Subsequently, a decision tree is used to classify children's developmentally-delay levels according to their physical illness.
This study identifies which type of illness causes certain types of delays by the decision tree.
The results provide healthcare personnel with important references during diagnosis and evaluations.
Al-Diabat and Mofleh \cite{al2018fuzzy} analyse the autism behavioural characteristics data and investigate fuzzy rule-based data mining models to forecast children's autistic symptoms. 
The results demonstrate that the fuzzy data mining model with regard to predictive accuracy and sensitivity rates outperform the other rule-based data mining models.

Leroy et al. \cite{leroy2006data} applied the data mining technologies of decision trees and association rules to develop a digital library of coded video segments that contain data on the appropriate and inappropriate behaviour of autistic children in different social scenarios.
The results show that therapy is effective in reducing inappropriate behaviour and increasing appropriate behavior. 
Data mining technologies such as classification algorithms are widely applied to predict childhood obesity. 
Abdullah et al. \cite{abdullah2016data} investigate the classification of childhood obesity for children in Malaysia. 
The classification technologies, Bayesian Network, Decision Tree, Neural Networks, and SVM were implemented and compared on the data sets.
The results show that the classifiers of J48 and SMO outperform the other two classifiers for predicting childhood obesity. 
Zhang et al. \cite{zhang2009comparing} compared data mining methods with logistic regression in childhood obesity prediction.
The results show that SVM and Bayesian algorithms outperform other algorithms for predicting overweight and obese children on the Wirral database. 

In \cite{momand2020data}, a data mining approach is proposed to predict the malnutrition status of children in Afghanistan. 
The results show that the proposed data mining approach robustly predict the malnutrition status based on clinical sign and anthropometric parameters of Afghan children.
Krotova et al. \cite{krotova2020diagnostics} applied data mining methods to develop a diagnostic model for diabetes mellitus complications, which explores the possibility of diagnosing diabetic polyneuropathy by using machine learning methods.
Interestingly, El and Gamal et al. \cite{el2013application} investigate if a relationship exists between the occurrence of allergies in children and daily upper-air observations (e.g., temperature, relative humidity, dew point, mixing ratio) and air pollution (e.g., particulate matter, sulfur dioxide, nitrogen dioxide, carbon monoxide, and ozone).
The results show that the prevalence of allergies increased over the last few years. 
Monitoring upper-air observation and air pollution data over time is a reliable method for predicting outbreaks of allergies among elementary school children, which enables parents and school nurses to implement effective precautionary measures.

Yue et al. \cite{yue2018application} designed an intelligent system to obtain video clips in a kindergarten classroom and leverage emotional data to portray the mental states of children. 
Gathering and analyzing children's emotional features by data mining methods reached important conclusions, which are beneficial to developing educational policy and teaching practice.
Diagnostic information on the presence of infection, severity and aetiology (bacterial versus viral) is crucial for the appropriate treatment of childhood pneumonia. 
The work \cite{naydenova2016power} explores a suite of data mining tools to facilitate automated diagnosis through quantifiable features.
The results indicate that machine learning tools are successfully used for the multi-faceted diagnosis of childhood pneumonia in resource-constrained settings, compensating for the shortage of advanced equipment and clinical expertise.


Here, we review the studies of machine learning in ECE. 
Machine learning is the development of discovering algorithms that permit machines to learn relations without human intervention from the existing datasets.
To find appropriate classifiers, Delavarian et al. \cite{delavarian2011automatic} compare 16 linear and non-linear classifiers to distinguish and diagnose children with many similar symptoms and different behavioural disorders such as ADHD, depression, and anxiety.
The results show that the nearest mean classifier was selected as a relevant classifier by categorizing 96.92\% of the samples correctly.
The article \cite{lin2020zhorai} presents a conversational platform and curriculum designed to facilitate children's understanding of machine learning concepts. 
The results indicate that the conversational platform increased engagement during learning and the novel visualizations make machine-learning concepts understandable for children. 
Hitron et al. \cite{hitron2018introducing} studied if 10-12 years old children understand machine learning concepts through the experience with a digital stick-like device. 
The results indicate that children are able to understand basic machine learning concepts and even apply these concepts to daily life. 

Visual and audio data are gathered during the child-robot interaction and processed towards deciding an engaged state of children by AdaBoost decision tree \cite{papakostas2021estimating}.
Hagenbuchner et al. \cite{hagenbuchner2015prediction} develop machine learning models for predicting activity types in preschool-aged children.
The results contribute to the evidence supporting the application of machine learning approaches to accelerometry data analysis in children.
Ahmadi et al. \cite{ahmadi2018machine} develop machine learning models to automatically identify the physical activity types in ambulant children with Cerebral palsy.
The random forest and support vector machine classifiers consistently outperform the binary decision tree classifiers. 
The results demonstrate that machine learning approaches with accelerometer data processing are feasible for identifying children with Cerebral palsy.

Rasheed et al. \cite{rasheed2021use} use machine learning technologies to examine the EEG data for predicting failure in the early years of school. 
The combination of the EEG data with sociodemographic and home environment variables can increase the specificity.
Carpenter et al. \cite{carpenter2016quantifying} apply machine learning tools to the PAPA data for identifying subsets of PAPA items, which could be developed into an efficient, reliable, and valid screening tool to assess the possibility of children with anxiety disorders.
The machine learning approaches identify children for both generalized anxiety disorder and separation anxiety disorder with an accuracy of over 96\%.
A supervised machine-learning method \cite{crippa2015use} is developed to correctly discriminate preschool children with ASD from the typically developing children by kinematic analysis of a simple reach-to-drop task. 
The findings offer insight into a possible motor signature of ASD that is useful for identifying a well-defined subset of patients and reducing the clinical heterogeneity within the broad behavioural phenotype.
Liu et al. \cite{liu2016identifying} proposed the face-scanning patterns to identify children with ASD by an SVM classifier with data-driven feature extraction.
Developing the AI-aid early detection and diagnosis system is helpful in screening and diagnosing ASD in children.

McGinnis et al. \cite{mcginnis2018wearable} proposed a new approach for diagnosing anxiety and depression in children, in which wearable sensors are used to monitor the participant's motion during the period.
The experimental results show that the approach performs a diagnostic accuracy of 75\%.
Similarly, McGinnis et al. \cite{mcginnis2019rapid} proposed a new approach to identify children with internalizing disorders using an instrumented 90-second mood induction task.
The data from a single wearable sensor during a 90-second fear induction task and a machine learning approach are proposed to fulfil this need.
This work provides an important step for overlooked children to both mitigate their distress and prevent subsequent comorbid emotional disorders and additional negative sequels.
Su et al. \cite{su2020machine} develop machine learning models to predict suicidal behavior among children and adolescents based on their longitudinal clinical records, and determine short- and long-term risk factors. 
The findings demonstrate that routinely collected electronic health records can be used to develop accurate predictive models for preventing suicide risk among children and adolescents.


Deep learning has been widely applied in various applications, including ECE. 
In this section, we review deep learning-based studies in ECE. 
Di Nuovo et al. \cite{di2018deep} proposed a novel deep-learning neural network to automatically estimate if the children focus on the robot during a therapy session.
The results show that CNN-based approaches significantly outperform the benchmark algorithms for estimating children's attention.
Rudovic et al. \cite{rudovic2018culturenet} applied deep learning models to the task of automated engagement estimation by using face images of children with autism.
The authors claim that it is the first study to investigate the effects of individual and cultural differences in children
with autism in the context of deep learning performed directly from face images.
Kyung-Min et al. \cite{kim2015pororobot} developed a deep learning-based system of a video question and answering game robot for early childhood interactive education in real-world environments.
She and Ren \cite{she2021enhance} proposed a chat robot with the use of a deep neural network model as the generative conversational agent, which generates meaningful and coherent dialogue responses to improve the context sensitivity of early children.

In \cite{liu2018detecting}, a deep learning approach was applied to detect premature ventricular contractions in children automatically. 
The AI-aided diagnosis model achieved high accuracy while sustaining stable performance.
Lempereur et al. \cite{lempereur2020new} study the deep learning-based method to detect gait events in children with gait disorders.
A long short-term memory recurrent neural network, called DeepEvent, is proposed to detect children's gait disorders.
The results show that DeepEvent outperforms the existing well-known approaches for detecting children's foot strikes and foot-off.
Kumar and Senthil \cite{kumar2021construction} used deep learning classifiers to predict children’s behavior by considering their emotional features.
Chatzimichail et al. \cite{chatzimichail2010artificial} present an effective deep neural network to predict persistent asthma in children. 
The results demonstrate that deep neural networks predict the asthma of children successfully.
The interesting study \cite{yu2021associations} investigates the associations between trees and grass presence with childhood asthma prevalence using deep learning-based image segmentation.
The results indicate a role of vegetation in the association between greenness exposure and childhood asthma. 
The finding provides valuable information to reveal the effects of different green vegetation on childhood asthma and the underlying mechanisms. 


In addition, virtual reality, computer vision, and natural language processing are the other mainstream AI technologies that are applied in ECE.
In this section, we mainly review the existing works in aspects of virtual reality, computer vision, and natural language processing.


In the early works, the article \cite{mccomas1998current} provided an overview of virtual reality for children with disabilities.
This work investigates the benefits of VR for children with disabilities and explores how to apply VR to the needs of children with disabilities.
The results show that VR provides disabled children with practice skills.
Gershon et al. \cite{gershon2004pilot} conducted VR as a distraction for children with cancer to reduce anxiety and pain associated with an invasive medical procedure.
The results found that using VR distraction reduces the pain and anxiety of children compared to the no-distraction scenes.
The findings suggest that VR is a useful tool for distraction during painful medical procedures.
Parsons and Cobb \cite{parsons2011state} addressed the state-of-the-art of VR technologies for autistic children to assess how VR can be used in practice.
Arane et al. \cite{arane2017virtual} investigate how VR work in reducing pain and anxiety in children patients.
The preliminary results show that VR is effective in reducing the pain and anxiety children patients experience compared with standard care or other distraction methods.

Foley and Maddison \cite{foley2010use} assess active video games to increase energy expenditure and physical activity behaviour in children.
The findings indicate that playing active games results in greater energy expenditure compared with nonactive video games, and is approximate to moderate-intensity physical activity. 
VR was applied to be a promising and motivating approach to practice and rehearse social skills for children with ASD \cite{didehbani2016virtual}.
The findings suggest that VR offers an effective treatment for improving social impairments in children with ASD.
Vishav and Uttama \cite{jyoti2019virtual} developed a VR-based joint attention system of varying difficulty levels coupled with a hierarchical prompt protocol for children with ASD.
The results indicate that the VR-based joint attention system was able to estimate the joint attention level of a group of children with ASD.

Josman et al. \cite{josman2008effectiveness} investigated whether children with ASD are capable of learning the skills needed to cross a street safely via a street-crossing VR and whether these skills can be transferred to real life.
The experimental results show a significant improvement in children with ASD crossing a real-street setting after learning and considerable improvement in the virtual street.
Similarly, Schwebel and McClure \cite{schwebel2010using} use VR to train children in safe street-crossing skills.
The results demonstrate the efficacy of VR to train child pedestrians in safe street crossings.
A VR system is applied to enhance emotional and social skills for children with ASD \cite{ip2018enhance}.
The study demonstrated the clear feasibility of VR for enhancing the emotional and social skills of children with ASD.


Computer vision approaches are mainly applied to recognize children's faces for ECE.
Marco et al. \cite{del2017computer} demonstrate that computer vision-based approaches for facial feature analysis can be used to understand emotional behaviours for the assessment and diagnosis of ASD in preschool children.
Dongming et al. \cite{dongming2020intelligent} introduced a face-tracking pan–tilt–zoom solution with an intelligent robot to track children's faces and record videos in ECE.
The convolutional neural network and k-nearest neighbour classification are applied to recognize children's faces.
Kamble et al. \cite{kamble2021face} addressed the various face recognition technologies using different classifications in children.
The results show that the approaches of machine learning and deep learning can be used to track facial changes in childhood with remarkable recognition accuracy. 
Xia et al. \cite{xia2017detecting} studied smile detection across the difference between children and adults.
The state-of-the-art transfer learning methods are applied to the discrepancy in the well-known deep neural networks of AlexNet and ResNet.
The results demonstrate the effectiveness of the proposed approach to smile detection across such a difference.


Druga et al. \cite{druga2017hey} investigate how children perceive natural language processing by exploring children's interaction with the agents of Amazon Alexa, Google Home, Cozmo, and Julie Chatbot. 
This work suggests a series of design considerations for future child-agent interaction from voice/prosody, interactive engagement and facilitating understanding.
Shahi et al. \cite{shahi2021using} applied natural language processing models to detect child physical abuse.
The results indicate that deep learning-based natural language processing for clinician judgement improves the recognition of physical abuse.



AI curriculum for children is an interactive and playful manner through engaging AI projects such as recognizing faces, VR training, and language learning.
Children could have a basic understanding of AI concepts with an AI curriculum.
Brownlee and Berthelsen \cite{brownlee2006personal} provide a conceptual framework by analysing the current studies on epistemological beliefs and tertiary learning.
The framework can be used in childhood teachers' education programs/curricula to explore how teachers implement AI concepts in ECE.

Williams et al. \cite{williams2019artificial} developed an AI platform (PopBots) for designing an AI curriculum in ECE, where preschool children interact with social robots to learn basic AI concepts.
In the PhD thesis \cite{williams2019popbots}, Williams explores how children explore and create with AI, and how the activities influence children’s perceptions of AI.
Specifically, PopBots is used to develop a novel developmentally appropriate Preschool-Oriented Programming curriculum. 
The experimental results show that the social robot as a learning companion and programmable artefact is effective in helping children understand basic AI concepts. 
In addition to the AI curriculum at preschool, family education about AI is also crucial.
Unlike unified childhood education in classes, family education is more flexible and targeted.
Jin \cite{jin2019study} studies the effects of AI on childhood family education. 
The results show that AI could help to provide better family education for children.

\begin{table*}[!ht] \centering \footnotesize
\renewcommand{\arraystretch}{0.8}
\setlength\tabcolsep{3pt}
\begin{tabular}{l l l l}  \toprule
AI technologies & Studies & Methods & \makecell[l]{Applications in ECE} \\ \midrule
Data mining & \makecell[l]{\cite{leroy2006data,chang2007study,zhang2009comparing,el2013application,naydenova2016power,abdullah2016data,yue2018application,al2018fuzzy,momand2020data,krotova2020diagnostics}} & \makecell[l]{$\bullet$ fuzzy data mining \\ $\bullet$ logistic regression, SVM \\ $\bullet$ Bayesian algorithm \\ $\bullet$ decision tree} & \makecell[l]{$\bullet$ detect autistic symptoms \\ $\bullet$ behavioral therapy \\ $\bullet$ predict allergies\\ $\bullet$ \makecell[l]{predict childhood \\ obesity/overweight}} \\ \midrule
Machine learning & \makecell[l]{\cite{delavarian2011automatic,hagenbuchner2015prediction,crippa2015use,liu2016identifying,carpenter2016quantifying,hitron2018introducing,ahmadi2018machine,mcginnis2018wearable,mcginnis2019rapid,lin2020zhorai,rasheed2021use,papakostas2021estimating}} & \makecell[l]{$\bullet$ AdaBoost decision tree \\ $\bullet$ neural networks \\ $\bullet$ random forest, SVM} & \makecell[l]{$\bullet$ machine learning concepts \\ $\bullet$ predicting activity types \\ $\bullet$ predicting suicide \\ $\bullet$ diagose anxiety \& depression} \\ \midrule
Deep learning & \makecell[l]{\cite{di2018deep,rudovic2018culturenet,kim2015pororobot,she2021enhance,liu2018detecting,lempereur2020new,kumar2021construction,chatzimichail2010artificial,yu2021associations}} & \makecell[l]{$\bullet$ deep neural networks \\ $\bullet$ image segmentation \\ $\bullet$ cascaded convolutional networks} & \makecell[l]{$\bullet$ estimate children's attention \\ $\bullet$ detect premature \\ $\bullet$ detect gait disorders \\ $\bullet$ predict persistant asthma} \\ \midrule
Virtual reality & \makecell[l]{ \cite{mccomas1998current,gershon2004pilot,parsons2011state,arane2017virtual,foley2010use,didehbani2016virtual}\\ \cite{josman2008effectiveness,schwebel2010using,ip2018enhance}\\ \cite{strickland1996brief,strickland1996virtual}} & $\bullet$ VR-based system & \makecell[l]{$\bullet$ reduce pain \& anxiety \\ $\bullet$ improve energy expenditure \\ $\bullet$ practice emotional \& social skills} \\ \midrule
Computer vision & \makecell[l]{\cite{xia2017detecting,del2017computer} \\ \cite{dongming2020intelligent,kamble2021face}} & \makecell[l]{$\bullet$ convolutional neural network \& \\  k-nearest neighbor classification \\ $\bullet$ transfer learning} & \makecell[l]{$\bullet$ face recognition \& tracking \\ $\bullet$ smile detection \\ $\bullet$ emotion recognition} \\ \midrule
\makecell[l]{Natural Language \\ Processing} & \makecell[l]{\cite{druga2017hey,shahi2021using}} & $\bullet$ NLP models & \makecell[l]{$\bullet$ exploring children's interaction \\ $\bullet$ detect children's physical abuse} \\ 
\bottomrule
\end{tabular}
\caption{The summary of AI technologies in ECE, including the studies, their methods, and their specific applications.}
\label{tab:ai_technologies}  
\end{table*}

\subsection{Discussion}
\label{subsec:discussionAI}

We summarize the key AI technologies in early childhood education in \autoref{tab:ai_technologies}, including the studies of AI technologies in ECE and their specific methods and applications. 
Data mining technologies are generally used to analyse the data and predict the potential issues during the age of early childhood education and provide insightful recommendations for designing personal education and childcare.
For example, data mining technologies were used to extract useful information from the behavioural characteristics data \cite{al2018fuzzy}, the digital library of coded video \cite{leroy2006data}, the occurrence of allergies in children and daily upper-air observations \cite{el2013application} and then decision tree \cite{leroy2006data,chang2007study}, fuzzy data mining models \cite{al2018fuzzy}, and classification algorithms \cite{abdullah2016data,zhang2009comparing} are used to detect/predict children's symptoms and design personal education and childcare.
Machine learning technologies developed algorithms (e.g., AdaBoost \cite{papakostas2021estimating}, random forest \cite{ahmadi2018machine}, support vector machine \cite{liu2016identifying}) to identify children's activities such as attention \cite{papakostas2021estimating}, failure \cite{rasheed2021use}, anxiety disorder \cite{carpenter2016quantifying,mcginnis2018wearable}, ASD \cite{crippa2015use,liu2016identifying}, suicidal behavior \cite{su2020machine}.
Note that although data mining and machine learning in ECE have a few overlaps of data, there are a considerable number of differences between them.
Data mining technologies aim to extract useful information from a vast amount of data and discover useful patterns in the ECE data such as interaction data.
Machine learning develops algorithms to learn relations from the existing ECE datasets.

\begin{table*}[!ht] \centering \footnotesize
\renewcommand{\arraystretch}{0.8}
\setlength\tabcolsep{1pt}
\begin{tabular}{l l l m{6.3cm}}  \toprule
\makecell[l]{AI \\ tech.} & Datasets & Methods & Dataset information \\ \midrule
\multirow{12}{*}{\makecell[l]{Data \\ mining}} & \makecell[l]{Medical records \cite{chang2007study}} & Decision tree & 605 pieces of developmentally-delayed children data \\ \cmidrule(lr){2-4}
& AQ-10-child \cite{al2018fuzzy} & \makecell[l]{FURIA} & Autism behavioural characteristics data contains 252 No-ASD instances and 257 ASD instances \\ \cmidrule(lr){2-4}
& SEGAK data \cite{abdullah2016data} & \makecell[l]{SVM, etc} & 4245 childhood obesity data from 153 schools \\ \cmidrule(lr){2-4}
& \makecell[l]{Wirral  child \\ database \cite{zhang2009comparing}} & \makecell[l]{SVM, etc} & 16653 children with 56 attributes for each sample including weight, height, age and sex \\ \cmidrule(lr){2-4}
& \makecell[l]{Medical records \cite{krotova2020diagnostics}} & SVM & 3204 children suffering from type 1 diabetes, including age, height and weight, etc. \\ \cmidrule(lr){2-4}
& \makecell[l]{Upper-air data \& \\ health records\cite{el2013application}} & \makecell[l]{Binary \\ logistic regression} & 168825 children with occurrence of allergies and upper-air observations. \\ \cmidrule(lr){2-4}
& \makecell[l]{Emotional face \cite{yue2018application}} & k-means & 14049 sets of data, including 38 children in 4 courses and each lasting 30 minutes, covering 7 expressions \\ \cmidrule(lr){2-4}
& \makecell[l]{Health records \cite{naydenova2016power}} & \makecell[l]{Random forest,etc} & 780 children diagnosed with pneumonia and 801 age-matched healthy controls \\ \midrule
\multirow{11}{*}{\makecell[l]{Machine \\ learning}} & \makecell[l]{Behavioral \\ disorders \cite{delavarian2011automatic}} & 16 classifiers & 306 children, 70 ADHD, 36 conduct disorder, 54 anxiety disorder, 38 depression, 34 comorbid depression and anxiety and 74 normal children \\ \cmidrule(lr){2-4}
& Visual and audio \cite{papakostas2021estimating} & AdaBoost & 819 samples with 11 features for each sample during the child-robot interaction, including 99 engaged samples, 720 not engaged samples \\ \cmidrule(lr){2-4}
& \makecell[l]{Accelerometer data \cite{hagenbuchner2015prediction}} & NNs & Accelerometer data of 12 activities including five activity classes (walking, running, etc) by 11 children in preschool \\ \cmidrule(lr){2-4}
& EEG data \cite{rasheed2021use} & K-NN & EEG data from 96 children at 8 years old \\ \cmidrule(lr){2-4}
& \makecell[l]{Psychiatric \\ Assessment \cite{carpenter2016quantifying}} & decision tree & diagnostic parent-report for assessing psychopathology in 2 to 5 years old children \\ \cmidrule(lr){2-4}
& \makecell[l]{Activities \cite{mcginnis2018wearable,mcginnis2019rapid}} & \makecell[l]{K-NN \\ logistic regression} & wearable sensor data included 63 children (57\% girls) with anxiety and depression \\ \cmidrule(lr){2-4}
& \makecell[l]{Health records \cite{su2020machine}} & logistic regression & 41721 childhood records within the Connecticut Children’s Medical Center from 2011 to 2016 \\ \midrule
\multirow{9}{*}{\makecell[l]{Deep \\ learning}} & \makecell[l]{Video data \cite{rudovic2018culturenet}} & CNN & audio-visual and autonomic physiological recordings of children in a robot-assisted education \\ \cmidrule(lr){2-4}
& \makecell[l]{Cartoon video \\ Q\&A dataset \cite{kim2015pororobot}} & CNN \& RNN & 1232 minutes and 183 episodes cartoon video, approximately 1200 question \& answer pairs \\ \cmidrule(lr){2-4}
& \makecell[l]{Dialogue \cite{she2021enhance}} & LSTM RNN & 352256 dialogue data between children and their parents, teachers, and friends \\ \cmidrule(lr){2-4}
& \makecell[l]{ECG data
\cite{liu2018detecting}} & CNN & Over 30000 normal controls and around 1200 PVC cases of children. \\ \cmidrule(lr){2-4}
& 3D gait \cite{lempereur2020new} & LSTM RNN & 10526 foot strike  (5247 left and 5279 right) and 9375 foot off (4654 left and 4721 right) from 226 children \\ \midrule
\multirow{3}{*}{Others} & Childhood smile \cite{xia2017detecting} & \makecell[l]{Adap. networks} & 17517 images of children under 5 years old, including 6171 training, 3086 for validation, and 8260 for test \\ \cmidrule(lr){2-4}
& \makecell[l]{Trauma data \cite{shahi2021using}} & DNN \& NLP & abused children and non-abusive trauma children \\
\bottomrule 
\end{tabular}
\caption{The summary of the datasets used in the studies of AI technologies in ECE, including the studies, topics, AI technologies, and dataset information.}
\label{tab:ai_datasets}  
\end{table*}

Deep learning technologies generally train neural networks on datasets to estimate association for ECE. 
For example, deep neural networks are applied to estimate children's attention \cite{di2018deep,rudovic2018culturenet,jyoti2019virtual}, develop video question-answer games \cite{kim2015pororobot} and conversational agents \cite{she2021enhance} for ECE.
In addition, virtual reality has been applied in the practical education of children with disabilities \cite{mccomas1998current}, reduce anxiety of autistic children \cite{gershon2004pilot,parsons2011state,arane2017virtual,didehbani2016virtual}, develop children's activity \cite{foley2010use}, practice skills \cite{schwebel2010using,ip2018enhance}.
Finally, computer vision and natural language processing are used to recognize children's faces and children's speech respectively to provide interactive information for early childhood education.

Although AI technologies have been applied to early childhood education, the current studies of AI in ECE still have various open issues that need to be further investigated.
First, AI technologies have not been investigated in many ECE tasks.
For example, using data mining and machine learning technologies to discover the proper learning ways for various children, and using virtual reality and computer vision to design interesting AI programs for ECE.
Furthermore, the current studies generally apply AI technologies to ECE at the level of scientific research.
However, the ECE need more AI educational tools to show children visual AI concepts and applications, particularly interactive and practical AI educational systems.
Second, most current studies applied typical and outdated AI technologies to ECE.
We assume that the researchers in this domain usually do not follow state-of-the-art AI technologies. 
Therefore, advanced AI technologies such as ChatGPT need to be further investigated in ECE.

Finally, datasets are generally crucial for current AI technologies, particularly data-driven AI approaches such as deep learning.
The available datasets are scarce since the useful data in ECE is hardly collected.
It is significantly hard to collect quality data from children, and the data in ECE is generally sparse even untrustworthy.
Therefore, data mining technologies are crucial to extract useful and large-scale data and build various datasets for supporting AI applications in various ECE tasks.
We summarize the used datasets in these studies of AI in ECE, as shown in \autoref{tab:ai_datasets} where we addressed the dataset information.

\section{Challenges and Trends}
\label{sec:challenges_trends}

Research trends and challenges are crucial for researchers to find interesting ideas.
We aim to provide organized insights for the future development of ECE.
In this section, we analyse the research trends and discuss the open challenges for shaping future research via bibliometric analysis. 
We retrieve the publication for discovering and demonstrating the trends of the key AI technologies in ECE. 

\subsection{Challenges}
\label{subsec:challenges}


Although AI technologies have been widely applied in ECE, current studies generally use weak AI technologies in ECE.
The AI technologies in ECE need to be further developed. 
There are many challenges to the further development of AI technologies in ECE.
In this section, we discuss the challenges of the key AI technologies in ECE, including the challenges of data mining, machine learning, deep learning, and the challenges towards educators.

\subsubsection{Challenges for Data Mining in ECE}
Although data mining technologies have been applied in ECE, there are still many open challenges \cite{baker2019challenges}.

\begin{itemize} [noitemsep]
    \item First, the transferability of the data mining model is a common challenge, which transfers the data mining models for various learning systems in ECE, e.g., the mentioned issues in \cite{chang2007study,krotova2020diagnostics}.
    \item Second, the durability of the data mining model is a challenge for the different children in different years \cite{yue2018application}. 
    \item Third, the generalizability of data mining is a significant challenge.
    The current data mining models have to be rebuilt basically from scratch for various different children learning systems in ECE \cite{al2018fuzzy}.
\end{itemize}

\subsubsection{Challenges for Machine Learning in ECE}

The technical challenges for machine learning in ECE mainly involve the data and models. 
\begin{itemize}[noitemsep]
\item Overfitting of training data is a common challenge for machine learning in ECE when data is massive amounts of noisy and biased data, which can generally be solved by removing outliers data and longer training time \cite{rasheed2021use,mcginnis2018wearable,mcginnis2019rapid}.

\item The absence of good-quality data in ECE is a current challenge for machine learning since it is hard to collect good-quality data from young children \cite{papakostas2021estimating,hagenbuchner2015prediction}. 
The operation of removing outliers, filtering missing values, and removing unwanted features can be used for data preprocessing.

\item Explainability and accountability are two major issues of using machine learning in education \cite{razaulla2022integration,webb2021machine}.
Explainability is described as the ability to understand and explain “in human terms” what is happening with the model.
Accountability refers to the ability to explain and justify methods, actions and decisions.

\end{itemize}

\subsubsection{Challenges for Deep Learning in ECE}

Although deep learning has proven remarkable performance in ECE, its open challenges are also well-known \cite{li2022interpretable}.

\begin{itemize}[noitemsep]
\item A well-known common challenge of deep learning in ECE is the trustworthiness of deep learning.
The results of black box models in deep learning are usually unknown to the trustworthy.

\item Unsupervised/semi-supervised learning with a few labelled data is a reasonable approach and a significant challenge as well.
Collecting and labelling large-scale data for supervised learning is a significant challenge that is even harder than collecting large-scale \cite{kim2015pororobot,xia2017detecting}.
Especially, there is a lack of labelled large-scale data in ECE.

\item 
The state-of-the-art deep learning models generally require various computing costs and powerful GPUs for the iterative process \cite{li2022interpretable}.
The optimization of deep learning models with smaller sizes for less computing cost is a difficult challenge \cite{rudovic2018culturenet,chatzimichail2010artificial}.

\item Establishing a clear interpretable model is a significant challenge in ECE \cite{li2022interpretable}.
Current deep learning models are hard to be explained how they generate specific results.
Interpretable techniques such as rule-based approaches, linear regression and tree-based algorithms could be fit to the interpretability in ECE \cite{delavarian2011automatic,su2020machine}.

\end{itemize}

\subsubsection{Challenges for Other AI technologies in ECE}

\begin{itemize} [noitemsep]
\item Although other AI technologies have been applied in ECE, there are still certain challenges that need to be investigated.
For example, one of the biggest challenges faced by virtual reality in ECE is the lack of content \cite{jyoti2019virtual,ip2018enhance}. 
However, developing VR content can be very expensive, which is a significant issue for ECE.

\item The challenges of natural language processing in ECE are mainly data- and semantic-related \cite{druga2017hey}.
In particular, preparing high-quality data from children is a challenging issue.
The semantic meaning of words from children is another common challenge for natural language processing in ECE \cite{shahi2021using}. 
In summary, most of the challenges of AI technologies in ECE are due to data quality such as sparsity, diversity and dimensionality. 

\end{itemize}

\subsubsection{Challenges Towards Educators}


To date, various funds have been invested in AI technologies to equip schools with educational technology tools, but the vast majority of teachers do not use AI technologies in meaningful ways \cite{national2008access,office1995teachers}.
Only a few teachers allow students to use educational technology tools to solve problems, analyze data, present information graphically, or participate in distance learning \cite{us2003federal}.
Large-scale studies have shown a significant increase in achievement scores of students using technology as a learning tool \cite{lei2007technology}.
The studies \cite{dani2008technology,schroeder2007meta,songer2007digital} demonstrate that the use of AI technologies has a positive influence on a wide variety of student learning, including the understanding of science concepts and the development of scientific reasoning skills.

Even though there have been available educational AI technologies tools in the classroom, the integration of AI technologies in ECE is still challenging for teachers.
With the use of AI technologies, teachers can prune curriculums and provide personalized learning options. 
Guzey and Roehrig \cite{guzey2012integrating} integrated educational technology into secondary science teaching to develop the foundation for understanding teachers’ use of technology.
The findings demonstrate that all the participating teachers were motivated to use AI technology in their teaching.
Many educators have brought AI concepts into ECE. 
We summarize four basic things that educators should know.
\begin{itemize}[noitemsep]
    \item Educators need a basic understanding of AI concepts.
    \item AI technologies are advancing rapidly that teaching curricula are irrelevant and lack proper AI instruction. 
    \item Children are intimately familiar with AI-driven technology. 
    Teachers can help children understand AI applications by connecting AI concepts to products.
    \item Coding is the baseline of AI instruction but not the goal. 
\end{itemize}

Aside from the classroom and children’s experience, AI can be used to improve the quality of education by empowering childhood practitioners. 
For example, AI technologies could be used for automating standardized procedures, gaining insights from data analytics, or providing AI-based professional development programs. 
AI educational technologies provide benefits compared to traditional education, including inspiring students' motives for creative activities \cite{prentzas2013artificial}, assisting teachers to provide students with personalized self-adaptive learning \cite{xu2020dilemma} and remove repetitive tasks and value-add to meaningful teaching tasks such as interactions with children.
Therefore, using AI technologies to develop programs, curricula, and design novel meaningful teaching tasks to meet the requirements of early childhood education is a big challenge for educators even parents in the AI era.
Finally, we assure that the role of teachers will never be replaced by AI. 
The human element is crucial in ECE.
Students are not able to learn purely through AI-assist. 

\subsection{Trends}
\label{subsec:trends}

In this section, we analyse the trends of AI in ECE from the aspects of intelligent robots and AI technologies by bibliometrics.

\subsubsection{Intelligent robots}

\begin{figure*}[!ht] \centering
    \includegraphics[width=0.7\textwidth,trim={15 0 20 20},clip]{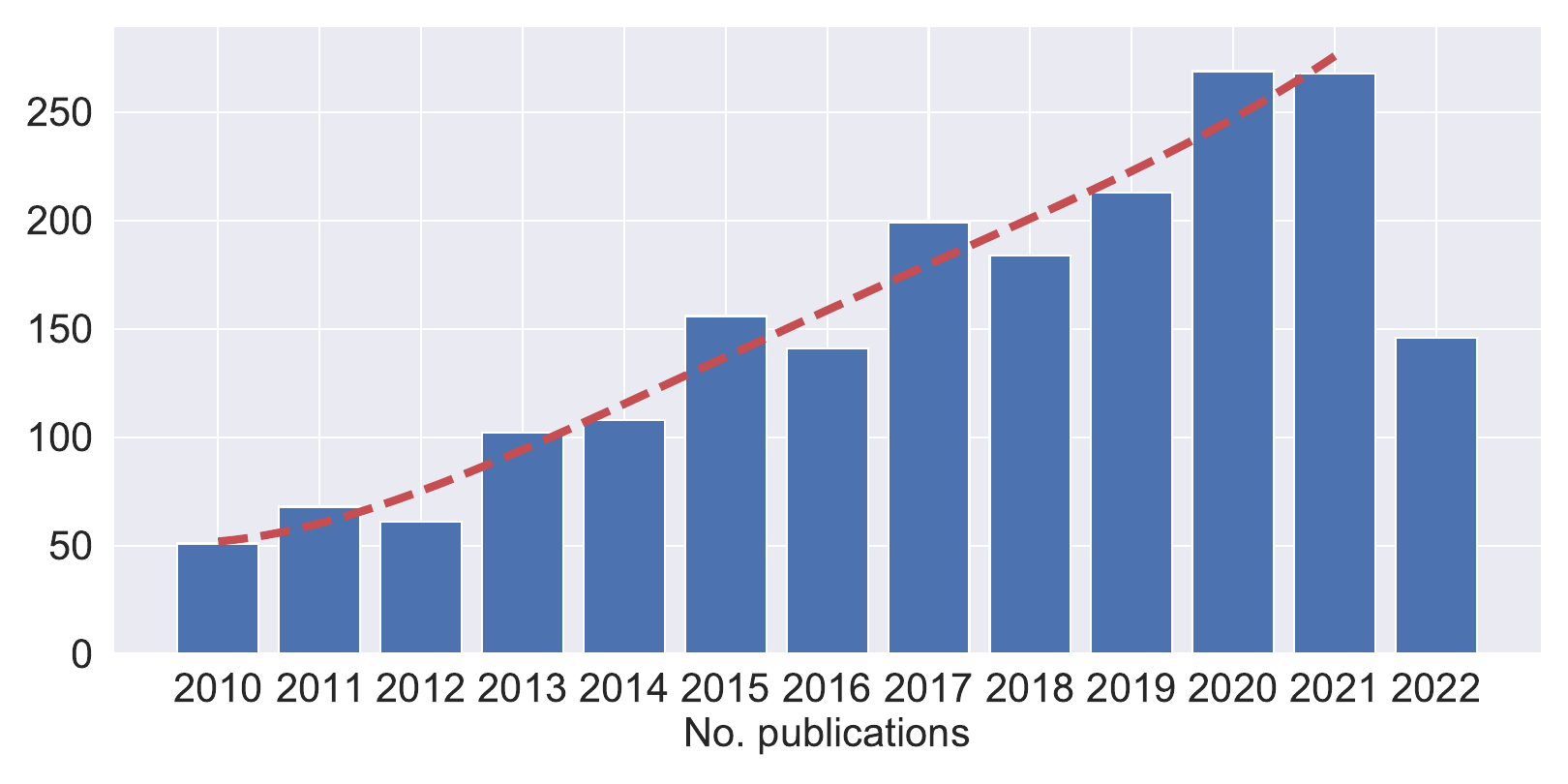}
    \caption{The number of publications per year (from 2010 to 2022) on the topic of intelligent robots in ECE. The red dashed line is the cubic polynomial fit of the number of publications. Scopus database returned 2169 results until 10/10/2022.}\label{fig:intelligent_years}
\end{figure*}

Intelligent robots have demonstrated good performance in ECE.
In particular, many commercialized robots (see \autoref{fig:robots}) have been widely applied in ECE. 
We retrieve the literature on intelligent robots in ECE with the keywords by the search code of title-abs-key (("robot" OR "robots") AND ("early childhood education" OR "autism" OR "autism spectrum disorder" OR "ASD" OR "preschool" OR "kindergarten")).
The number of publications over the years is shown in \autoref{fig:intelligent_years}.
The number of publications on the topic of intelligent robots in ECE shows a significantly increasing trend.
Although more and more works apply intelligent robots to ECE, current intelligent robots generally integrate weak AI even non-AI (computer-based learning systems).
We conclude that intelligent robots, in particular AI-based intelligent robots, will be an interesting research trend in ECE.

We summarize the specific research trends of intelligent robots in ECE that also respond to the discussion and research challenges.
\begin{itemize}[noitemsep]
\item Integrating state-of-the-art AI technologies into intelligent robots for ECE, such as the AI-powered language model ChatGPT, to generate human-like interactive education with children.
\item Investigating novel applications of AI-based intelligent robots to assist early childhood education, particularly the education of children with interaction disorders.
\item Developing more AI educational tools with intelligent robots to show children visual AI concepts and applications, particularly interactive and practical AI educational systems.
\end{itemize}

\begin{figure*}[!ht] \centering
    \includegraphics[width=0.7\textwidth,trim={10 0 20 18},clip]{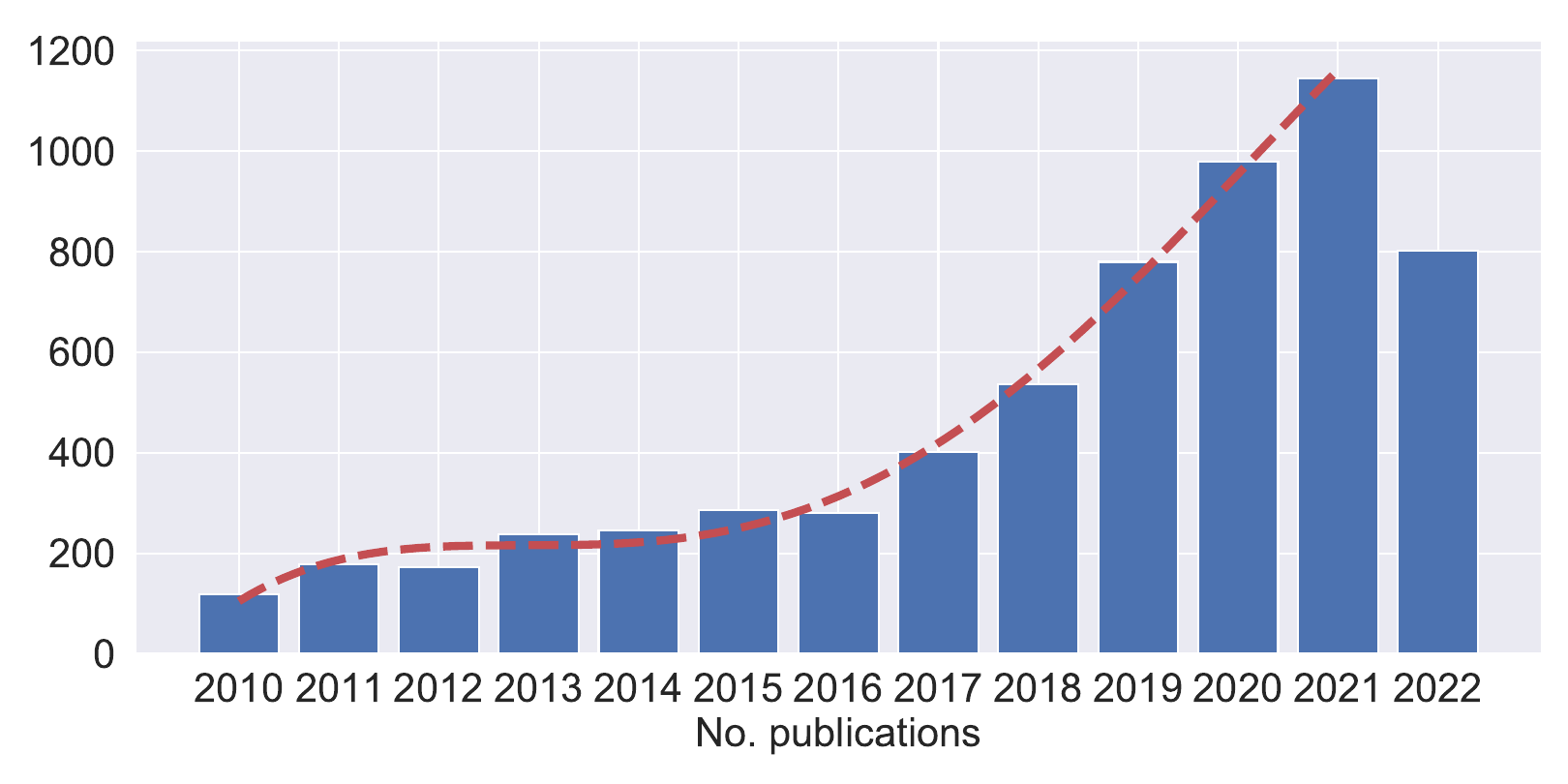}
    \caption{The number of publications per year (from 2010 to 2022) on the topic of the key AI technologies in ECE. The red dashed line is the cubic polynomial fit of the number of publications. Scopus database returned 7134 results until 10/10/2022.}\label{fig:ai_years}
\end{figure*} 

\subsubsection{AI technologies}
In the past years, AI technology is significantly developed in many domains, e.g., ECE. 
We retrieve the literature on AI-based ECE with the keywords by the search code of title-abs-key(("artificial intelligence" OR "AI" OR "machine learning" OR "deep learning" OR "data mining" OR "virtual reality" OR "natural language processing") AND ("early childhood education" OR "autism" OR "autism spectrum disorder" OR "ASD" OR "preschool" OR "kindergarten")).
The number of publications over the years is shown in \autoref{fig:ai_years}.
More and more studies apply AI technologies to ECE.
The number of publications shows a growth trend from 2010.
In particular, it shows a significant increase since 2017 which is a period of the explosive development of AI and its applications.
In summary, various AI technologies are significant research trends in ECE.

We summarize the specific research trends of AI in ECE that are imminent and significant to be investigated.  
\begin{itemize}[noitemsep]
    \item Developing useful and large-scale data and building various datasets for supporting AI applications in various ECE tasks due to the absence of good-quality data for AI in ECE.
    \item Developing real-time AI-based ECE models for childhood education on common computing systems.  
    \item Applying state-of-the-art AI technologies to the ECE tasks for better early childhood education systems, such as interaction with ChatGPT.
    \item Extending novel applications of AI in ECE for improving early childhood education, e.g., using generative deep neural networks to generate ECE contents for AI-designed curriculum.
\end{itemize}


\section{Conclusion}
\label{sec:conclusion}
%
In this paper, we provide an up-to-date and in-depth overview of the studies of the key AI technologies in ECE by discussing the representative studies, delineating the historical perspective, and outlining open questions.
We review the studies that apply AI-based robots and the key AI technologies to ECE, including improving the social interaction of children with an autism spectrum disorder.
We discuss the challenges and trends through a detailed bibliometric analysis and provide insightful recommendations for future research.
This paper significantly contributes to comprehensively reviewing the up-to-date studies, which are suitable as introductory material for beginners to AI technologies in ECE, as well as supplementary material for advanced users.


Although we present a complete overview of the existing literature on \ai in \ece, there are still multiple open issues and possible future work. 
Here, we outline some of them and highlight possible extensions of our work.
First, we review the limited number of representative studies due to our limited knowledge and space limitations, some meaningful research findings may not be discussed in this paper.
Second, this paper focuses on reviewing the studies of AI in ECE, social and economic factors were not discussed. 
We believe that the social and economic factors of AI in ECE are interesting for many readers.
Finally, in the further, we will review the studies by cross-discussing our studies about AI in ECE for more technical details. 






\bmhead{Acknowledgments}

This work was partially supported by the GuangDong Basic and Applied Basic Research Foundation (No: 2021A1515110641).

\section*{Declarations}

\bmhead{Confict of interest}
The authors have no competing interests to declare that are relevant to the content of this article.

\bibliography{sn-bibliography}


\begin{thebibliography}{160}
\ifx \bisbn   \undefined \def \bisbn  #1{ISBN #1}\fi
\ifx \binits  \undefined \def \binits#1{#1}\fi
\ifx \bauthor  \undefined \def \bauthor#1{#1}\fi
\ifx \batitle  \undefined \def \batitle#1{#1}\fi
\ifx \bjtitle  \undefined \def \bjtitle#1{#1}\fi
\ifx \bvolume  \undefined \def \bvolume#1{\textbf{#1}}\fi
\ifx \byear  \undefined \def \byear#1{#1}\fi
\ifx \bissue  \undefined \def \bissue#1{#1}\fi
\ifx \bfpage  \undefined \def \bfpage#1{#1}\fi
\ifx \blpage  \undefined \def \blpage #1{#1}\fi
\ifx \burl  \undefined \def \burl#1{\textsf{#1}}\fi
\ifx \doiurl  \undefined \def \doiurl#1{\url{https://doi.org/#1}}\fi
\ifx \betal  \undefined \def \betal{\textit{et al.}}\fi
\ifx \binstitute  \undefined \def \binstitute#1{#1}\fi
\ifx \binstitutionaled  \undefined \def \binstitutionaled#1{#1}\fi
\ifx \bctitle  \undefined \def \bctitle#1{#1}\fi
\ifx \beditor  \undefined \def \beditor#1{#1}\fi
\ifx \bpublisher  \undefined \def \bpublisher#1{#1}\fi
\ifx \bbtitle  \undefined \def \bbtitle#1{#1}\fi
\ifx \bedition  \undefined \def \bedition#1{#1}\fi
\ifx \bseriesno  \undefined \def \bseriesno#1{#1}\fi
\ifx \blocation  \undefined \def \blocation#1{#1}\fi
\ifx \bsertitle  \undefined \def \bsertitle#1{#1}\fi
\ifx \bsnm \undefined \def \bsnm#1{#1}\fi
\ifx \bsuffix \undefined \def \bsuffix#1{#1}\fi
\ifx \bparticle \undefined \def \bparticle#1{#1}\fi
\ifx \barticle \undefined \def \barticle#1{#1}\fi
\bibcommenthead
\ifx \bconfdate \undefined \def \bconfdate #1{#1}\fi
\ifx \botherref \undefined \def \botherref #1{#1}\fi
\ifx \url \undefined \def \url#1{\textsf{#1}}\fi
\ifx \bchapter \undefined \def \bchapter#1{#1}\fi
\ifx \bbook \undefined \def \bbook#1{#1}\fi
\ifx \bcomment \undefined \def \bcomment#1{#1}\fi
\ifx \oauthor \undefined \def \oauthor#1{#1}\fi
\ifx \citeauthoryear \undefined \def \citeauthoryear#1{#1}\fi
\ifx \endbibitem  \undefined \def \endbibitem {}\fi
\ifx \bconflocation  \undefined \def \bconflocation#1{#1}\fi
\ifx \arxivurl  \undefined \def \arxivurl#1{\textsf{#1}}\fi
\csname PreBibitemsHook\endcsname

\bibitem[\protect\citeauthoryear{Williams et~al.}{2018}]{williams2018popbots}
\begin{botherref}
\oauthor{\bsnm{Williams}, \binits{R.}}, et al.:
{PopBots}: leveraging social robots to aid preschool children's artificial
  intelligence education.
PhD thesis,
Massachusetts Institute of Technology
(2018)
\end{botherref}
\endbibitem

\bibitem[\protect\citeauthoryear{Prentzas}{2013}]{prentzas2013artificial}
\begin{bchapter}
\bauthor{\bsnm{Prentzas}, \binits{J.}}:
\bctitle{Artificial intelligence methods in early childhood education}.
In: \bbtitle{Artificial Intelligence, Evolutionary Computing and
  Metaheuristics},
pp. \bfpage{169}--\blpage{199}
(\byear{2013})
\end{bchapter}
\endbibitem

\bibitem[\protect\citeauthoryear{Roblyer and
  Doering}{2007}]{roblyer2007integrating}
\begin{bbook}
\bauthor{\bsnm{Roblyer}, \binits{M.}},
\bauthor{\bsnm{Doering}, \binits{A.H.}}:
\bbtitle{Integrating Educational Technology Into Teaching},
(\byear{2007})
\end{bbook}
\endbibitem

\bibitem[\protect\citeauthoryear{Xu}{2020}]{xu2020dilemma}
\begin{bchapter}
\bauthor{\bsnm{Xu}, \binits{L.}}:
\bctitle{The dilemma and countermeasures of ai in educational application}.
In: \bbtitle{2020 4th International Conference on Computer Science and
  Artificial Intelligence},
pp. \bfpage{289}--\blpage{294}
(\byear{2020})
\end{bchapter}
\endbibitem

\bibitem[\protect\citeauthoryear{Su and Yang}{2022}]{su2022artificial}
\begin{botherref}
\oauthor{\bsnm{Su}, \binits{J.}},
\oauthor{\bsnm{Yang}, \binits{W.}}:
Artificial intelligence in early childhood education: A scoping review.
Computers and Education: Artificial Intelligence,
100049
(2022)
\end{botherref}
\endbibitem

\bibitem[\protect\citeauthoryear{Gammage}{2006}]{gammage2006early}
\begin{barticle}
\bauthor{\bsnm{Gammage}, \binits{P.}}:
\batitle{Early childhood education and care: politics, policies and
  possibilities}.
\bjtitle{Early years}
\bvolume{26}(\bissue{3}),
\bfpage{235}--\blpage{248}
(\byear{2006})
\end{barticle}
\endbibitem

\bibitem[\protect\citeauthoryear{Holland}{2000}]{holland2000artificial}
\begin{barticle}
\bauthor{\bsnm{Holland}, \binits{S.}}:
\batitle{Artificial intelligence in music education: A critical review}.
\bjtitle{Readings in Music and Artificial Intelligence}
\bvolume{20},
\bfpage{239}--\blpage{274}
(\byear{2000})
\end{barticle}
\endbibitem

\bibitem[\protect\citeauthoryear{Chassignol
  et~al.}{2012}]{drigas2012artificial}
\begin{barticle}
\bauthor{\bsnm{Chassignol}, \binits{M.}},
\bauthor{\bsnm{Khoroshavin}, \binits{A.}},
\bauthor{\bsnm{Klimova}, \binits{A.}},
\bauthor{\bsnm{Bilyatdinova}, \binits{A.}}:
\batitle{Artificial intelligence in special education: A decade review}.
\bjtitle{International Journal of Engineering Education}
\bvolume{28}(\bissue{6}),
\bfpage{1366}
(\byear{2012})
\end{barticle}
\endbibitem

\bibitem[\protect\citeauthoryear{Drigas and Ioannidou}{2011}]{drigas2011review}
\begin{bchapter}
\bauthor{\bsnm{Drigas}, \binits{A.S.}},
\bauthor{\bsnm{Ioannidou}, \binits{R.-E.}}:
\bctitle{A review on artificial intelligence in special education}.
In: \bbtitle{World Summit on Knowledge Society},
pp. \bfpage{385}--\blpage{391}
(\byear{2011}).
\bcomment{Springer}
\end{bchapter}
\endbibitem

\bibitem[\protect\citeauthoryear{Huijnen et~al.}{2017}]{huijnen2017robots}
\begin{barticle}
\bauthor{\bsnm{Huijnen}, \binits{C.}},
\bauthor{\bsnm{Lexis}, \binits{M.}},
\bauthor{\bsnm{De~Witte}, \binits{L.}}:
\batitle{Robots as new tools in therapy and education for children with
  autism}.
\bjtitle{International Journal of Neurorehabilitation}
\bvolume{4}(\bissue{4}),
\bfpage{1}--\blpage{4}
(\byear{2017})
\end{barticle}
\endbibitem

\bibitem[\protect\citeauthoryear{Chassignol
  et~al.}{2018}]{chassignol2018artificial}
\begin{barticle}
\bauthor{\bsnm{Chassignol}, \binits{M.}},
\bauthor{\bsnm{Khoroshavin}, \binits{A.}},
\bauthor{\bsnm{Klimova}, \binits{A.}},
\bauthor{\bsnm{Bilyatdinova}, \binits{A.}}:
\batitle{Artificial intelligence trends in education: a narrative overview}.
\bjtitle{Procedia Computer Science}
\bvolume{136},
\bfpage{16}--\blpage{24}
(\byear{2018})
\end{barticle}
\endbibitem

\bibitem[\protect\citeauthoryear{Zawacki-Richter
  et~al.}{2019}]{zawacki2019systematic}
\begin{barticle}
\bauthor{\bsnm{Zawacki-Richter}, \binits{O.}},
\bauthor{\bsnm{Mar{\'\i}n}, \binits{V.I.}},
\bauthor{\bsnm{Bond}, \binits{M.}},
\bauthor{\bsnm{Gouverneur}, \binits{F.}}:
\batitle{Systematic review of research on artificial intelligence applications
  in higher education--where are the educators?}
\bjtitle{International Journal of Educational Technology in Higher Education}
\bvolume{16}(\bissue{1}),
\bfpage{1}--\blpage{27}
(\byear{2019})
\end{barticle}
\endbibitem

\bibitem[\protect\citeauthoryear{Zaidi et~al.}{2019}]{zaidi2019review}
\begin{botherref}
\oauthor{\bsnm{Zaidi}, \binits{A.}},
\oauthor{\bsnm{Beadle}, \binits{S.}},
\oauthor{\bsnm{Hannah}, \binits{A.}}:
Review of the online learning and artificial intelligence education market.
Department for Education, ICF Consulting Services Ltd
(2019)
\end{botherref}
\endbibitem

\bibitem[\protect\citeauthoryear{Chan and Zary}{2019}]{chan2019applications}
\begin{barticle}
\bauthor{\bsnm{Chan}, \binits{K.S.}},
\bauthor{\bsnm{Zary}, \binits{N.}}:
\batitle{Applications and challenges of implementing artificial intelligence in
  medical education: integrative review}.
\bjtitle{JMIR medical education}
\bvolume{5}(\bissue{1}),
\bfpage{13930}
(\byear{2019})
\end{barticle}
\endbibitem

\bibitem[\protect\citeauthoryear{Chen et~al.}{2020}]{chen2020artificial}
\begin{barticle}
\bauthor{\bsnm{Chen}, \binits{L.}},
\bauthor{\bsnm{Chen}, \binits{P.}},
\bauthor{\bsnm{Lin}, \binits{Z.}}:
\batitle{Artificial intelligence in education: a review}.
\bjtitle{IEEE Access}
\bvolume{8},
\bfpage{75264}--\blpage{75278}
(\byear{2020})
\end{barticle}
\endbibitem

\bibitem[\protect\citeauthoryear{Sapci and Sapci}{2020}]{sapci2020artificial}
\begin{barticle}
\bauthor{\bsnm{Sapci}, \binits{A.H.}},
\bauthor{\bsnm{Sapci}, \binits{H.A.}}:
\batitle{Artificial intelligence education and tools for medical and health
  informatics students: systematic review}.
\bjtitle{JMIR Medical Education}
\bvolume{6}(\bissue{1}),
\bfpage{19285}
(\byear{2020})
\end{barticle}
\endbibitem

\bibitem[\protect\citeauthoryear{Chen et~al.}{2020}]{chen2020application}
\begin{barticle}
\bauthor{\bsnm{Chen}, \binits{X.}},
\bauthor{\bsnm{Xie}, \binits{H.}},
\bauthor{\bsnm{Zou}, \binits{D.}},
\bauthor{\bsnm{Hwang}, \binits{G.-J.}}:
\batitle{Application and theory gaps during the rise of artificial intelligence
  in education}.
\bjtitle{Computers and Education: Artificial Intelligence}
\bvolume{1},
\bfpage{100002}
(\byear{2020})
\end{barticle}
\endbibitem

\bibitem[\protect\citeauthoryear{Ahmad et~al.}{2020}]{ahmad2020artificial}
\begin{botherref}
\oauthor{\bsnm{Ahmad}, \binits{K.}},
\oauthor{\bsnm{Qadir}, \binits{J.}},
\oauthor{\bsnm{Al-Fuqaha}, \binits{A.}},
\oauthor{\bsnm{Iqbal}, \binits{W.}},
\oauthor{\bsnm{El-Hassan}, \binits{A.}},
\oauthor{\bsnm{Benhaddou}, \binits{D.}},
\oauthor{\bsnm{Ayyash}, \binits{M.}}:
Artificial intelligence in education: A panoramic review
(2020)
\end{botherref}
\endbibitem

\bibitem[\protect\citeauthoryear{Buchanan et~al.}{2021}]{buchanan2021predicted}
\begin{barticle}
\bauthor{\bsnm{Buchanan}, \binits{C.}},
\bauthor{\bsnm{Howitt}, \binits{M.L.}},
\bauthor{\bsnm{Wilson}, \binits{R.}},
\bauthor{\bsnm{Booth}, \binits{R.G.}},
\bauthor{\bsnm{Risling}, \binits{T.}},
\bauthor{\bsnm{Bamford}, \binits{M.}}:
\batitle{Predicted influences of artificial intelligence on nursing education:
  Scoping review}.
\bjtitle{JMIR Nursing}
\bvolume{4}(\bissue{1}),
\bfpage{23933}
(\byear{2021})
\end{barticle}
\endbibitem

\bibitem[\protect\citeauthoryear{Maghsudi
  et~al.}{2021}]{maghsudi2021personalized}
\begin{barticle}
\bauthor{\bsnm{Maghsudi}, \binits{S.}},
\bauthor{\bsnm{Lan}, \binits{A.}},
\bauthor{\bsnm{Xu}, \binits{J.}},
\bauthor{\bsnm{Der~Schaar}, \binits{M.}}:
\batitle{Personalized education in the artificial intelligence era: What to
  expect next}.
\bjtitle{IEEE Signal Processing Magazine}
\bvolume{38}(\bissue{3}),
\bfpage{37}--\blpage{50}
(\byear{2021})
\end{barticle}
\endbibitem

\bibitem[\protect\citeauthoryear{Iftikhar
  et~al.}{2019}]{iftikhar2019bibliometric}
\begin{botherref}
\oauthor{\bsnm{Iftikhar}, \binits{P.M.}},
\oauthor{\bsnm{Ali}, \binits{F.}},
\oauthor{\bsnm{Faisaluddin}, \binits{M.}},
\oauthor{\bsnm{Khayyat}, \binits{A.}},
\oauthor{\bsnm{De~Sa}, \binits{M.D.G.}},
\oauthor{\bsnm{Rao}, \binits{T.}}:
A bibliometric analysis of the top 30 most-cited articles in gestational
  diabetes mellitus literature (1946-2019).
Cureus
\textbf{11}(2)
(2019)
\end{botherref}
\endbibitem

\bibitem[\protect\citeauthoryear{Falagas et~al.}{2008}]{falagas2008comparison}
\begin{barticle}
\bauthor{\bsnm{Falagas}, \binits{M.E.}},
\bauthor{\bsnm{Pitsouni}, \binits{E.I.}},
\bauthor{\bsnm{Malietzis}, \binits{G.A.}},
\bauthor{\bsnm{Pappas}, \binits{G.}}:
\batitle{Comparison of pubmed, scopus, web of science, and google scholar:
  strengths and weaknesses}.
\bjtitle{The FASEB journal}
\bvolume{22}(\bissue{2}),
\bfpage{338}--\blpage{342}
(\byear{2008})
\end{barticle}
\endbibitem

\bibitem[\protect\citeauthoryear{Rose and
  Kitchin}{2019}]{rose2019pybliometrics}
\begin{barticle}
\bauthor{\bsnm{Rose}, \binits{M.E.}},
\bauthor{\bsnm{Kitchin}, \binits{J.R.}}:
\batitle{Pybliometrics: Scriptable bibliometrics using a python interface to
  scopus}.
\bjtitle{SoftwareX}
\bvolume{10},
\bfpage{100263}
(\byear{2019})
\end{barticle}
\endbibitem

\bibitem[\protect\citeauthoryear{Ismail et~al.}{2018}]{ismail2018application}
\begin{barticle}
\bauthor{\bsnm{Ismail}, \binits{M.}},
\bauthor{\bsnm{Azaman}, \binits{N.}},
\bauthor{\bsnm{Khalid}, \binits{N.}}:
\batitle{Application of robots to improve social and communication skills among
  autistic children}.
\bjtitle{Journal of Advanced Manufacturing Technology (JAMT)}
\bvolume{12}(\bissue{1 (1)}),
\bfpage{421}--\blpage{430}
(\byear{2018})
\end{barticle}
\endbibitem

\bibitem[\protect\citeauthoryear{Haibin}{2012}]{haibin2012development}
\begin{botherref}
\oauthor{\bsnm{Haibin}, \binits{Y.}}:
Development of a robotic nanny for children and a case study of emotion
  recognition in human-robotic interaction.
Department o mechanical engineering, National University of Singapore
(2012)
\end{botherref}
\endbibitem

\bibitem[\protect\citeauthoryear{Koay et~al.}{2007}]{koay2007living}
\begin{bchapter}
\bauthor{\bsnm{Koay}, \binits{K.L.}},
\bauthor{\bsnm{Syrdal}, \binits{D.S.}},
\bauthor{\bsnm{Walters}, \binits{M.L.}},
\bauthor{\bsnm{Dautenhahn}, \binits{K.}}:
\bctitle{Living with robots: Investigating the habituation effect in
  participants' preferences during a longitudinal human-robot interaction
  study}.
In: \bbtitle{RO-MAN 2007-The 16th IEEE International Symposium on Robot and
  Human Interactive Communication},
pp. \bfpage{564}--\blpage{569}
(\byear{2007}).
\bcomment{IEEE}
\end{bchapter}
\endbibitem

\bibitem[\protect\citeauthoryear{Austermann and
  Yamada}{2008}]{austermann2008good}
\begin{bchapter}
\bauthor{\bsnm{Austermann}, \binits{A.}},
\bauthor{\bsnm{Yamada}, \binits{S.}}:
\bctitle{“good robot”,“bad robot”—analyzing users’ feedback in a
  human-robot teaching task}.
In: \bbtitle{RO-MAN 2008-The 17th IEEE International Symposium on Robot and
  Human Interactive Communication},
pp. \bfpage{41}--\blpage{46}
(\byear{2008}).
\bcomment{IEEE}
\end{bchapter}
\endbibitem

\bibitem[\protect\citeauthoryear{Hyun et~al.}{2010}]{hyun2010usability}
\begin{barticle}
\bauthor{\bsnm{Hyun}, \binits{E.-J.}},
\bauthor{\bsnm{Park}, \binits{H.-K.}},
\bauthor{\bsnm{Jang}, \binits{S.-K.}},
\bauthor{\bsnm{Yeon}, \binits{H.-M.}}:
\batitle{The usability of a robot as an educational assistant in a kindergarten
  and young children's perceptions of their relationship with the robot}.
\bjtitle{Korean Journal of Child Studies}
\bvolume{31}(\bissue{1}),
\bfpage{267}--\blpage{282}
(\byear{2010})
\end{barticle}
\endbibitem

\bibitem[\protect\citeauthoryear{Hyun et~al.}{2012}]{hyun2012young}
\begin{bchapter}
\bauthor{\bsnm{Hyun}, \binits{E.}},
\bauthor{\bsnm{Lee}, \binits{H.}},
\bauthor{\bsnm{Yeon}, \binits{H.}}:
\bctitle{Young children's perception of irobiq, the teacher assistive robot,
  with reference to speech register}.
In: \bbtitle{2012 8th International Conference on Computing Technology and
  Information Management (NCM and ICNIT)},
vol. \bseriesno{1},
pp. \bfpage{366}--\blpage{369}
(\byear{2012}).
\bcomment{IEEE}
\end{bchapter}
\endbibitem

\bibitem[\protect\citeauthoryear{Han et~al.}{2015}]{han2015examining}
\begin{barticle}
\bauthor{\bsnm{Han}, \binits{J.}},
\bauthor{\bsnm{Jo}, \binits{M.}},
\bauthor{\bsnm{Hyun}, \binits{E.}},
\bauthor{\bsnm{So}, \binits{H.-J.}}:
\batitle{Examining young children’s perception toward augmented
  reality-infused dramatic play}.
\bjtitle{Educational Technology Research and Development}
\bvolume{63}(\bissue{3}),
\bfpage{455}--\blpage{474}
(\byear{2015})
\end{barticle}
\endbibitem

\bibitem[\protect\citeauthoryear{Hyun and Yoon}{2009}]{hyun2009characteristics}
\begin{bchapter}
\bauthor{\bsnm{Hyun}, \binits{E.}},
\bauthor{\bsnm{Yoon}, \binits{H.}}:
\bctitle{Characteristics of young children's utilization of a robot during play
  time: A case study}.
In: \bbtitle{RO-MAN 2009-The 18th IEEE International Symposium on Robot and
  Human Interactive Communication},
pp. \bfpage{675}--\blpage{680}
(\byear{2009}).
\bcomment{IEEE}
\end{bchapter}
\endbibitem

\bibitem[\protect\citeauthoryear{Hyun et~al.}{2010}]{hyun2010relationships}
\begin{bchapter}
\bauthor{\bsnm{Hyun}, \binits{E.}},
\bauthor{\bsnm{Yoon}, \binits{H.}},
\bauthor{\bsnm{Son}, \binits{S.}}:
\bctitle{Relationships between user experiences and children's perceptions of
  the education robot}.
In: \bbtitle{2010 5th ACM/IEEE International Conference on Human-Robot
  Interaction (HRI)},
pp. \bfpage{199}--\blpage{200}
(\byear{2010}).
\bcomment{IEEE}
\end{bchapter}
\endbibitem

\bibitem[\protect\citeauthoryear{Neumann}{2020}]{neumann2020social}
\begin{barticle}
\bauthor{\bsnm{Neumann}, \binits{M.M.}}:
\batitle{Social robots and young children’s early language and literacy
  learning}.
\bjtitle{Early Childhood Education Journal}
\bvolume{48}(\bissue{2}),
\bfpage{157}--\blpage{170}
(\byear{2020})
\end{barticle}
\endbibitem

\bibitem[\protect\citeauthoryear{Lee and Hyun}{2015}]{lee2015intelligent}
\begin{barticle}
\bauthor{\bsnm{Lee}, \binits{H.}},
\bauthor{\bsnm{Hyun}, \binits{E.}}:
\batitle{The intelligent robot contents for children with speech-language
  disorder}.
\bjtitle{Journal of Educational Technology \& Society}
\bvolume{18}(\bissue{3}),
\bfpage{100}--\blpage{113}
(\byear{2015})
\end{barticle}
\endbibitem

\bibitem[\protect\citeauthoryear{Hsiao et~al.}{2015}]{hsiao2015irobiq}
\begin{barticle}
\bauthor{\bsnm{Hsiao}, \binits{H.-S.}},
\bauthor{\bsnm{Chang}, \binits{C.-S.}},
\bauthor{\bsnm{Lin}, \binits{C.-Y.}},
\bauthor{\bsnm{Hsu}, \binits{H.-L.}}:
\batitle{{iRobiQ}: the influence of bidirectional interaction on
  kindergarteners’ reading motivation, literacy, and behavior}.
\bjtitle{Interactive Learning Environments}
\bvolume{23}(\bissue{3}),
\bfpage{269}--\blpage{292}
(\byear{2015})
\end{barticle}
\endbibitem

\bibitem[\protect\citeauthoryear{Gouaillier
  et~al.}{2009}]{gouaillier2009mechatronic}
\begin{bchapter}
\bauthor{\bsnm{Gouaillier}, \binits{D.}},
\bauthor{\bsnm{Hugel}, \binits{V.}},
\bauthor{\bsnm{Blazevic}, \binits{P.}},
\bauthor{\bsnm{Kilner}, \binits{C.}},
\bauthor{\bsnm{Monceaux}, \binits{J.}},
\bauthor{\bsnm{Lafourcade}, \binits{P.}},
\bauthor{\bsnm{Marnier}, \binits{B.}},
\bauthor{\bsnm{Serre}, \binits{J.}},
\bauthor{\bsnm{Maisonnier}, \binits{B.}}:
\bctitle{Mechatronic design of nao humanoid}.
In: \bbtitle{2009 IEEE International Conference on Robotics and Automation},
pp. \bfpage{769}--\blpage{774}
(\byear{2009}).
\bcomment{IEEE}
\end{bchapter}
\endbibitem

\bibitem[\protect\citeauthoryear{Ahmad et~al.}{2016}]{ahmad2016children}
\begin{bchapter}
\bauthor{\bsnm{Ahmad}, \binits{M.I.}},
\bauthor{\bsnm{Mubin}, \binits{O.}},
\bauthor{\bsnm{Orlando}, \binits{J.}}:
\bctitle{Children views' on social robot's adaptations in education}.
In: \bbtitle{Proceedings of the 28th Australian Conference on Computer-Human
  Interaction},
pp. \bfpage{145}--\blpage{149}
(\byear{2016})
\end{bchapter}
\endbibitem

\bibitem[\protect\citeauthoryear{Rosi et~al.}{2016}]{rosi2016use}
\begin{barticle}
\bauthor{\bsnm{Rosi}, \binits{A.}},
\bauthor{\bsnm{Dall’Asta}, \binits{M.}},
\bauthor{\bsnm{Brighenti}, \binits{F.}},
\bauthor{\bsnm{Del~Rio}, \binits{D.}},
\bauthor{\bsnm{Volta}, \binits{E.}},
\bauthor{\bsnm{Baroni}, \binits{I.}},
\bauthor{\bsnm{Nalin}, \binits{M.}},
\bauthor{\bsnm{Zelati}, \binits{M.C.}},
\bauthor{\bsnm{Sanna}, \binits{A.}},
\bauthor{\bsnm{Scazzina}, \binits{F.}}:
\batitle{The use of new technologies for nutritional education in primary
  schools: a pilot study}.
\bjtitle{Public health}
\bvolume{140},
\bfpage{50}--\blpage{55}
(\byear{2016})
\end{barticle}
\endbibitem

\bibitem[\protect\citeauthoryear{Vrochidou et~al.}{2018}]{vrochidou2018social}
\begin{bchapter}
\bauthor{\bsnm{Vrochidou}, \binits{E.}},
\bauthor{\bsnm{Najoua}, \binits{A.}},
\bauthor{\bsnm{Lytridis}, \binits{C.}},
\bauthor{\bsnm{Salonidis}, \binits{M.}},
\bauthor{\bsnm{Ferelis}, \binits{V.}},
\bauthor{\bsnm{Papakostas}, \binits{G.A.}}:
\bctitle{Social robot {NAO} as a self-regulating didactic mediator: A case
  study of teaching/learning numeracy}.
In: \bbtitle{2018 26th International Conference on Software, Telecommunications
  and Computer Networks (SoftCOM)},
pp. \bfpage{1}--\blpage{5}
(\byear{2018}).
\bcomment{IEEE}
\end{bchapter}
\endbibitem

\bibitem[\protect\citeauthoryear{Ioannou et~al.}{2015}]{ioannou2015pre}
\begin{barticle}
\bauthor{\bsnm{Ioannou}, \binits{A.}},
\bauthor{\bsnm{Andreou}, \binits{E.}},
\bauthor{\bsnm{Christofi}, \binits{M.}}:
\batitle{Pre-schoolers’ interest and caring behaviour around a humanoid
  robot}.
\bjtitle{TechTrends}
\bvolume{59}(\bissue{2}),
\bfpage{23}--\blpage{26}
(\byear{2015})
\end{barticle}
\endbibitem

\bibitem[\protect\citeauthoryear{Depe{\v{s}}ov{\'a}
  et~al.}{2018}]{depevsova2018search}
\begin{barticle}
\bauthor{\bsnm{Depe{\v{s}}ov{\'a}}, \binits{J.}},
\bauthor{\bsnm{Noga}, \binits{H.}},
\bauthor{\bsnm{Migo}, \binits{P.}}:
\batitle{In search of modern teaching methods-humanoid nao robot, as help in
  the realization of it subjects}.
\bjtitle{TEM Journal}
\bvolume{7}(\bissue{2}),
\bfpage{250}
(\byear{2018})
\end{barticle}
\endbibitem

\bibitem[\protect\citeauthoryear{Alkhalifah et~al.}{2015}]{alkhalifah2015using}
\begin{bchapter}
\bauthor{\bsnm{Alkhalifah}, \binits{A.}},
\bauthor{\bsnm{Alsalman}, \binits{B.}},
\bauthor{\bsnm{Alnuhait}, \binits{D.}},
\bauthor{\bsnm{Meldah}, \binits{O.}},
\bauthor{\bsnm{Aloud}, \binits{S.}},
\bauthor{\bsnm{Al-Khalifa}, \binits{H.S.}},
\bauthor{\bsnm{Al-Otaibi}, \binits{H.M.}}:
\bctitle{Using {NAO} humanoid robot in kindergarten: a proposed system}.
In: \bbtitle{2015 IEEE 15th International Conference on Advanced Learning
  Technologies},
pp. \bfpage{166}--\blpage{167}
(\byear{2015})
\end{bchapter}
\endbibitem

\bibitem[\protect\citeauthoryear{Suzuki et~al.}{2017}]{suzuki2017nao}
\begin{bchapter}
\bauthor{\bsnm{Suzuki}, \binits{R.}},
\bauthor{\bsnm{Lee}, \binits{J.}},
\bauthor{\bsnm{Rudovic}, \binits{O.}}:
\bctitle{{NAO}-dance therapy for children with {ASD}}.
In: \bbtitle{Proceedings of the Companion of the 2017 ACM/IEEE International
  Conference on Human-Robot Interaction},
pp. \bfpage{295}--\blpage{296}
(\byear{2017})
\end{bchapter}
\endbibitem

\bibitem[\protect\citeauthoryear{Amanatiadis
  et~al.}{2017}]{amanatiadis2017interactive}
\begin{bchapter}
\bauthor{\bsnm{Amanatiadis}, \binits{A.}},
\bauthor{\bsnm{Kaburlasos}, \binits{V.G.}},
\bauthor{\bsnm{Dardani}, \binits{C.}},
\bauthor{\bsnm{Chatzichristofis}, \binits{S.A.}}:
\bctitle{Interactive social robots in special education}.
In: \bbtitle{2017 IEEE 7th International Conference on Consumer
  electronics-Berlin},
pp. \bfpage{126}--\blpage{129}
(\byear{2017})
\end{bchapter}
\endbibitem

\bibitem[\protect\citeauthoryear{Gao}{2016}]{gao2016improvements}
\begin{botherref}
\oauthor{\bsnm{Gao}, \binits{X.}}:
The improvements of nao robots in education
(2016)
\end{botherref}
\endbibitem

\bibitem[\protect\citeauthoryear{Jim{\'e}nez
  et~al.}{2019}]{jimenez2019recognition}
\begin{botherref}
\oauthor{\bsnm{Jim{\'e}nez}, \binits{M.}},
\oauthor{\bsnm{Ochoa}, \binits{A.}},
\oauthor{\bsnm{Escobedo}, \binits{D.}},
\oauthor{\bsnm{Estrada}, \binits{R.}},
\oauthor{\bsnm{Martinez}, \binits{E.}},
\oauthor{\bsnm{Maciel}, \binits{R.}},
\oauthor{\bsnm{Larios}, \binits{V.}}:
Recognition of colors through use of a humanoid nao robot in therapies for
  children with down syndrome in a smart city.
Research in Computing Science
\textbf{148}(6)
(2019)
\end{botherref}
\endbibitem

\bibitem[\protect\citeauthoryear{Wood et~al.}{2017}]{wood2017iterative}
\begin{bchapter}
\bauthor{\bsnm{Wood}, \binits{L.J.}},
\bauthor{\bsnm{Zaraki}, \binits{A.}},
\bauthor{\bsnm{Walters}, \binits{M.L.}},
\bauthor{\bsnm{Novanda}, \binits{O.}},
\bauthor{\bsnm{Robins}, \binits{B.}},
\bauthor{\bsnm{Dautenhahn}, \binits{K.}}:
\bctitle{The iterative development of the humanoid robot kaspar: an assistive
  robot for children with autism}.
In: \bbtitle{International Conference on Social Robotics},
pp. \bfpage{53}--\blpage{63}
(\byear{2017}).
\bcomment{Springer}
\end{bchapter}
\endbibitem

\bibitem[\protect\citeauthoryear{Wood et~al.}{2021}]{wood2021developing}
\begin{barticle}
\bauthor{\bsnm{Wood}, \binits{L.J.}},
\bauthor{\bsnm{Zaraki}, \binits{A.}},
\bauthor{\bsnm{Robins}, \binits{B.}},
\bauthor{\bsnm{Dautenhahn}, \binits{K.}}:
\batitle{Developing kaspar: a humanoid robot for children with autism}.
\bjtitle{International Journal of Social Robotics}
\bvolume{13}(\bissue{3}),
\bfpage{491}--\blpage{508}
(\byear{2021})
\end{barticle}
\endbibitem

\bibitem[\protect\citeauthoryear{Wainer et~al.}{2010}]{wainer2010collaborating}
\begin{bchapter}
\bauthor{\bsnm{Wainer}, \binits{J.}},
\bauthor{\bsnm{Dautenhahn}, \binits{K.}},
\bauthor{\bsnm{Robins}, \binits{B.}},
\bauthor{\bsnm{Amirabdollahian}, \binits{F.}}:
\bctitle{Collaborating with kaspar: Using an autonomous humanoid robot to
  foster cooperative dyadic play among children with autism}.
In: \bbtitle{2010 10th IEEE-RAS International Conference on Humanoid Robots},
pp. \bfpage{631}--\blpage{638}
(\byear{2010}).
\bcomment{IEEE}
\end{bchapter}
\endbibitem

\bibitem[\protect\citeauthoryear{Wainer et~al.}{2014a}]{wainer2014using}
\begin{barticle}
\bauthor{\bsnm{Wainer}, \binits{J.}},
\bauthor{\bsnm{Robins}, \binits{B.}},
\bauthor{\bsnm{Amirabdollahian}, \binits{F.}},
\bauthor{\bsnm{Dautenhahn}, \binits{K.}}:
\batitle{Using the humanoid robot kaspar to autonomously play triadic games and
  facilitate collaborative play among children with autism}.
\bjtitle{IEEE Transactions on Autonomous Mental Development}
\bvolume{6}(\bissue{3}),
\bfpage{183}--\blpage{199}
(\byear{2014})
\end{barticle}
\endbibitem

\bibitem[\protect\citeauthoryear{Wainer et~al.}{2014b}]{wainer2014pilot}
\begin{barticle}
\bauthor{\bsnm{Wainer}, \binits{J.}},
\bauthor{\bsnm{Dautenhahn}, \binits{K.}},
\bauthor{\bsnm{Robins}, \binits{B.}},
\bauthor{\bsnm{Amirabdollahian}, \binits{F.}}:
\batitle{A pilot study with a novel setup for collaborative play of the
  humanoid robot kaspar with children with autism}.
\bjtitle{International journal of social robotics}
\bvolume{6}(\bissue{1}),
\bfpage{45}--\blpage{65}
(\byear{2014})
\end{barticle}
\endbibitem

\bibitem[\protect\citeauthoryear{Huijnen et~al.}{2017}]{huijnen2017implement}
\begin{barticle}
\bauthor{\bsnm{Huijnen}, \binits{C.A.}},
\bauthor{\bsnm{Lexis}, \binits{M.A.}},
\bauthor{\bsnm{Jansens}, \binits{R.}},
\bauthor{\bsnm{Witte}, \binits{L.P.}}:
\batitle{How to implement robots in interventions for children with autism? a
  co-creation study involving people with autism, parents and professionals}.
\bjtitle{Journal of autism and developmental disorders}
\bvolume{47}(\bissue{10}),
\bfpage{3079}--\blpage{3096}
(\byear{2017})
\end{barticle}
\endbibitem

\bibitem[\protect\citeauthoryear{Zaraki et~al.}{2018}]{zaraki2018novel}
\begin{bchapter}
\bauthor{\bsnm{Zaraki}, \binits{A.}},
\bauthor{\bsnm{Khamassi}, \binits{M.}},
\bauthor{\bsnm{Wood}, \binits{L.}},
\bauthor{\bsnm{Lakatos}, \binits{G.}},
\bauthor{\bsnm{Tzafestas}, \binits{C.}},
\bauthor{\bsnm{Robins}, \binits{B.}},
\bauthor{\bsnm{Dautenhahn}, \binits{K.}}:
\bctitle{A novel paradigm for children as teachers to the kaspar robot
  learner}.
In: \bbtitle{BAILAR Workshop at the 27th International Symposium on Robot and
  Human Interactive Communication (RO-MAN 2018). Nanjing, China}
(\byear{2018})
\end{bchapter}
\endbibitem

\bibitem[\protect\citeauthoryear{Costa et~al.}{2013}]{costa2013your}
\begin{botherref}
\oauthor{\bsnm{Costa}, \binits{S.}},
\oauthor{\bsnm{Lehmann}, \binits{H.}},
\oauthor{\bsnm{Robins}, \binits{B.}},
\oauthor{\bsnm{Dautenhahn}, \binits{K.}},
\oauthor{\bsnm{Soares}, \binits{F.}}:
Where is your nose?: developing body awareness skills among children with
  autism using a humanoid robot
(2013)
\end{botherref}
\endbibitem

\bibitem[\protect\citeauthoryear{Costa et~al.}{2015}]{costa2015using}
\begin{barticle}
\bauthor{\bsnm{Costa}, \binits{S.}},
\bauthor{\bsnm{Lehmann}, \binits{H.}},
\bauthor{\bsnm{Dautenhahn}, \binits{K.}},
\bauthor{\bsnm{Robins}, \binits{B.}},
\bauthor{\bsnm{Soares}, \binits{F.}}:
\batitle{Using a humanoid robot to elicit body awareness and appropriate
  physical interaction in children with autism}.
\bjtitle{International journal of social robotics}
\bvolume{7}(\bissue{2}),
\bfpage{265}--\blpage{278}
(\byear{2015})
\end{barticle}
\endbibitem

\bibitem[\protect\citeauthoryear{Kozima et~al.}{2005}]{kozima2005using}
\begin{bchapter}
\bauthor{\bsnm{Kozima}, \binits{H.}},
\bauthor{\bsnm{Nakagawa}, \binits{C.}},
\bauthor{\bsnm{Yano}, \binits{H.}}:
\bctitle{Using robots for the study of human social development}.
In: \bbtitle{AAAI Spring Symposium on Developmental Robotics},
vol. \bseriesno{2005}
(\byear{2005}).
\bcomment{Citeseer}
\end{bchapter}
\endbibitem

\bibitem[\protect\citeauthoryear{Kozima et~al.}{2009}]{kozima2009keepon}
\begin{barticle}
\bauthor{\bsnm{Kozima}, \binits{H.}},
\bauthor{\bsnm{Michalowski}, \binits{M.P.}},
\bauthor{\bsnm{Nakagawa}, \binits{C.}}:
\batitle{Keepon}.
\bjtitle{International Journal of Social Robotics}
\bvolume{1}(\bissue{1}),
\bfpage{3}--\blpage{18}
(\byear{2009})
\end{barticle}
\endbibitem

\bibitem[\protect\citeauthoryear{Costescu et~al.}{2015}]{costescu2015reversal}
\begin{barticle}
\bauthor{\bsnm{Costescu}, \binits{C.A.}},
\bauthor{\bsnm{Vanderborght}, \binits{B.}},
\bauthor{\bsnm{David}, \binits{D.O.}}:
\batitle{Reversal learning task in children with autism spectrum disorder: a
  robot-based approach}.
\bjtitle{Journal of autism and developmental disorders}
\bvolume{45}(\bissue{11}),
\bfpage{3715}--\blpage{3725}
(\byear{2015})
\end{barticle}
\endbibitem

\bibitem[\protect\citeauthoryear{Kozima and Nakagawa}{2007}]{kozima2007robot}
\begin{bchapter}
\bauthor{\bsnm{Kozima}, \binits{H.}},
\bauthor{\bsnm{Nakagawa}, \binits{C.}}:
\bctitle{A robot in a playroom with preschool children: Longitudinal field
  practice}.
In: \bbtitle{RO-MAN 2007-The 16th IEEE International Symposium on Robot and
  Human Interactive Communication},
pp. \bfpage{1058}--\blpage{1059}
(\byear{2007}).
\bcomment{IEEE}
\end{bchapter}
\endbibitem

\bibitem[\protect\citeauthoryear{Kozima et~al.}{2007}]{kozima2007social}
\begin{bchapter}
\bauthor{\bsnm{Kozima}, \binits{H.}},
\bauthor{\bsnm{Yasuda}, \binits{Y.}},
\bauthor{\bsnm{Nakagawa}, \binits{C.}}:
\bctitle{Social interaction facilitated by a minimally-designed robot: Findings
  from longitudinal therapeutic practices for autistic children}.
In: \bbtitle{RO-MAN 2007-the 16th IEEE International Symposium on Robot and
  Human Interactive Communication},
pp. \bfpage{599}--\blpage{604}
(\byear{2007}).
\bcomment{IEEE}
\end{bchapter}
\endbibitem

\bibitem[\protect\citeauthoryear{Kozima and Nakagawa}{2006}]{kozima2006social}
\begin{bchapter}
\bauthor{\bsnm{Kozima}, \binits{H.}},
\bauthor{\bsnm{Nakagawa}, \binits{C.}}:
\bctitle{Social robots for children: Practice in communication-care}.
In: \bbtitle{9th IEEE International Workshop on Advanced Motion Control,
  2006.},
pp. \bfpage{768}--\blpage{773}
(\byear{2006}).
\bcomment{IEEE}
\end{bchapter}
\endbibitem

\bibitem[\protect\citeauthoryear{Peca et~al.}{2016}]{peca2016infants}
\begin{barticle}
\bauthor{\bsnm{Peca}, \binits{A.}},
\bauthor{\bsnm{Simut}, \binits{R.}},
\bauthor{\bsnm{Cao}, \binits{H.-L.}},
\bauthor{\bsnm{Vanderborght}, \binits{B.}}:
\batitle{Do infants perceive the social robot keepon as a communicative
  partner?}
\bjtitle{Infant Behavior and Development}
\bvolume{42},
\bfpage{157}--\blpage{167}
(\byear{2016})
\end{barticle}
\endbibitem

\bibitem[\protect\citeauthoryear{Batliner et~al.}{2004}]{batliner2004you}
\begin{botherref}
\oauthor{\bsnm{Batliner}, \binits{A.}},
\oauthor{\bsnm{Hacker}, \binits{C.}},
\oauthor{\bsnm{Steidl}, \binits{S.}},
\oauthor{\bsnm{N{\"o}th}, \binits{E.}},
\oauthor{\bsnm{D'Arcy}, \binits{S.}},
\oauthor{\bsnm{Russell}, \binits{M.J.}},
\oauthor{\bsnm{Wong}, \binits{M.}}:
"you stupid tin box"-children interacting with the aibo robot: A
  cross-linguistic emotional speech corpus
(2004)
\end{botherref}
\endbibitem

\bibitem[\protect\citeauthoryear{Bartlett et~al.}{2004}]{bartlett2004dogs}
\begin{bchapter}
\bauthor{\bsnm{Bartlett}, \binits{B.}},
\bauthor{\bsnm{Estivill-Castro}, \binits{V.}},
\bauthor{\bsnm{Seymon}, \binits{S.}}:
\bctitle{Dogs or robots: why do children see them as robotic pets rather than
  canine machines?}
In: \bbtitle{Proceedings of the Fifth Conference on Australasian User
  interface-Volume 28},
pp. \bfpage{7}--\blpage{14}
(\byear{2004})
\end{bchapter}
\endbibitem

\bibitem[\protect\citeauthoryear{Fior et~al.}{2010}]{fior2010children}
\begin{botherref}
\oauthor{\bsnm{Fior}, \binits{M.}},
\oauthor{\bsnm{Nugent}, \binits{S.}},
\oauthor{\bsnm{Beran}, \binits{T.N.}},
\oauthor{\bsnm{Ramirez-Serrano}, \binits{A.}},
\oauthor{\bsnm{Kuzyk}, \binits{R.}}:
Children’s relationships with robots: robot is child’s new friend
(2010)
\end{botherref}
\endbibitem

\bibitem[\protect\citeauthoryear{Wei et~al.}{2011}]{wei2011joyful}
\begin{barticle}
\bauthor{\bsnm{Wei}, \binits{C.-W.}},
\bauthor{\bsnm{Hung}, \binits{I.}}, \betal:
\batitle{A joyful classroom learning system with robot learning companion for
  children to learn mathematics multiplication.}
\bjtitle{Turkish Online Journal of Educational Technology-TOJET}
\bvolume{10}(\bissue{2}),
\bfpage{11}--\blpage{23}
(\byear{2011})
\end{barticle}
\endbibitem

\bibitem[\protect\citeauthoryear{Van Den~Heuvel et~al.}{2017a}]{van2017can}
\begin{barticle}
\bauthor{\bsnm{Van Den~Heuvel}, \binits{R.J.}},
\bauthor{\bsnm{Lexis}, \binits{M.A.}},
\bauthor{\bsnm{Witte}, \binits{L.P.}}:
\batitle{Can the iromec robot support play in children with severe physical
  disabilities? a pilot study}.
\bjtitle{International journal of rehabilitation research}
\bvolume{40}(\bissue{1}),
\bfpage{53}--\blpage{59}
(\byear{2017})
\end{barticle}
\endbibitem

\bibitem[\protect\citeauthoryear{Van Den~Heuvel et~al.}{2017b}]{van2017robots}
\begin{barticle}
\bauthor{\bsnm{Van Den~Heuvel}, \binits{R.J.}},
\bauthor{\bsnm{Lexis}, \binits{M.A.}},
\bauthor{\bsnm{Janssens}, \binits{R.M.}},
\bauthor{\bsnm{Marti}, \binits{P.}},
\bauthor{\bsnm{De~Witte}, \binits{L.P.}}:
\batitle{Robots supporting play for children with physical disabilities:
  exploring the potential of iromec}.
\bjtitle{Technology and Disability}
\bvolume{29}(\bissue{3}),
\bfpage{109}--\blpage{120}
(\byear{2017})
\end{barticle}
\endbibitem

\bibitem[\protect\citeauthoryear{Ferrari et~al.}{2009}]{ferrari2009therapeutic}
\begin{bchapter}
\bauthor{\bsnm{Ferrari}, \binits{E.}},
\bauthor{\bsnm{Robins}, \binits{B.}},
\bauthor{\bsnm{Dautenhahn}, \binits{K.}}:
\bctitle{Therapeutic and educational objectives in robot assisted play for
  children with autism}.
In: \bbtitle{RO-MAN 2009-The 18th IEEE International Symposium on Robot and
  Human Interactive Communication},
pp. \bfpage{108}--\blpage{114}
(\byear{2009}).
\bcomment{IEEE}
\end{bchapter}
\endbibitem

\bibitem[\protect\citeauthoryear{Robins et~al.}{2012}]{robins2012scenarios}
\begin{barticle}
\bauthor{\bsnm{Robins}, \binits{B.}},
\bauthor{\bsnm{Dautenhahn}, \binits{K.}},
\bauthor{\bsnm{Ferrari}, \binits{E.}},
\bauthor{\bsnm{Kronreif}, \binits{G.}},
\bauthor{\bsnm{Prazak-Aram}, \binits{B.}},
\bauthor{\bsnm{Marti}, \binits{P.}},
\bauthor{\bsnm{Iacono}, \binits{I.}},
\bauthor{\bsnm{Gelderblom}, \binits{G.J.}},
\bauthor{\bsnm{Bernd}, \binits{T.}},
\bauthor{\bsnm{Caprino}, \binits{F.}}, \betal:
\batitle{Scenarios of robot-assisted play for children with cognitive and
  physical disabilities}.
\bjtitle{Interaction Studies}
\bvolume{13}(\bissue{2}),
\bfpage{189}--\blpage{234}
(\byear{2012})
\end{barticle}
\endbibitem

\bibitem[\protect\citeauthoryear{van Breemen et~al.}{2005}]{van2005icat}
\begin{bchapter}
\bauthor{\bsnm{Breemen}, \binits{A.}},
\bauthor{\bsnm{Yan}, \binits{X.}},
\bauthor{\bsnm{Meerbeek}, \binits{B.}}:
\bctitle{{iCat}: an animated user-interface robot with personality}.
In: \bbtitle{Proceedings of the Fourth International Joint Conference on
  Autonomous Agents and Multiagent Systems},
pp. \bfpage{143}--\blpage{144}
(\byear{2005})
\end{bchapter}
\endbibitem

\bibitem[\protect\citeauthoryear{Leite et~al.}{2012}]{leite2012modelling}
\begin{bchapter}
\bauthor{\bsnm{Leite}, \binits{I.}},
\bauthor{\bsnm{Castellano}, \binits{G.}},
\bauthor{\bsnm{Pereira}, \binits{A.}},
\bauthor{\bsnm{Martinho}, \binits{C.}},
\bauthor{\bsnm{Paiva}, \binits{A.}}:
\bctitle{Modelling empathic behaviour in a robotic game companion for children:
  an ethnographic study in real-world settings}.
In: \bbtitle{Proceedings of the Seventh Annual ACM/IEEE International
  Conference on Human-Robot Interaction},
pp. \bfpage{367}--\blpage{374}
(\byear{2012})
\end{bchapter}
\endbibitem

\bibitem[\protect\citeauthoryear{Shahid et~al.}{2010}]{shahid2010child}
\begin{bchapter}
\bauthor{\bsnm{Shahid}, \binits{S.}},
\bauthor{\bsnm{Krahmer}, \binits{E.}},
\bauthor{\bsnm{Swerts}, \binits{M.}},
\bauthor{\bsnm{Mubin}, \binits{O.}}:
\bctitle{Child-robot interaction during collaborative game play: Effects of age
  and gender on emotion and experience}.
In: \bbtitle{Proceedings of the 22nd Conference of the Computer-Human
  Interaction},
pp. \bfpage{332}--\blpage{335}
(\byear{2010})
\end{bchapter}
\endbibitem

\bibitem[\protect\citeauthoryear{Palestra
  et~al.}{2017}]{palestra2017artificial}
\begin{bchapter}
\bauthor{\bsnm{Palestra}, \binits{G.}},
\bauthor{\bsnm{De~Carolis}, \binits{B.}},
\bauthor{\bsnm{Esposito}, \binits{F.}}:
\bctitle{Artificial intelligence for robot-assisted treatment of autism.}
In: \bbtitle{WAIAH@ AI* IA},
pp. \bfpage{17}--\blpage{24}
(\byear{2017})
\end{bchapter}
\endbibitem

\bibitem[\protect\citeauthoryear{Kasimoglu et~al.}{2020}]{kasimoglu2020robotic}
\begin{barticle}
\bauthor{\bsnm{Kasimoglu}, \binits{Y.}},
\bauthor{\bsnm{Kocaaydin}, \binits{S.}},
\bauthor{\bsnm{Karsli}, \binits{E.}},
\bauthor{\bsnm{Esen}, \binits{M.}},
\bauthor{\bsnm{Bektas}, \binits{I.}},
\bauthor{\bsnm{Ince}, \binits{G.}},
\bauthor{\bsnm{Tuna}, \binits{E.B.}}:
\batitle{Robotic approach to the reduction of dental anxiety in children}.
\bjtitle{Acta Odontologica Scandinavica}
\bvolume{78}(\bissue{6}),
\bfpage{474}--\blpage{480}
(\byear{2020})
\end{barticle}
\endbibitem

\bibitem[\protect\citeauthoryear{Sharkey and Wood}{2014}]{sharkey2014paro}
\begin{bchapter}
\bauthor{\bsnm{Sharkey}, \binits{A.}},
\bauthor{\bsnm{Wood}, \binits{N.}}:
\bctitle{The paro seal robot: demeaning or enabling}.
In: \bbtitle{Proceedings of AISB},
vol. \bseriesno{36},
p. \bfpage{2014}
(\byear{2014})
\end{bchapter}
\endbibitem

\bibitem[\protect\citeauthoryear{Cifuentes et~al.}{2020}]{cifuentes2020social}
\begin{botherref}
\oauthor{\bsnm{Cifuentes}, \binits{C.A.}},
\oauthor{\bsnm{Pinto}, \binits{M.J.}},
\oauthor{\bsnm{C{\'e}spedes}, \binits{N.}},
\oauthor{\bsnm{M{\'u}nera}, \binits{M.}}:
Social robots in therapy and care.
Current Robotics Reports,
1--16
(2020)
\end{botherref}
\endbibitem

\bibitem[\protect\citeauthoryear{Crossman et~al.}{2018}]{crossman2018influence}
\begin{barticle}
\bauthor{\bsnm{Crossman}, \binits{M.K.}},
\bauthor{\bsnm{Kazdin}, \binits{A.E.}},
\bauthor{\bsnm{Kitt}, \binits{E.R.}}:
\batitle{The influence of a socially assistive robot on mood, anxiety, and
  arousal in children.}
\bjtitle{Professional Psychology: Research and Practice}
\bvolume{49}(\bissue{1}),
\bfpage{48}
(\byear{2018})
\end{barticle}
\endbibitem

\bibitem[\protect\citeauthoryear{Pipitpukdee and
  Phantachat}{2011}]{pipitpukdee2011study}
\begin{bchapter}
\bauthor{\bsnm{Pipitpukdee}, \binits{J.}},
\bauthor{\bsnm{Phantachat}, \binits{W.}}:
\bctitle{The study of the pet robot therapy in thai autistic children}.
In: \bbtitle{Proceedings of the 5th International Conference on Rehabilitation
  Engineering \& Assistive Technology},
pp. \bfpage{1}--\blpage{4}
(\byear{2011})
\end{bchapter}
\endbibitem

\bibitem[\protect\citeauthoryear{Shibata and
  Coughlin}{2014}]{shibata2014trends}
\begin{barticle}
\bauthor{\bsnm{Shibata}, \binits{T.}},
\bauthor{\bsnm{Coughlin}, \binits{J.F.}}:
\batitle{Trends of robot therapy with neurological therapeutic seal robot,
  paro}.
\bjtitle{Journal of Robotics and Mechatronics}
\bvolume{26}(\bissue{4}),
\bfpage{418}--\blpage{425}
(\byear{2014})
\end{barticle}
\endbibitem

\bibitem[\protect\citeauthoryear{Goris et~al.}{2010}]{goris2010probo}
\begin{bchapter}
\bauthor{\bsnm{Goris}, \binits{K.}},
\bauthor{\bsnm{Saldien}, \binits{J.}},
\bauthor{\bsnm{Vanderborght}, \binits{B.}},
\bauthor{\bsnm{Lefeber}, \binits{D.}}:
\bctitle{Probo, an intelligent huggable robot for {HRI} studies with children}.
In: \bbtitle{Human-Robot Interaction},
pp. \bfpage{33}--\blpage{42}
(\byear{2010}).
\bcomment{Intech}
\end{bchapter}
\endbibitem

\bibitem[\protect\citeauthoryear{Goris et~al.}{2008}]{goris2008huggable}
\begin{bchapter}
\bauthor{\bsnm{Goris}, \binits{K.}},
\bauthor{\bsnm{Saldien}, \binits{J.}},
\bauthor{\bsnm{Vanderniepen}, \binits{I.}},
\bauthor{\bsnm{Lefeber}, \binits{D.}}:
\bctitle{The huggable robot {Probo}, a multi-disciplinary research platform}.
In: \bbtitle{International Conference on Research and Education in Robotics},
pp. \bfpage{29}--\blpage{41}
(\byear{2008}).
\bcomment{Springer}
\end{bchapter}
\endbibitem

\bibitem[\protect\citeauthoryear{Chevalier
  et~al.}{2017}]{chevalier2017dialogue}
\begin{bchapter}
\bauthor{\bsnm{Chevalier}, \binits{P.}},
\bauthor{\bsnm{Li}, \binits{J.J.}},
\bauthor{\bsnm{Ainger}, \binits{E.}},
\bauthor{\bsnm{Alcorn}, \binits{A.M.}},
\bauthor{\bsnm{Babovic}, \binits{S.}},
\bauthor{\bsnm{Charisi}, \binits{V.}},
\bauthor{\bsnm{Petrovic}, \binits{S.}},
\bauthor{\bsnm{Schadenberg}, \binits{B.R.}},
\bauthor{\bsnm{Pellicano}, \binits{E.}},
\bauthor{\bsnm{Evers}, \binits{V.}}:
\bctitle{Dialogue design for a robot-based face-mirroring game to engage
  autistic children with emotional expressions}.
In: \bbtitle{International Conference on Social Robotics},
pp. \bfpage{546}--\blpage{555}
(\byear{2017}).
\bcomment{Springer}
\end{bchapter}
\endbibitem

\bibitem[\protect\citeauthoryear{Vanderborght
  et~al.}{2012}]{vanderborght2012using}
\begin{barticle}
\bauthor{\bsnm{Vanderborght}, \binits{B.}},
\bauthor{\bsnm{Simut}, \binits{R.}},
\bauthor{\bsnm{Saldien}, \binits{J.}},
\bauthor{\bsnm{Pop}, \binits{C.}},
\bauthor{\bsnm{Rusu}, \binits{A.S.}},
\bauthor{\bsnm{Pintea}, \binits{S.}},
\bauthor{\bsnm{Lefeber}, \binits{D.}},
\bauthor{\bsnm{David}, \binits{D.O.}}:
\batitle{Using the social robot {Probo} as a social story telling agent for
  children with {ASD}}.
\bjtitle{Interaction Studies}
\bvolume{13}(\bissue{3}),
\bfpage{348}--\blpage{372}
(\byear{2012})
\end{barticle}
\endbibitem

\bibitem[\protect\citeauthoryear{Osada et~al.}{2006}]{osada2006scenario}
\begin{bchapter}
\bauthor{\bsnm{Osada}, \binits{J.}},
\bauthor{\bsnm{Ohnaka}, \binits{S.}},
\bauthor{\bsnm{Sato}, \binits{M.}}:
\bctitle{The scenario and design process of childcare robot, papero}.
In: \bbtitle{Proceedings of the 2006 ACM SIGCHI International Conference on
  Advances in Computer Entertainment Technology},
p. \bfpage{80}
(\byear{2006})
\end{bchapter}
\endbibitem

\bibitem[\protect\citeauthoryear{Yun et~al.}{2011}]{yun2011engkey}
\begin{bchapter}
\bauthor{\bsnm{Yun}, \binits{S.}},
\bauthor{\bsnm{Shin}, \binits{J.}},
\bauthor{\bsnm{Kim}, \binits{D.}},
\bauthor{\bsnm{Kim}, \binits{C.G.}},
\bauthor{\bsnm{Kim}, \binits{M.}},
\bauthor{\bsnm{Choi}, \binits{M.-T.}}:
\bctitle{Engkey: Tele-education robot}.
In: \bbtitle{International Conference on Social Robotics},
pp. \bfpage{142}--\blpage{152}
(\byear{2011}).
\bcomment{Springer}
\end{bchapter}
\endbibitem

\bibitem[\protect\citeauthoryear{Movellan et~al.}{2007}]{movellan2007rubi}
\begin{bchapter}
\bauthor{\bsnm{Movellan}, \binits{J.R.}},
\bauthor{\bsnm{Tanaka}, \binits{F.}},
\bauthor{\bsnm{Fasel}, \binits{I.R.}},
\bauthor{\bsnm{Taylor}, \binits{C.}},
\bauthor{\bsnm{Ruvolo}, \binits{P.}},
\bauthor{\bsnm{Eckhardt}, \binits{M.}}:
\bctitle{The {RUBI} project: a progress report}.
In: \bbtitle{2007 2nd ACM/IEEE International Conference on Human-Robot
  Interaction (HRI)},
pp. \bfpage{333}--\blpage{339}
(\byear{2007}).
\bcomment{IEEE}
\end{bchapter}
\endbibitem

\bibitem[\protect\citeauthoryear{Han et~al.}{2005}]{han2005educational}
\begin{bchapter}
\bauthor{\bsnm{Han}, \binits{J.}},
\bauthor{\bsnm{Jo}, \binits{M.}},
\bauthor{\bsnm{Park}, \binits{S.}},
\bauthor{\bsnm{Kim}, \binits{S.}}:
\bctitle{The educational use of home robots for children}.
In: \bbtitle{ROMAN 2005. IEEE International Workshop on Robot and Human
  Interactive Communication, 2005.},
pp. \bfpage{378}--\blpage{383}
(\byear{2005}).
\bcomment{IEEE}
\end{bchapter}
\endbibitem

\bibitem[\protect\citeauthoryear{Kawata et~al.}{2008}]{kawata2008field}
\begin{bchapter}
\bauthor{\bsnm{Kawata}, \binits{H.}},
\bauthor{\bsnm{Takano}, \binits{Y.}},
\bauthor{\bsnm{Iwata}, \binits{Y.}},
\bauthor{\bsnm{Kanamaru}, \binits{N.}},
\bauthor{\bsnm{Shimokura}, \binits{K.-i.}},
\bauthor{\bsnm{Fujita}, \binits{Y.}}:
\bctitle{Field trial of asynchronous communication using network-based
  interactive child watch system for the participation of parents in day-care
  activities}.
In: \bbtitle{2008 IEEE International Conference on Robotics and Automation},
pp. \bfpage{2558}--\blpage{2563}
(\byear{2008}).
\bcomment{IEEE}
\end{bchapter}
\endbibitem

\bibitem[\protect\citeauthoryear{Sharkey}{2016}]{sharkey2016should}
\begin{barticle}
\bauthor{\bsnm{Sharkey}, \binits{A.J.}}:
\batitle{Should we welcome robot teachers?}
\bjtitle{Ethics and Information Technology}
\bvolume{18}(\bissue{4}),
\bfpage{283}--\blpage{297}
(\byear{2016})
\end{barticle}
\endbibitem

\bibitem[\protect\citeauthoryear{Kim et~al.}{2012}]{kim2012gesture}
\begin{bchapter}
\bauthor{\bsnm{Kim}, \binits{J.}},
\bauthor{\bsnm{Chun}, \binits{K.S.}},
\bauthor{\bsnm{Kwon}, \binits{D.-S.}}:
\bctitle{Gesture motion programming by applying robot motion hierarchy
  structure for the educational/entertainment robot engkey}.
In: \bbtitle{2012 IEEE Workshop on Advanced Robotics and Its Social Impacts},
pp. \bfpage{36}--\blpage{39}
(\byear{2012})
\end{bchapter}
\endbibitem

\bibitem[\protect\citeauthoryear{Johnson et~al.}{2012}]{johnson2012design}
\begin{bchapter}
\bauthor{\bsnm{Johnson}, \binits{D.}},
\bauthor{\bsnm{Malmir}, \binits{M.}},
\bauthor{\bsnm{Forster}, \binits{D.}},
\bauthor{\bsnm{Alac}, \binits{M.}},
\bauthor{\bsnm{Movellan}, \binits{J.}}:
\bctitle{Design and early evaluation of the {RUBI}-5 sociable robots}.
In: \bbtitle{2012 IEEE International Conference on Development and Learning and
  Epigenetic Robotics (ICDL)},
pp. \bfpage{1}--\blpage{2}
(\byear{2012}).
\bcomment{IEEE}
\end{bchapter}
\endbibitem

\bibitem[\protect\citeauthoryear{Movellan et~al.}{2009}]{movellan2009sociable}
\begin{bchapter}
\bauthor{\bsnm{Movellan}, \binits{J.}},
\bauthor{\bsnm{Eckhardt}, \binits{M.}},
\bauthor{\bsnm{Virnes}, \binits{M.}},
\bauthor{\bsnm{Rodriguez}, \binits{A.}}:
\bctitle{Sociable robot improves toddler vocabulary skills}.
In: \bbtitle{Proceedings of the 4th ACM/IEEE International Conference on Human
  Robot Interaction},
pp. \bfpage{307}--\blpage{308}
(\byear{2009})
\end{bchapter}
\endbibitem

\bibitem[\protect\citeauthoryear{Malmir et~al.}{2013}]{malmir2013home}
\begin{bchapter}
\bauthor{\bsnm{Malmir}, \binits{M.}},
\bauthor{\bsnm{Forster}, \binits{D.}},
\bauthor{\bsnm{Youngstrom}, \binits{K.}},
\bauthor{\bsnm{Morrison}, \binits{L.}},
\bauthor{\bsnm{Movellan}, \binits{J.}}:
\bctitle{Home alone: Social robots for digital ethnography of toddler
  behavior}.
In: \bbtitle{Proceedings of the IEEE International Conference on Computer
  Vision Workshops},
pp. \bfpage{762}--\blpage{768}
(\byear{2013})
\end{bchapter}
\endbibitem

\bibitem[\protect\citeauthoryear{Del~Coco et~al.}{2017}]{del2017computer}
\begin{bchapter}
\bauthor{\bsnm{Del~Coco}, \binits{M.}},
\bauthor{\bsnm{Leo}, \binits{M.}},
\bauthor{\bsnm{Carcagni}, \binits{P.}},
\bauthor{\bsnm{Spagnolo}, \binits{P.}},
\bauthor{\bsnm{Luigi~Mazzeo}, \binits{P.}},
\bauthor{\bsnm{Bernava}, \binits{M.}},
\bauthor{\bsnm{Marino}, \binits{F.}},
\bauthor{\bsnm{Pioggia}, \binits{G.}},
\bauthor{\bsnm{Distante}, \binits{C.}}:
\bctitle{A computer vision based approach for understanding emotional
  involvements in children with autism spectrum disorders}.
In: \bbtitle{Proceedings of the IEEE International Conference on Computer
  Vision Workshops},
pp. \bfpage{1401}--\blpage{1407}
(\byear{2017})
\end{bchapter}
\endbibitem

\bibitem[\protect\citeauthoryear{Kamble and Dale}{2021}]{kamble2021face}
\begin{bchapter}
\bauthor{\bsnm{Kamble}, \binits{V.}},
\bauthor{\bsnm{Dale}, \binits{M.}}:
\bctitle{Face recognition of children using ai classification approaches}.
In: \bbtitle{2021 International Conference on Emerging Smart Computing and
  Informatics (ESCI)},
pp. \bfpage{248}--\blpage{251}
(\byear{2021}).
\bcomment{IEEE}
\end{bchapter}
\endbibitem

\bibitem[\protect\citeauthoryear{Chang}{2007}]{chang2007study}
\begin{barticle}
\bauthor{\bsnm{Chang}, \binits{C.-L.}}:
\batitle{A study of applying data mining to early intervention for
  developmentally-delayed children}.
\bjtitle{Expert Systems with Applications}
\bvolume{33}(\bissue{2}),
\bfpage{407}--\blpage{412}
(\byear{2007})
\end{barticle}
\endbibitem

\bibitem[\protect\citeauthoryear{Al-Diabat}{2018}]{al2018fuzzy}
\begin{barticle}
\bauthor{\bsnm{Al-Diabat}, \binits{M.}}:
\batitle{Fuzzy data mining for autism classification of children}.
\bjtitle{International Journal of Advanced Computer Science and Applications}
\bvolume{9}(\bissue{7}),
\bfpage{11}--\blpage{17}
(\byear{2018})
\end{barticle}
\endbibitem

\bibitem[\protect\citeauthoryear{Leroy et~al.}{June, 2006}]{leroy2006data}
\begin{bchapter}
\bauthor{\bsnm{Leroy}, \binits{G.A.}},
\bauthor{\bsnm{Irmscher}, \binits{A.}},
\bauthor{\bsnm{Charlop}, \binits{M.H.}}:
\bctitle{Data mining techniques to study therapy success with autistic
  children}.
In: \bbtitle{Proceedings of the 2006 International Conference on Data Mining},
pp. \bfpage{1}--\blpage{4}
(\byear{June, 2006})
\end{bchapter}
\endbibitem

\bibitem[\protect\citeauthoryear{Abdullah et~al.}{2016}]{abdullah2016data}
\begin{bchapter}
\bauthor{\bsnm{Abdullah}, \binits{F.S.}},
\bauthor{\bsnm{Abd~Manan}, \binits{N.S.}},
\bauthor{\bsnm{Ahmad}, \binits{A.}},
\bauthor{\bsnm{Wafa}, \binits{S.W.}},
\bauthor{\bsnm{Shahril}, \binits{M.R.}},
\bauthor{\bsnm{Zulaily}, \binits{N.}},
\bauthor{\bsnm{Amin}, \binits{R.M.}},
\bauthor{\bsnm{Ahmed}, \binits{A.}}:
\bctitle{Data mining techniques for classification of childhood obesity among
  year 6 school children}.
In: \bbtitle{International Conference on Soft Computing and Data Mining},
pp. \bfpage{465}--\blpage{474}
(\byear{2016}).
\bcomment{Springer}
\end{bchapter}
\endbibitem

\bibitem[\protect\citeauthoryear{Zhang et~al.}{2009}]{zhang2009comparing}
\begin{barticle}
\bauthor{\bsnm{Zhang}, \binits{S.}},
\bauthor{\bsnm{Tjortjis}, \binits{C.}},
\bauthor{\bsnm{Zeng}, \binits{X.}},
\bauthor{\bsnm{Qiao}, \binits{H.}},
\bauthor{\bsnm{Buchan}, \binits{I.}},
\bauthor{\bsnm{Keane}, \binits{J.}}:
\batitle{Comparing data mining methods with logistic regression in childhood
  obesity prediction}.
\bjtitle{Information Systems Frontiers}
\bvolume{11}(\bissue{4}),
\bfpage{449}--\blpage{460}
(\byear{2009})
\end{barticle}
\endbibitem

\bibitem[\protect\citeauthoryear{Momand et~al.}{2020}]{momand2020data}
\begin{bchapter}
\bauthor{\bsnm{Momand}, \binits{Z.}},
\bauthor{\bsnm{Mongkolnam}, \binits{P.}},
\bauthor{\bsnm{Kositpanthavong}, \binits{P.}},
\bauthor{\bsnm{Chan}, \binits{J.H.}}:
\bctitle{Data mining based prediction of malnutrition in afghan children}.
In: \bbtitle{2020 12th International Conference on Knowledge and Smart
  Technology (KST)},
pp. \bfpage{12}--\blpage{17}
(\byear{2020}).
\bcomment{IEEE}
\end{bchapter}
\endbibitem

\bibitem[\protect\citeauthoryear{Krotova et~al.}{2020}]{krotova2020diagnostics}
\begin{bchapter}
\bauthor{\bsnm{Krotova}, \binits{O.}},
\bauthor{\bsnm{Moskalev}, \binits{I.}},
\bauthor{\bsnm{Nazarkina}, \binits{O.}},
\bauthor{\bsnm{Khvorova}, \binits{L.}}:
\bctitle{Diagnostics of diabetic polyneuropathy in children and adolescents
  using data mining methods}.
In: \bbtitle{Journal of Physics: Conference Series},
vol. \bseriesno{1615},
p. \bfpage{012015}
(\byear{2020})
\end{bchapter}
\endbibitem

\bibitem[\protect\citeauthoryear{El~Afandi}{2013}]{el2013application}
\begin{barticle}
\bauthor{\bsnm{El~Afandi}, \binits{G.}}:
\batitle{Application of data mining techniques to predict allergy outbreaks
  among elementary school children}.
\bjtitle{Journal of Communication and Computer}
\bvolume{10},
\bfpage{451}--\blpage{460}
(\byear{2013})
\end{barticle}
\endbibitem

\bibitem[\protect\citeauthoryear{Yue et~al.}{2018}]{yue2018application}
\begin{bchapter}
\bauthor{\bsnm{Yue}, \binits{L.}},
\bauthor{\bsnm{Chunhong}, \binits{Z.}},
\bauthor{\bsnm{Chujie}, \binits{T.}},
\bauthor{\bsnm{Xiaomeng}, \binits{Z.}},
\bauthor{\bsnm{Ruizhi}, \binits{Z.}},
\bauthor{\bsnm{Yang}, \binits{J.}}:
\bctitle{Application of data mining for young children education using emotion
  information}.
In: \bbtitle{Proceedings of the 2018 International Conference on Data Science
  and Information Technology},
pp. \bfpage{96}--\blpage{104}
(\byear{2018})
\end{bchapter}
\endbibitem

\bibitem[\protect\citeauthoryear{Naydenova et~al.}{2016}]{naydenova2016power}
\begin{barticle}
\bauthor{\bsnm{Naydenova}, \binits{E.}},
\bauthor{\bsnm{Tsanas}, \binits{A.}},
\bauthor{\bsnm{Howie}, \binits{S.}},
\bauthor{\bsnm{Casals-Pascual}, \binits{C.}},
\bauthor{\bsnm{De~Vos}, \binits{M.}}:
\batitle{The power of data mining in diagnosis of childhood pneumonia}.
\bjtitle{Journal of The Royal Society Interface}
\bvolume{13}(\bissue{120}),
\bfpage{20160266}
(\byear{2016})
\end{barticle}
\endbibitem

\bibitem[\protect\citeauthoryear{Delavarian
  et~al.}{2011}]{delavarian2011automatic}
\begin{barticle}
\bauthor{\bsnm{Delavarian}, \binits{M.}},
\bauthor{\bsnm{Towhidkhah}, \binits{F.}},
\bauthor{\bsnm{Gharibzadeh}, \binits{S.}},
\bauthor{\bsnm{Dibajnia}, \binits{P.}}:
\batitle{Automatic classification of hyperactive children: Comparing multiple
  artificial intelligence approaches}.
\bjtitle{Neuroscience letters}
\bvolume{498}(\bissue{3}),
\bfpage{190}--\blpage{193}
(\byear{2011})
\end{barticle}
\endbibitem

\bibitem[\protect\citeauthoryear{Lin et~al.}{2020}]{lin2020zhorai}
\begin{bchapter}
\bauthor{\bsnm{Lin}, \binits{P.}},
\bauthor{\bsnm{Van~Brummelen}, \binits{J.}},
\bauthor{\bsnm{Lukin}, \binits{G.}},
\bauthor{\bsnm{Williams}, \binits{R.}},
\bauthor{\bsnm{Breazeal}, \binits{C.}}:
\bctitle{Zhorai: Designing a conversational agent for children to explore
  machine learning concepts}.
In: \bbtitle{Proceedings of the AAAI Conference on Artificial Intelligence},
vol. \bseriesno{34},
pp. \bfpage{13381}--\blpage{13388}
(\byear{2020})
\end{bchapter}
\endbibitem

\bibitem[\protect\citeauthoryear{Hitron et~al.}{2018}]{hitron2018introducing}
\begin{bchapter}
\bauthor{\bsnm{Hitron}, \binits{T.}},
\bauthor{\bsnm{Wald}, \binits{I.}},
\bauthor{\bsnm{Erel}, \binits{H.}},
\bauthor{\bsnm{Zuckerman}, \binits{O.}}:
\bctitle{Introducing children to machine learning concepts through hands-on
  experience}.
In: \bbtitle{Proceedings of the 17th ACM Conference on Interaction Design and
  Children},
pp. \bfpage{563}--\blpage{568}
(\byear{2018})
\end{bchapter}
\endbibitem

\bibitem[\protect\citeauthoryear{Papakostas
  et~al.}{2021}]{papakostas2021estimating}
\begin{botherref}
\oauthor{\bsnm{Papakostas}, \binits{G.A.}},
\oauthor{\bsnm{Sidiropoulos}, \binits{G.K.}},
\oauthor{\bsnm{Lytridis}, \binits{C.}},
\oauthor{\bsnm{Bazinas}, \binits{C.}},
\oauthor{\bsnm{Kaburlasos}, \binits{V.G.}},
\oauthor{\bsnm{Kourampa}, \binits{E.}},
\oauthor{\bsnm{Karageorgiou}, \binits{E.}},
\oauthor{\bsnm{Kechayas}, \binits{P.}},
\oauthor{\bsnm{Papadopoulou}, \binits{M.T.}}:
Estimating children engagement interacting with robots in special education
  using machine learning.
Mathematical Problems in Engineering
\textbf{2021}
(2021)
\end{botherref}
\endbibitem

\bibitem[\protect\citeauthoryear{Hagenbuchner
  et~al.}{2015}]{hagenbuchner2015prediction}
\begin{barticle}
\bauthor{\bsnm{Hagenbuchner}, \binits{M.}},
\bauthor{\bsnm{Cliff}, \binits{D.P.}},
\bauthor{\bsnm{Trost}, \binits{S.G.}},
\bauthor{\bsnm{Van~Tuc}, \binits{N.}},
\bauthor{\bsnm{Peoples}, \binits{G.E.}}:
\batitle{Prediction of activity type in preschool children using machine
  learning techniques}.
\bjtitle{Journal of Science and Medicine in Sport}
\bvolume{18}(\bissue{4}),
\bfpage{426}--\blpage{431}
(\byear{2015})
\end{barticle}
\endbibitem

\bibitem[\protect\citeauthoryear{Ahmadi et~al.}{2018}]{ahmadi2018machine}
\begin{barticle}
\bauthor{\bsnm{Ahmadi}, \binits{M.}},
\bauthor{\bsnm{O’Neil}, \binits{M.}},
\bauthor{\bsnm{Fragala-Pinkham}, \binits{M.}},
\bauthor{\bsnm{Lennon}, \binits{N.}},
\bauthor{\bsnm{Trost}, \binits{S.}}:
\batitle{Machine learning algorithms for activity recognition in ambulant
  children and adolescents with cerebral palsy}.
\bjtitle{Journal of neuroengineering and rehabilitation}
\bvolume{15}(\bissue{1}),
\bfpage{1}--\blpage{9}
(\byear{2018})
\end{barticle}
\endbibitem

\bibitem[\protect\citeauthoryear{Rasheed et~al.}{2021}]{rasheed2021use}
\begin{barticle}
\bauthor{\bsnm{Rasheed}, \binits{M.A.}},
\bauthor{\bsnm{Chand}, \binits{P.}},
\bauthor{\bsnm{Ahmed}, \binits{S.}},
\bauthor{\bsnm{Sharif}, \binits{H.}},
\bauthor{\bsnm{Hoodbhoy}, \binits{Z.}},
\bauthor{\bsnm{Siddiqui}, \binits{A.}},
\bauthor{\bsnm{Hasan}, \binits{B.S.}}:
\batitle{Use of artificial intelligence on electroencephalogram (eeg) waveforms
  to predict failure in early school grades in children from a rural cohort in
  pakistan}.
\bjtitle{PLoS One}
\bvolume{16}(\bissue{2}),
\bfpage{0246236}
(\byear{2021})
\end{barticle}
\endbibitem

\bibitem[\protect\citeauthoryear{Carpenter
  et~al.}{2016}]{carpenter2016quantifying}
\begin{barticle}
\bauthor{\bsnm{Carpenter}, \binits{K.L.}},
\bauthor{\bsnm{Sprechmann}, \binits{P.}},
\bauthor{\bsnm{Calderbank}, \binits{R.}},
\bauthor{\bsnm{Sapiro}, \binits{G.}},
\bauthor{\bsnm{Egger}, \binits{H.L.}}:
\batitle{Quantifying risk for anxiety disorders in preschool children: a
  machine learning approach}.
\bjtitle{PLoS One}
\bvolume{11}(\bissue{11}),
\bfpage{0165524}
(\byear{2016})
\end{barticle}
\endbibitem

\bibitem[\protect\citeauthoryear{Crippa et~al.}{2015}]{crippa2015use}
\begin{barticle}
\bauthor{\bsnm{Crippa}, \binits{A.}},
\bauthor{\bsnm{Salvatore}, \binits{C.}},
\bauthor{\bsnm{Perego}, \binits{P.}},
\bauthor{\bsnm{Forti}, \binits{S.}},
\bauthor{\bsnm{Nobile}, \binits{M.}},
\bauthor{\bsnm{Molteni}, \binits{M.}},
\bauthor{\bsnm{Castiglioni}, \binits{I.}}:
\batitle{Use of machine learning to identify children with autism and their
  motor abnormalities}.
\bjtitle{Journal of autism and developmental disorders}
\bvolume{45}(\bissue{7}),
\bfpage{2146}--\blpage{2156}
(\byear{2015})
\end{barticle}
\endbibitem

\bibitem[\protect\citeauthoryear{Liu et~al.}{2016}]{liu2016identifying}
\begin{barticle}
\bauthor{\bsnm{Liu}, \binits{W.}},
\bauthor{\bsnm{Li}, \binits{M.}},
\bauthor{\bsnm{Yi}, \binits{L.}}:
\batitle{Identifying children with autism spectrum disorder based on their face
  processing abnormality: A machine learning framework}.
\bjtitle{Autism Research}
\bvolume{9}(\bissue{8}),
\bfpage{888}--\blpage{898}
(\byear{2016})
\end{barticle}
\endbibitem

\bibitem[\protect\citeauthoryear{McGinnis et~al.}{2018}]{mcginnis2018wearable}
\begin{bchapter}
\bauthor{\bsnm{McGinnis}, \binits{R.S.}},
\bauthor{\bsnm{McGinnis}, \binits{E.W.}},
\bauthor{\bsnm{Hruschak}, \binits{J.}},
\bauthor{\bsnm{Lopez-Duran}, \binits{N.L.}},
\bauthor{\bsnm{Fitzgerald}, \binits{K.}},
\bauthor{\bsnm{Rosenblum}, \binits{K.L.}},
\bauthor{\bsnm{Muzik}, \binits{M.}}:
\bctitle{Wearable sensors and machine learning diagnose anxiety and depression
  in young children}.
In: \bbtitle{2018 IEEE EMBS International Conference on Biomedical \& Health
  Informatics (BHI)},
pp. \bfpage{410}--\blpage{413}
(\byear{2018}).
\bcomment{IEEE}
\end{bchapter}
\endbibitem

\bibitem[\protect\citeauthoryear{McGinnis et~al.}{2019}]{mcginnis2019rapid}
\begin{barticle}
\bauthor{\bsnm{McGinnis}, \binits{R.S.}},
\bauthor{\bsnm{McGinnis}, \binits{E.W.}},
\bauthor{\bsnm{Hruschak}, \binits{J.}},
\bauthor{\bsnm{Lopez-Duran}, \binits{N.L.}},
\bauthor{\bsnm{Fitzgerald}, \binits{K.}},
\bauthor{\bsnm{Rosenblum}, \binits{K.L.}},
\bauthor{\bsnm{Muzik}, \binits{M.}}:
\batitle{Rapid detection of internalizing diagnosis in young children enabled
  by wearable sensors and machine learning}.
\bjtitle{PLoS One}
\bvolume{14}(\bissue{1}),
\bfpage{0210267}
(\byear{2019})
\end{barticle}
\endbibitem

\bibitem[\protect\citeauthoryear{Su et~al.}{2020}]{su2020machine}
\begin{barticle}
\bauthor{\bsnm{Su}, \binits{C.}},
\bauthor{\bsnm{Aseltine}, \binits{R.}},
\bauthor{\bsnm{Doshi}, \binits{R.}},
\bauthor{\bsnm{Chen}, \binits{K.}},
\bauthor{\bsnm{Rogers}, \binits{S.C.}},
\bauthor{\bsnm{Wang}, \binits{F.}}:
\batitle{Machine learning for suicide risk prediction in children and
  adolescents with electronic health records}.
\bjtitle{Translational psychiatry}
\bvolume{10}(\bissue{1}),
\bfpage{1}--\blpage{10}
(\byear{2020})
\end{barticle}
\endbibitem

\bibitem[\protect\citeauthoryear{Di~Nuovo et~al.}{2018}]{di2018deep}
\begin{barticle}
\bauthor{\bsnm{Di~Nuovo}, \binits{A.}},
\bauthor{\bsnm{Conti}, \binits{D.}},
\bauthor{\bsnm{Trubia}, \binits{G.}},
\bauthor{\bsnm{Buono}, \binits{S.}},
\bauthor{\bsnm{Di~Nuovo}, \binits{S.}}:
\batitle{Deep learning systems for estimating visual attention in
  robot-assisted therapy of children with autism and intellectual disability}.
\bjtitle{Robotics}
\bvolume{7}(\bissue{2}),
\bfpage{25}
(\byear{2018})
\end{barticle}
\endbibitem

\bibitem[\protect\citeauthoryear{Rudovic et~al.}{2018}]{rudovic2018culturenet}
\begin{bchapter}
\bauthor{\bsnm{Rudovic}, \binits{O.}},
\bauthor{\bsnm{Utsumi}, \binits{Y.}},
\bauthor{\bsnm{Lee}, \binits{J.}},
\bauthor{\bsnm{Hernandez}, \binits{J.}},
\bauthor{\bsnm{Ferrer}, \binits{E.C.}},
\bauthor{\bsnm{Schuller}, \binits{B.}},
\bauthor{\bsnm{Picard}, \binits{R.W.}}:
\bctitle{Culturenet: a deep learning approach for engagement intensity
  estimation from face images of children with autism}.
In: \bbtitle{2018 IEEE/RSJ International Conference on Intelligent Robots and
  Systems (IROS)},
pp. \bfpage{339}--\blpage{346}
(\byear{2018}).
\bcomment{IEEE}
\end{bchapter}
\endbibitem

\bibitem[\protect\citeauthoryear{Kim et~al.}{2015}]{kim2015pororobot}
\begin{bchapter}
\bauthor{\bsnm{Kim}, \binits{K.-M.}},
\bauthor{\bsnm{Nan}, \binits{C.-J.}},
\bauthor{\bsnm{Ha}, \binits{J.-W.}},
\bauthor{\bsnm{Heo}, \binits{Y.-J.}},
\bauthor{\bsnm{Zhang}, \binits{B.-T.}}:
\bctitle{Pororobot: A deep learning robot that plays video q\&a games}.
In: \bbtitle{2015 AAAI Fall Symposium Series}
(\byear{2015})
\end{bchapter}
\endbibitem

\bibitem[\protect\citeauthoryear{She and Ren}{2021}]{she2021enhance}
\begin{barticle}
\bauthor{\bsnm{She}, \binits{T.}},
\bauthor{\bsnm{Ren}, \binits{F.}}:
\batitle{Enhance the language ability of humanoid robot nao through deep
  learning to interact with autistic children}.
\bjtitle{Electronics}
\bvolume{10}(\bissue{19}),
\bfpage{2393}
(\byear{2021})
\end{barticle}
\endbibitem

\bibitem[\protect\citeauthoryear{Liu et~al.}{2018}]{liu2018detecting}
\begin{barticle}
\bauthor{\bsnm{Liu}, \binits{Y.}},
\bauthor{\bsnm{Huang}, \binits{Y.}},
\bauthor{\bsnm{Wang}, \binits{J.}},
\bauthor{\bsnm{Liu}, \binits{L.}},
\bauthor{\bsnm{Luo}, \binits{J.}}:
\batitle{Detecting premature ventricular contraction in children with deep
  learning}.
\bjtitle{Journal of Shanghai Jiaotong University (Science)}
\bvolume{23}(\bissue{1}),
\bfpage{66}--\blpage{73}
(\byear{2018})
\end{barticle}
\endbibitem

\bibitem[\protect\citeauthoryear{Lempereur et~al.}{2020}]{lempereur2020new}
\begin{barticle}
\bauthor{\bsnm{Lempereur}, \binits{M.}},
\bauthor{\bsnm{Rousseau}, \binits{F.}},
\bauthor{\bsnm{R{\'e}my-N{\'e}ris}, \binits{O.}},
\bauthor{\bsnm{Pons}, \binits{C.}},
\bauthor{\bsnm{Houx}, \binits{L.}},
\bauthor{\bsnm{Quellec}, \binits{G.}},
\bauthor{\bsnm{Brochard}, \binits{S.}}:
\batitle{A new deep learning-based method for the detection of gait events in
  children with gait disorders: Proof-of-concept and concurrent validity}.
\bjtitle{Journal of biomechanics}
\bvolume{98},
\bfpage{109490}
(\byear{2020})
\end{barticle}
\endbibitem

\bibitem[\protect\citeauthoryear{Kumar and
  Senthil}{2021}]{kumar2021construction}
\begin{barticle}
\bauthor{\bsnm{Kumar}, \binits{T.S.}},
\bauthor{\bsnm{Senthil}, \binits{T.}}:
\batitle{Construction of hybrid deep learning model for predicting children
  behavior based on their emotional reaction}.
\bjtitle{Journal of Information Technology}
\bvolume{3}(\bissue{01}),
\bfpage{29}--\blpage{43}
(\byear{2021})
\end{barticle}
\endbibitem

\bibitem[\protect\citeauthoryear{Chatzimichail
  et~al.}{2010}]{chatzimichail2010artificial}
\begin{bchapter}
\bauthor{\bsnm{Chatzimichail}, \binits{E.A.}},
\bauthor{\bsnm{Rigas}, \binits{A.G.}},
\bauthor{\bsnm{Paraskakis}, \binits{E.N.}}:
\bctitle{An artificial intelligence technique for the prediction of persistent
  asthma in children}.
In: \bbtitle{Proceedings of the 10th IEEE International Conference on
  Information Technology and Applications in Biomedicine},
pp. \bfpage{1}--\blpage{4}
(\byear{2010}).
\bcomment{IEEE}
\end{bchapter}
\endbibitem

\bibitem[\protect\citeauthoryear{Yu et~al.}{2021}]{yu2021associations}
\begin{botherref}
\oauthor{\bsnm{Yu}, \binits{H.}},
\oauthor{\bsnm{Zhou}, \binits{Y.}},
\oauthor{\bsnm{Wang}, \binits{R.}},
\oauthor{\bsnm{Qian}, \binits{Z.}},
\oauthor{\bsnm{Knibbs}, \binits{L.D.}},
\oauthor{\bsnm{Jalaludin}, \binits{B.}},
\oauthor{\bsnm{Schootman}, \binits{M.}},
\oauthor{\bsnm{McMillin}, \binits{S.E.}},
\oauthor{\bsnm{Howard}, \binits{S.W.}},
\oauthor{\bsnm{Lin}, \binits{L.-Z.}}, et al.:
Associations between trees and grass presence with childhood asthma prevalence
  using deep learning image segmentation and a novel green view index.
Environmental Pollution,
117582
(2021)
\end{botherref}
\endbibitem

\bibitem[\protect\citeauthoryear{McComas et~al.}{1998}]{mccomas1998current}
\begin{botherref}
\oauthor{\bsnm{McComas}, \binits{J.}},
\oauthor{\bsnm{Pivik}, \binits{P.}},
\oauthor{\bsnm{Laflamme}, \binits{M.}}:
Current uses of virtual reality for children with disabilities.
Studies in health technology and informatics,
161--169
(1998)
\end{botherref}
\endbibitem

\bibitem[\protect\citeauthoryear{Gershon et~al.}{2004}]{gershon2004pilot}
\begin{barticle}
\bauthor{\bsnm{Gershon}, \binits{J.}},
\bauthor{\bsnm{Zimand}, \binits{E.}},
\bauthor{\bsnm{Pickering}, \binits{M.}},
\bauthor{\bsnm{Rothbaum}, \binits{B.O.}},
\bauthor{\bsnm{Hodges}, \binits{L.}}:
\batitle{A pilot and feasibility study of virtual reality as a distraction for
  children with cancer}.
\bjtitle{Journal of the American Academy of Child \& Adolescent Psychiatry}
\bvolume{43}(\bissue{10}),
\bfpage{1243}--\blpage{1249}
(\byear{2004})
\end{barticle}
\endbibitem

\bibitem[\protect\citeauthoryear{Parsons and Cobb}{2011}]{parsons2011state}
\begin{barticle}
\bauthor{\bsnm{Parsons}, \binits{S.}},
\bauthor{\bsnm{Cobb}, \binits{S.}}:
\batitle{State-of-the-art of virtual reality technologies for children on the
  autism spectrum}.
\bjtitle{European Journal of Special Needs Education}
\bvolume{26}(\bissue{3}),
\bfpage{355}--\blpage{366}
(\byear{2011})
\end{barticle}
\endbibitem

\bibitem[\protect\citeauthoryear{Arane et~al.}{2017}]{arane2017virtual}
\begin{barticle}
\bauthor{\bsnm{Arane}, \binits{K.}},
\bauthor{\bsnm{Behboudi}, \binits{A.}},
\bauthor{\bsnm{Goldman}, \binits{R.D.}}:
\batitle{Virtual reality for pain and anxiety management in children}.
\bjtitle{Canadian Family Physician}
\bvolume{63}(\bissue{12}),
\bfpage{932}--\blpage{934}
(\byear{2017})
\end{barticle}
\endbibitem

\bibitem[\protect\citeauthoryear{Foley and Maddison}{2010}]{foley2010use}
\begin{barticle}
\bauthor{\bsnm{Foley}, \binits{L.}},
\bauthor{\bsnm{Maddison}, \binits{R.}}:
\batitle{Use of active video games to increase physical activity in children: a
  (virtual) reality?}
\bjtitle{Pediatric exercise science}
\bvolume{22}(\bissue{1}),
\bfpage{7}--\blpage{20}
(\byear{2010})
\end{barticle}
\endbibitem

\bibitem[\protect\citeauthoryear{Didehbani et~al.}{2016}]{didehbani2016virtual}
\begin{barticle}
\bauthor{\bsnm{Didehbani}, \binits{N.}},
\bauthor{\bsnm{Allen}, \binits{T.}},
\bauthor{\bsnm{Kandalaft}, \binits{M.}},
\bauthor{\bsnm{Krawczyk}, \binits{D.}},
\bauthor{\bsnm{Chapman}, \binits{S.}}:
\batitle{Virtual reality social cognition training for children with high
  functioning autism}.
\bjtitle{Computers in human behavior}
\bvolume{62},
\bfpage{703}--\blpage{711}
(\byear{2016})
\end{barticle}
\endbibitem

\bibitem[\protect\citeauthoryear{Jyoti and Lahiri}{2019}]{jyoti2019virtual}
\begin{barticle}
\bauthor{\bsnm{Jyoti}, \binits{V.}},
\bauthor{\bsnm{Lahiri}, \binits{U.}}:
\batitle{Virtual reality based joint attention task platform for children with
  autism}.
\bjtitle{IEEE Transactions on Learning Technologies}
\bvolume{13}(\bissue{1}),
\bfpage{198}--\blpage{210}
(\byear{2019})
\end{barticle}
\endbibitem

\bibitem[\protect\citeauthoryear{Josman et~al.}{2008}]{josman2008effectiveness}
\begin{barticle}
\bauthor{\bsnm{Josman}, \binits{N.}},
\bauthor{\bsnm{Ben-Chaim}, \binits{H.M.}},
\bauthor{\bsnm{Friedrich}, \binits{S.}},
\bauthor{\bsnm{Weiss}, \binits{P.L.}}:
\batitle{Effectiveness of virtual reality for teaching street-crossing skills
  to children and adolescents with autism}.
\bjtitle{International Journal on Disability and Human Development}
\bvolume{7}(\bissue{1}),
\bfpage{49}--\blpage{56}
(\byear{2008})
\end{barticle}
\endbibitem

\bibitem[\protect\citeauthoryear{Schwebel and
  McClure}{2010}]{schwebel2010using}
\begin{barticle}
\bauthor{\bsnm{Schwebel}, \binits{D.C.}},
\bauthor{\bsnm{McClure}, \binits{L.A.}}:
\batitle{Using virtual reality to train children in safe street-crossing
  skills}.
\bjtitle{Injury prevention}
\bvolume{16}(\bissue{1}),
\bfpage{1}--\blpage{1}
(\byear{2010})
\end{barticle}
\endbibitem

\bibitem[\protect\citeauthoryear{Ip et~al.}{2018}]{ip2018enhance}
\begin{barticle}
\bauthor{\bsnm{Ip}, \binits{H.H.}},
\bauthor{\bsnm{Wong}, \binits{S.W.}},
\bauthor{\bsnm{Chan}, \binits{D.F.}},
\bauthor{\bsnm{Byrne}, \binits{J.}},
\bauthor{\bsnm{Li}, \binits{C.}},
\bauthor{\bsnm{Yuan}, \binits{V.S.}},
\bauthor{\bsnm{Lau}, \binits{K.S.}},
\bauthor{\bsnm{Wong}, \binits{J.Y.}}:
\batitle{Enhance emotional and social adaptation skills for children with
  autism spectrum disorder: A virtual reality enabled approach}.
\bjtitle{Computers \& Education}
\bvolume{117},
\bfpage{1}--\blpage{15}
(\byear{2018})
\end{barticle}
\endbibitem

\bibitem[\protect\citeauthoryear{Dongming
  et~al.}{2020}]{dongming2020intelligent}
\begin{bchapter}
\bauthor{\bsnm{Dongming}, \binits{L.}},
\bauthor{\bsnm{Wanjing}, \binits{L.}},
\bauthor{\bsnm{Shuang}, \binits{C.}},
\bauthor{\bsnm{Shuying}, \binits{Z.}}:
\bctitle{Intelligent robot for early childhood education}.
In: \bbtitle{Proceedings of the 2020 8th International Conference on
  Information and Education Technology},
pp. \bfpage{142}--\blpage{146}
(\byear{2020})
\end{bchapter}
\endbibitem

\bibitem[\protect\citeauthoryear{Xia et~al.}{2017}]{xia2017detecting}
\begin{bchapter}
\bauthor{\bsnm{Xia}, \binits{Y.}},
\bauthor{\bsnm{Huang}, \binits{D.}},
\bauthor{\bsnm{Wang}, \binits{Y.}}:
\bctitle{Detecting smiles of young children via deep transfer learning}.
In: \bbtitle{Proceedings of the IEEE International Conference on Computer
  Vision Workshops},
pp. \bfpage{1673}--\blpage{1681}
(\byear{2017})
\end{bchapter}
\endbibitem

\bibitem[\protect\citeauthoryear{Druga et~al.}{2017}]{druga2017hey}
\begin{bchapter}
\bauthor{\bsnm{Druga}, \binits{S.}},
\bauthor{\bsnm{Williams}, \binits{R.}},
\bauthor{\bsnm{Breazeal}, \binits{C.}},
\bauthor{\bsnm{Resnick}, \binits{M.}}:
\bctitle{Hey google is it ok if i eat you? initial explorations in child-agent
  interaction}.
In: \bbtitle{Proceedings of the 2017 Conference on Interaction Design and
  Children},
pp. \bfpage{595}--\blpage{600}
(\byear{2017})
\end{bchapter}
\endbibitem

\bibitem[\protect\citeauthoryear{Shahi et~al.}{2021}]{shahi2021using}
\begin{botherref}
\oauthor{\bsnm{Shahi}, \binits{N.}},
\oauthor{\bsnm{Shahi}, \binits{A.K.}},
\oauthor{\bsnm{Phillips}, \binits{R.}},
\oauthor{\bsnm{Shirek}, \binits{G.}},
\oauthor{\bsnm{Lindberg}, \binits{D.M.}},
\oauthor{\bsnm{Moulton}, \binits{S.L.}}:
Using deep learning and natural language processing models to detect child
  physical abuse.
Journal of Pediatric Surgery
(2021)
\end{botherref}
\endbibitem

\bibitem[\protect\citeauthoryear{Brownlee and
  Berthelsen}{2006}]{brownlee2006personal}
\begin{barticle}
\bauthor{\bsnm{Brownlee}, \binits{J.}},
\bauthor{\bsnm{Berthelsen}, \binits{D.}}:
\batitle{Personal epistemology and relational pedagogy in early childhood
  teacher education programs}.
\bjtitle{Early Years}
\bvolume{26}(\bissue{1}),
\bfpage{17}--\blpage{29}
(\byear{2006})
\end{barticle}
\endbibitem

\bibitem[\protect\citeauthoryear{Williams
  et~al.}{2019a}]{williams2019artificial}
\begin{bchapter}
\bauthor{\bsnm{Williams}, \binits{R.}},
\bauthor{\bsnm{Park}, \binits{H.W.}},
\bauthor{\bsnm{Breazeal}, \binits{C.}}:
\bctitle{A is for artificial intelligence: the impact of artificial
  intelligence activities on young children's perceptions of robots}.
In: \bbtitle{Proceedings of the 2019 CHI Conference on Human Factors in
  Computing Systems},
pp. \bfpage{1}--\blpage{11}
(\byear{2019})
\end{bchapter}
\endbibitem

\bibitem[\protect\citeauthoryear{Williams et~al.}{2019b}]{williams2019popbots}
\begin{bchapter}
\bauthor{\bsnm{Williams}, \binits{R.}},
\bauthor{\bsnm{Park}, \binits{H.W.}},
\bauthor{\bsnm{Oh}, \binits{L.}},
\bauthor{\bsnm{Breazeal}, \binits{C.}}:
\bctitle{Popbots: Designing an artificial intelligence curriculum for early
  childhood education}.
In: \bbtitle{Proceedings of the AAAI Conference on Artificial Intelligence},
vol. \bseriesno{33},
pp. \bfpage{9729}--\blpage{9736}
(\byear{2019})
\end{bchapter}
\endbibitem

\bibitem[\protect\citeauthoryear{Jin}{2019}]{jin2019study}
\begin{bchapter}
\bauthor{\bsnm{Jin}, \binits{L.}}:
\bctitle{Study on influences of artificial intelligence era on early childhood
  family education in china}.
In: \bbtitle{Journal of Physics: Conference Series},
vol. \bseriesno{1302},
p. \bfpage{032043}
(\byear{2019})
\end{bchapter}
\endbibitem

\bibitem[\protect\citeauthoryear{Strickland et~al.}{1996}]{strickland1996brief}
\begin{barticle}
\bauthor{\bsnm{Strickland}, \binits{D.}},
\bauthor{\bsnm{Marcus}, \binits{L.M.}},
\bauthor{\bsnm{Mesibov}, \binits{G.B.}},
\bauthor{\bsnm{Hogan}, \binits{K.}}:
\batitle{Brief report: Two case studies using virtual reality as a learning
  tool for autistic children}.
\bjtitle{Journal of autism and developmental disorders}
\bvolume{26}(\bissue{6}),
\bfpage{651}--\blpage{659}
(\byear{1996})
\end{barticle}
\endbibitem

\bibitem[\protect\citeauthoryear{Strickland}{1996}]{strickland1996virtual}
\begin{barticle}
\bauthor{\bsnm{Strickland}, \binits{D.}}:
\batitle{A virtual reality application with autistic children}.
\bjtitle{Presence: Teleoperators \& Virtual Environments}
\bvolume{5}(\bissue{3}),
\bfpage{319}--\blpage{329}
(\byear{1996})
\end{barticle}
\endbibitem

\bibitem[\protect\citeauthoryear{Baker et~al.}{2019}]{baker2019challenges}
\begin{barticle}
\bauthor{\bsnm{Baker}, \binits{R.S.}}, \betal:
\batitle{Challenges for the future of educational data mining: The baker
  learning analytics prizes}.
\bjtitle{Journal of Educational Data Mining}
\bvolume{11}(\bissue{1}),
\bfpage{1}--\blpage{17}
(\byear{2019})
\end{barticle}
\endbibitem

\bibitem[\protect\citeauthoryear{Razaulla
  et~al.}{2022}]{razaulla2022integration}
\begin{bchapter}
\bauthor{\bsnm{Razaulla}, \binits{S.M.}},
\bauthor{\bsnm{Pasha}, \binits{M.}},
\bauthor{\bsnm{Farooq}, \binits{M.U.}}:
\bctitle{Integration of machine learning in education: Challenges, issues and
  trends}.
In: \bbtitle{Machine Learning and Internet of Things for Societal Issues},
pp. \bfpage{23}--\blpage{34}
(\byear{2022})
\end{bchapter}
\endbibitem

\bibitem[\protect\citeauthoryear{Webb et~al.}{2021}]{webb2021machine}
\begin{barticle}
\bauthor{\bsnm{Webb}, \binits{M.E.}},
\bauthor{\bsnm{Fluck}, \binits{A.}},
\bauthor{\bsnm{Magenheim}, \binits{J.}},
\bauthor{\bsnm{Malyn-Smith}, \binits{J.}},
\bauthor{\bsnm{Waters}, \binits{J.}},
\bauthor{\bsnm{Desch{\^e}nes}, \binits{M.}},
\bauthor{\bsnm{Zagami}, \binits{J.}}:
\batitle{Machine learning for human learners: opportunities, issues, tensions
  and threats}.
\bjtitle{Educational Technology Research and Development}
\bvolume{69}(\bissue{4}),
\bfpage{2109}--\blpage{2130}
(\byear{2021})
\end{barticle}
\endbibitem

\bibitem[\protect\citeauthoryear{Li et~al.}{2022}]{li2022interpretable}
\begin{barticle}
\bauthor{\bsnm{Li}, \binits{X.}},
\bauthor{\bsnm{Xiong}, \binits{H.}},
\bauthor{\bsnm{Li}, \binits{X.}},
\bauthor{\bsnm{Wu}, \binits{X.}},
\bauthor{\bsnm{Zhang}, \binits{X.}},
\bauthor{\bsnm{Liu}, \binits{J.}},
\bauthor{\bsnm{Bian}, \binits{J.}},
\bauthor{\bsnm{Dou}, \binits{D.}}:
\batitle{Interpretable deep learning: Interpretation, interpretability,
  trustworthiness, and beyond}.
\bjtitle{Knowledge and Information Systems}
\bvolume{64}(\bissue{12}),
\bfpage{3197}--\blpage{3234}
(\byear{2022})
\end{barticle}
\endbibitem

\bibitem[\protect\citeauthoryear{{National Education Association and
  others}}{2008}]{national2008access}
\begin{botherref}
\oauthor{\bsnm{{National Education Association and others}}}:
Access, adequacy, and equity in education technology: Results of a survey of
  america’s teachers and support professionals on technology in public
  schools and classrooms.
National Education Association
(2008)
\end{botherref}
\endbibitem

\bibitem[\protect\citeauthoryear{{Office of Technology
  Assessment}}{1995}]{office1995teachers}
\begin{botherref}
\oauthor{\bsnm{{Office of Technology Assessment}}}:
Teachers and technology: Making the connection.
Report OTA-EHR-616
(1995)
\end{botherref}
\endbibitem

\bibitem[\protect\citeauthoryear{{US Department of
  Education}}{2003}]{us2003federal}
\begin{botherref}
\oauthor{\bsnm{{US Department of Education}}}:
Federal funding for educational technology and how it is used in the classroom:
  A summary of findings from the Integrated Studies of Educational Technology.
Office of the Under Secretary, Policy and Program Studies Service Washington,
  DC
(2003)
\end{botherref}
\endbibitem

\bibitem[\protect\citeauthoryear{Lei and Zhao}{2007}]{lei2007technology}
\begin{barticle}
\bauthor{\bsnm{Lei}, \binits{J.}},
\bauthor{\bsnm{Zhao}, \binits{Y.}}:
\batitle{Technology uses and student achievement: A longitudinal study}.
\bjtitle{Computers \& Education}
\bvolume{49}(\bissue{2}),
\bfpage{284}--\blpage{296}
(\byear{2007})
\end{barticle}
\endbibitem

\bibitem[\protect\citeauthoryear{Dani and Koenig}{2008}]{dani2008technology}
\begin{barticle}
\bauthor{\bsnm{Dani}, \binits{D.E.}},
\bauthor{\bsnm{Koenig}, \binits{K.M.}}:
\batitle{Technology and reform-based science education}.
\bjtitle{Theory into Practice}
\bvolume{47}(\bissue{3}),
\bfpage{204}--\blpage{211}
(\byear{2008})
\end{barticle}
\endbibitem

\bibitem[\protect\citeauthoryear{Schroeder et~al.}{2007}]{schroeder2007meta}
\begin{barticle}
\bauthor{\bsnm{Schroeder}, \binits{C.M.}},
\bauthor{\bsnm{Scott}, \binits{T.P.}},
\bauthor{\bsnm{Tolson}, \binits{H.}},
\bauthor{\bsnm{Huang}, \binits{T.-Y.}},
\bauthor{\bsnm{Lee}, \binits{Y.-H.}}:
\batitle{A meta-analysis of national research: Effects of teaching strategies
  on student achievement in science in the united states}.
\bjtitle{Journal of Research in Science Teaching}
\bvolume{44}(\bissue{10}),
\bfpage{1436}--\blpage{1460}
(\byear{2007})
\end{barticle}
\endbibitem

\bibitem[\protect\citeauthoryear{Songer}{2007}]{songer2007digital}
\begin{botherref}
\oauthor{\bsnm{Songer}, \binits{N.B.}}:
Digital resources versus cognitive tools: A discussion of learning science with
  technology.
Handbook of research on science education,
471--491
(2007)
\end{botherref}
\endbibitem

\bibitem[\protect\citeauthoryear{Guzey and
  Roehrig}{2012}]{guzey2012integrating}
\begin{barticle}
\bauthor{\bsnm{Guzey}, \binits{S.S.}},
\bauthor{\bsnm{Roehrig}, \binits{G.H.}}:
\batitle{Integrating educational technology into the secondary science
  teaching}.
\bjtitle{Contemporary Issues in Technology and Teacher Education}
\bvolume{12}(\bissue{2}),
\bfpage{162}--\blpage{183}
(\byear{2012})
\end{barticle}
\endbibitem

\end{thebibliography}

\end{document}